\documentclass[10pt,journal,compsoc]{IEEEtran}
\ifCLASSOPTIONcompsoc
  \usepackage[nocompress]{cite}
\else
  \usepackage{cite}
\fi
\ifCLASSINFOpdf
  \usepackage[pdftex]{graphicx}
  \graphicspath{{figures/}{pictures/}{images/}{./}} % where to search for the images
\else
\fi
\usepackage{custom}
\usepackage{mathtools}
\usepackage{hyperref}
\usepackage{dblfloatfix}    % To enable figures at the bottom of page
\usepackage{float}
\floatstyle{plaintop}
\restylefloat{table}
\usepackage{multirow}
\usepackage{tabu}

\newcommand{\hl}{\ell}

\DeclareMathAlphabet{\mathcal}{OMS}{cmsy}{m}{n}
\def \smag {$\mathcal{S}_\text{mag}$\xspace}
\def \slvl {$\mathcal{S}_\text{lvl}$\xspace}
\def \sbit {$\mathcal{S}_\text{bit}$\xspace}
\def \swav {$\mathcal{S}_\text{wav}$\xspace}

\def \scen {$\mathcal{S}_\text{cen}$\xspace}
\def \ssbr {$\mathcal{S}_\text{lcf}$\xspace}

\def \wsb {\textbf{\textit{S}}}
\def \wwb {\textbf{\textit{W}}}

\newcommand{\mesh}{\textsc{AMM}\xspace}
\newcommand{\zfp}{\textsc{zfp}\xspace}

\usepackage{subcaption}
\usepackage{xspace}

\newcommand{\etal} {{\textit{et al.}}\xspace}

\newcommand{\ie} {{\textit{i.e.},}\xspace}
\newcommand{\eg} {{\textit{e.g.},}\xspace}

\begin{document}
%\title{Bare Advanced Demo of IEEEtran.cls for\\ IEEE Computer Society Journals}
\title{\mesh: Adaptive Multilinear Meshes}

\author{%
 Harsh Bhatia, 
 Duong Hoang, 
 Nate Morrical,
 Valerio Pascucci, 
 Peer-Timo Bremer, 
 and Peter Lindstrom \vspace{-0.3em}
\IEEEcompsocitemizethanks{\IEEEcompsocthanksitem Bhatia, Bremer, and 
Lindstrom are with the Center for Applied Scientific Computing, Lawrence 
Livermore National Lab., Livermore, CA, 94550.%\protect\\
\IEEEcompsocthanksitem Hoang, Morrical, and Pascucci are with the
Scientific Computing and Imaging Institute, The University of Utah, 
Salt Lake City, UT, 
84112.}}

\markboth{Bhatia \MakeLowercase{\etal}: \mesh: Adaptive Multilinear Meshes}%
{Bhatia \MakeLowercase{\etal}: \mesh: Adaptive Multilinear Meshes}



\IEEEtitleabstractindextext{%
\begin{abstract}
Adaptive representations are increasingly indispensable for reducing the
in-memory and on-disk footprints of large-scale data.
Usual solutions are designed broadly along two themes: reducing data precision,
\eg through compression, or adapting data resolution, \eg using spatial
hierarchies.
Recent research suggests that combining the two approaches, \ie adapting both
resolution and precision simultaneously, can offer significant gains over using
them individually.
However, there currently exist no practical solutions to creating and
evaluating such representations at scale.
In this work, we present a new \textit{resolution-precision-adaptive
representation} to support hybrid data reduction schemes and offer an interface
to existing tools and algorithms.
Through novelties in spatial hierarchy, our representation, \emph{Adaptive
Multilinear Meshes} (\mesh), provides considerable reduction in the mesh size.
\mesh creates a piecewise multilinear representation of uniformly sampled
scalar data and can selectively relax or enforce constraints on conformity,
continuity, and coverage, delivering a flexible adaptive representation.
\mesh also supports representing the function using mixed-precision values to
further the achievable gains in data reduction.
We describe a practical approach to creating \mesh incrementally using
arbitrary orderings of data and demonstrate \mesh on six types of resolution
and precision datastreams.
By interfacing with state-of-the-art rendering tools through VTK, we
demonstrate the practical and computational advantages of our representation
for visualization techniques.
With an open-source release of our tool to create \mesh, we make such
evaluation of data reduction accessible to the community, which we hope will
foster new opportunities and future data reduction schemes.

% ----------------------------------------------------------------------------

\vspace{-0.3em}
\end{abstract}

}

\maketitle

% is not selected <OR> if conference mode is selected - because compsoc
% conference papers position the abstract like regular (non-compsoc)
\IEEEdisplaynontitleabstractindextext

% \fi
\IEEEpeerreviewmaketitle

% ---------------------------------------------------------------------------
\ifCLASSOPTIONcompsoc
\IEEEraisesectionheading{\section{Introduction}\label{sec:introduction}}
\else
\section{Introduction}
\label{sec:introduction}
\fi

% ---------------------------------------------------------------------------
\IEEEPARstart{A}{s} scientific datasets continue to grow in size and
complexity, adaptive representations have become key to enabling interactive
analysis and visualization~\cite{DOE:2013}.  Such representations can reduce
the memory footprint and processing costs of large-scale data by orders of
magnitude, often without perceptible degradation of visualization quality or
analysis results~\cite{Laney13}.  However, existing approaches are limited to
either compressed representations of regular grids~\cite{Iverson12, zfp} or
multiresolution structures, such as octrees and k-d
trees~\cite{Shekhar:1996:ODM, Pascucci:2001:GSI}.  The former typically provide
little spatial adaptivity and, thus, do not benefit from the sparsity that is
common in scientific datasets, where often only small portions of space are of
interest.
With few exceptions~\cite{zfp}, data is usually stored uncompressed in memory,
limiting the overall grid resolution.  Spatially adaptive structures overcome
this problem by selectively refining regions of interest, resulting in a
smaller memory footprint.  However, many multiresolution representations imply
structural constraints, causing unnecessary refinement in unimportant regions,
especially for odd-sized domains like skinny rectangles or L-shapes.  Finally,
both approaches are limited to their respective notions of fidelity, adapting
either numerical precision or spatial resolution.

The recent work of Hoang \etal~\cite{Hoang:2019, Hoang2021} has demonstrated
that combining both concepts --- adapting both resolution and precision
simultaneously --- can provide significant advantages in reduced storage and/or
improved accuracy.  Unfortunately, there currently exists no data structure
that can easily exploit this idea, as precision-based compression methods do
not provide spatial adaptivity, whereas multiresolution grids do not generally
adapt in precision.

%\teaser{
\begin{figure*}[ht]
\centering
\begin{picture}(170,115)
\put(0,0){\includegraphics[height=0.8in]
 {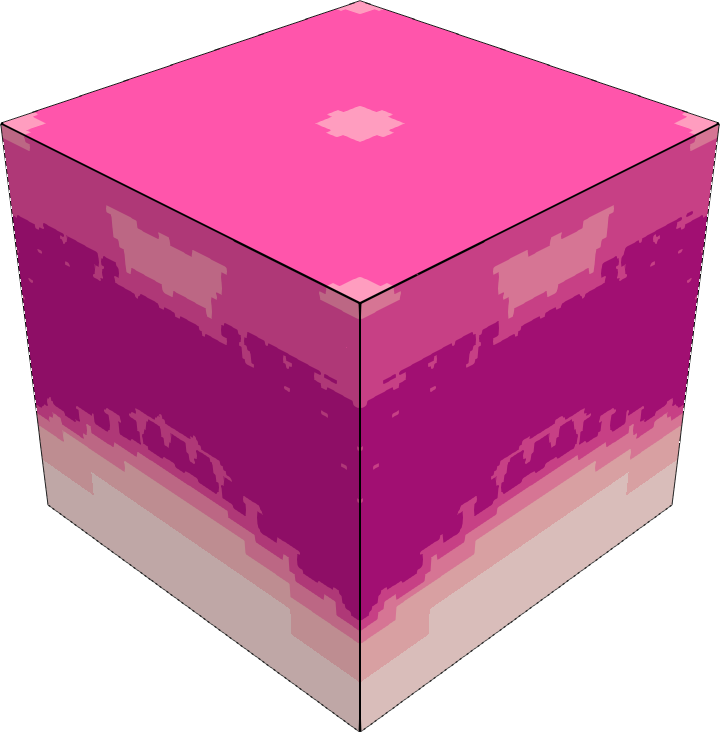}}
\put(0,60){\includegraphics[height=0.8in]
{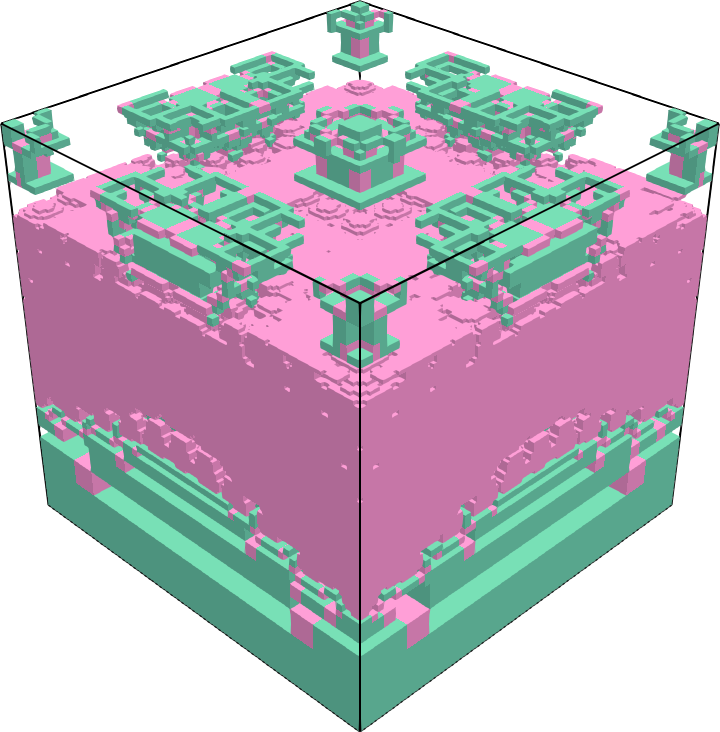}}
\put(58,4){\includegraphics[height=1.6in]{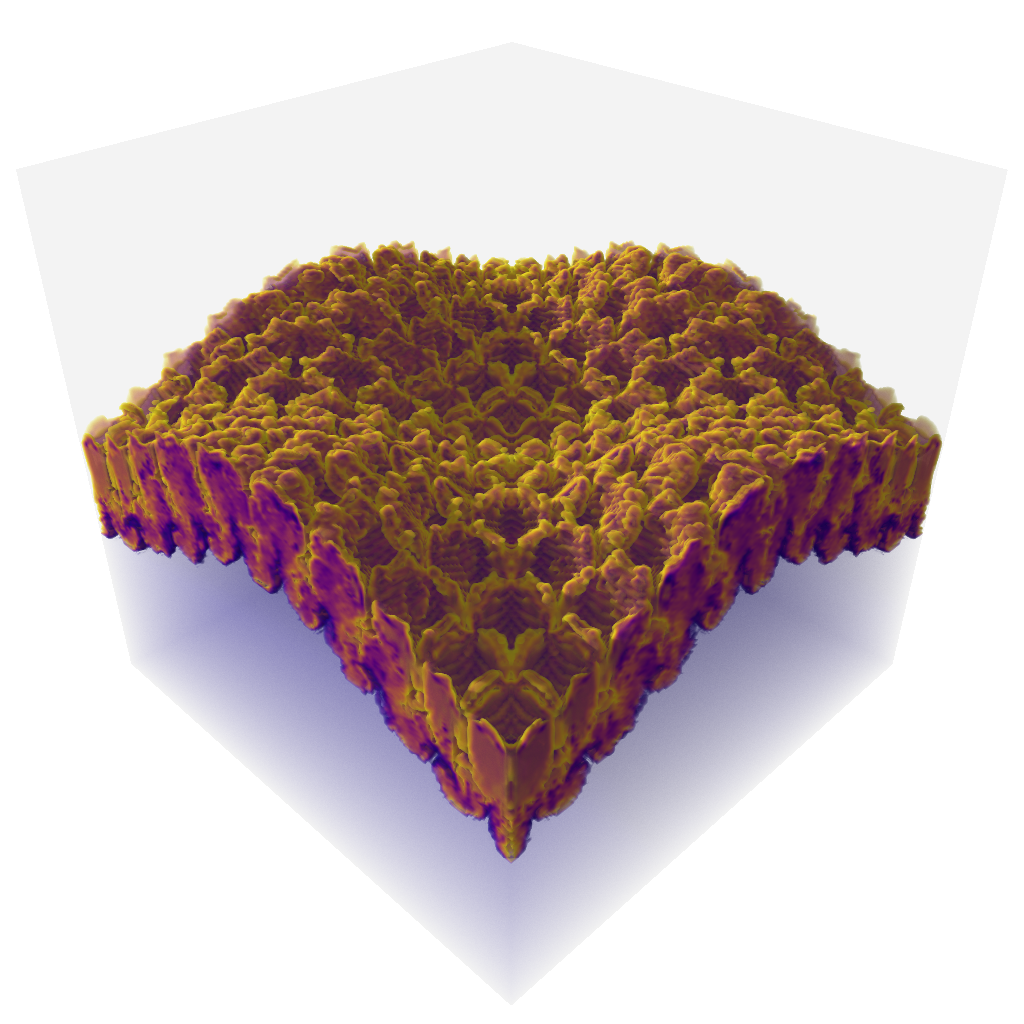}}
\put(45,2){Reduced with \ssbr (41 MB)}
\end{picture}
% --------------------------------------------------------------------------------------------
\hspace{0.5em}
\begin{picture}(118,115)
\put(0,4){\includegraphics[height=1.6in]{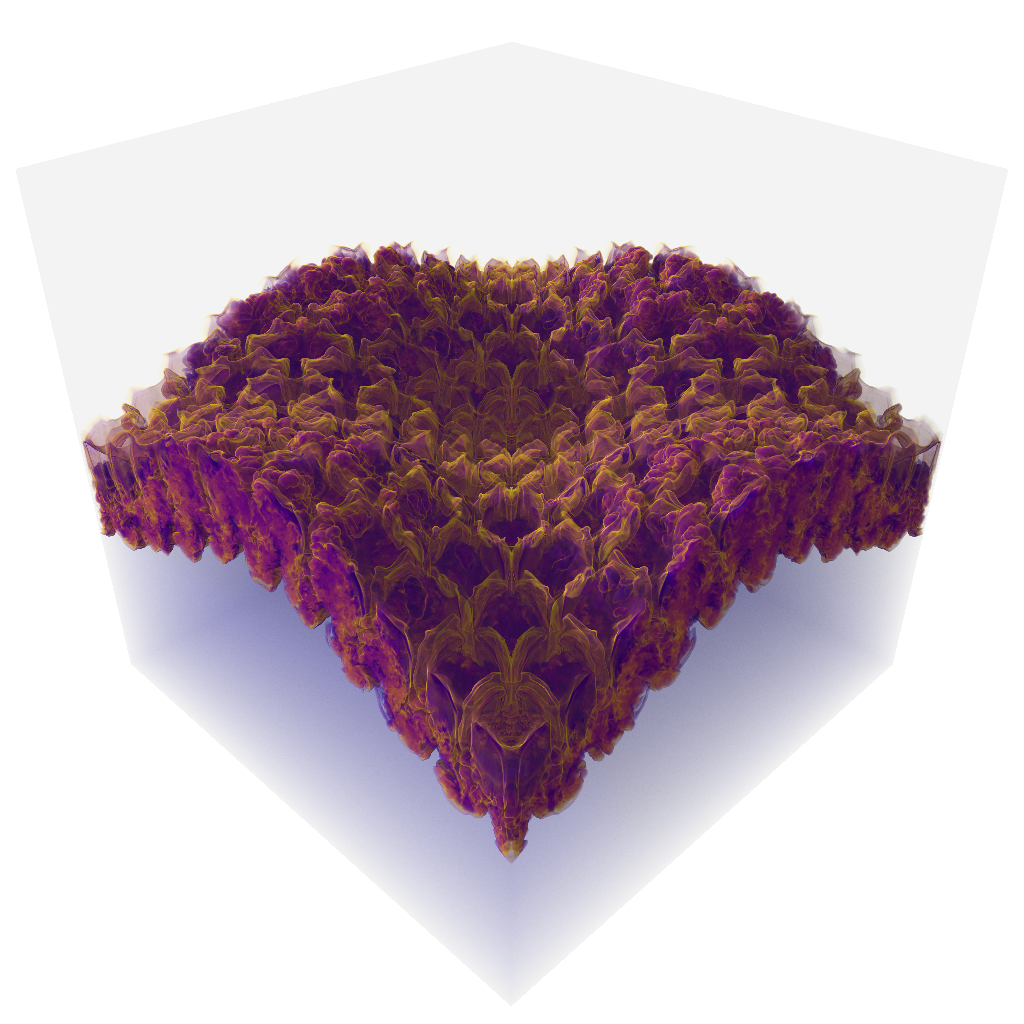}}
\put(18,2){Full data (30 GB)}
\end{picture}
\hspace{0.05em}
% --------------------------------------------------------------------------------------------
\begin{picture}(170,115)
\put(0,4){\includegraphics[height=1.6in]{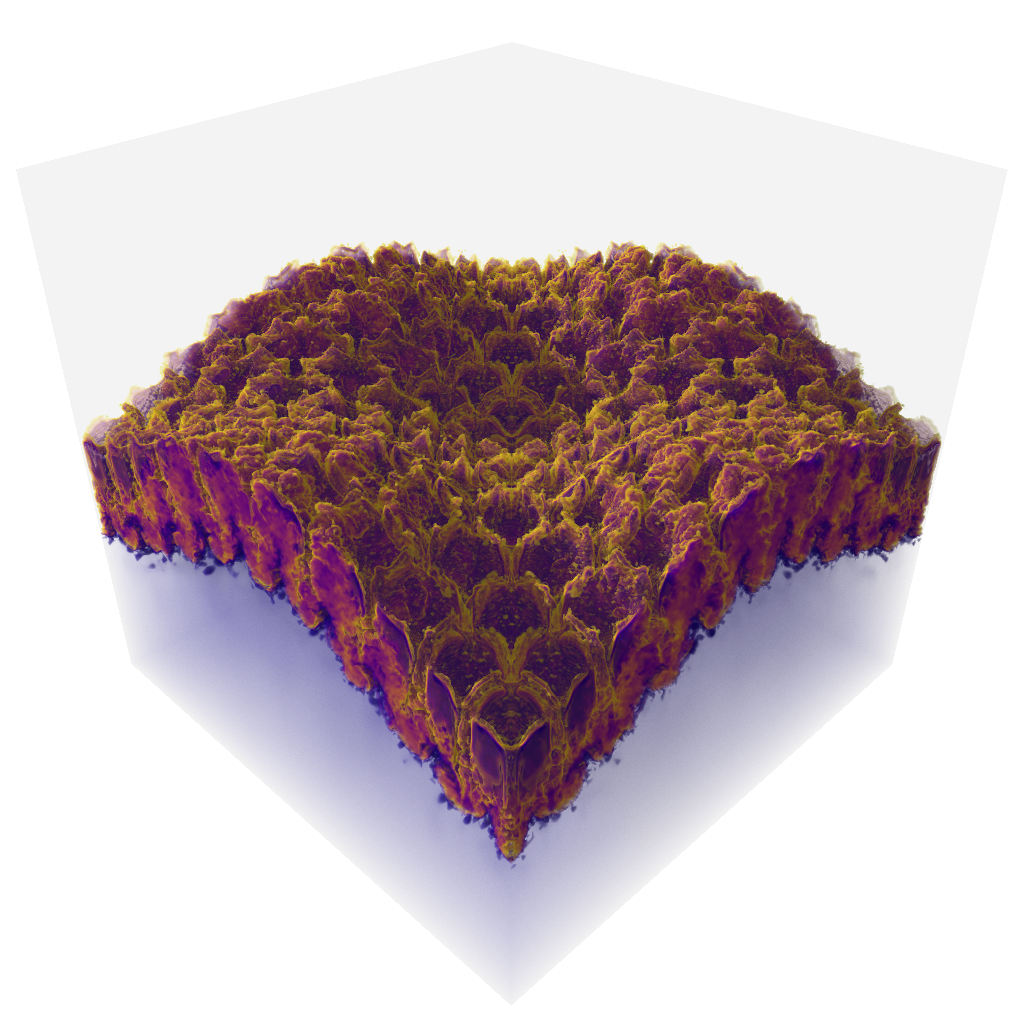}}
\put(116,0){\includegraphics[height=0.8in]
 {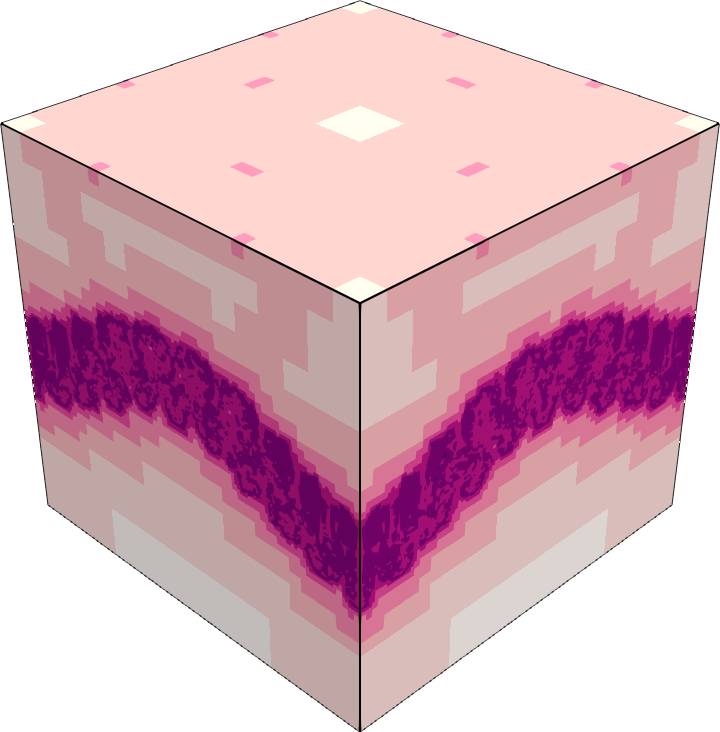}}
\put(116,60){\includegraphics[height=0.8in]
 {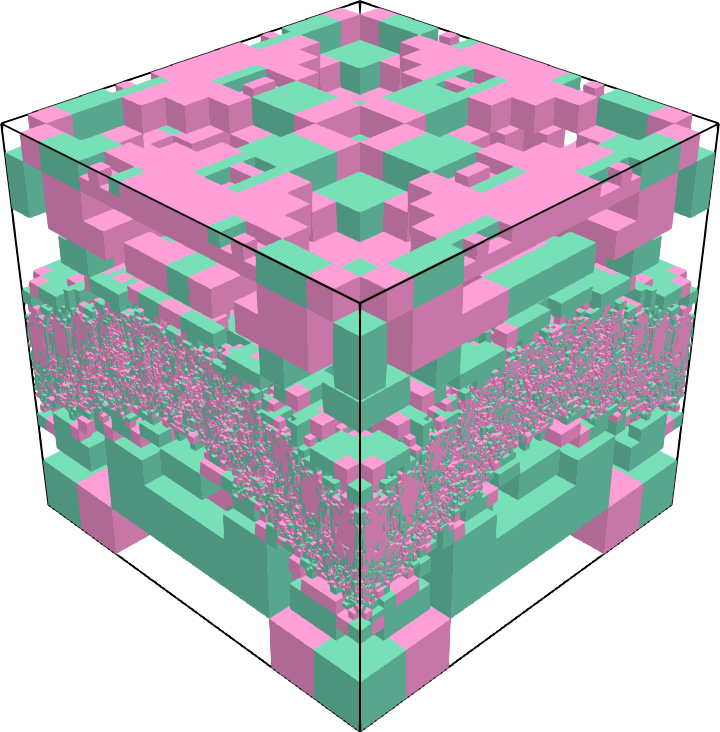}}
\put(8,2){Reduced with \scen (159 MB)}
\end{picture}
\caption{\emph{Adaptive Multilinear Meshes} (\mesh) are a
\emph{resolution-precision-adaptive} representation of uniformly sampled scalar
fields. \mesh facilitates a flexible spatial hierarchy --- to adapt in space
--- and mixed-precision storage --- to adapt in precision. \mesh allows
creating and comparing data reduction through arbitrary resolution-precision
streams and facilitates interfacing with state-of-the-art tools for
visualization and analysis.  The figure visualizes data reduction after
ingesting 8 MB of data through two streams, \ssbr and \scen.  The smaller
renderings on the sides visualize the types of cells (top; type-1 in green and
type-2 in pink) and depths of cells (bottom; deeper cells are darker) in
\mesh's spatial hierarchy.}
\label{fig:spatial:rm}
\label{fig:teaser}
\vspace{-1em}
\end{figure*}
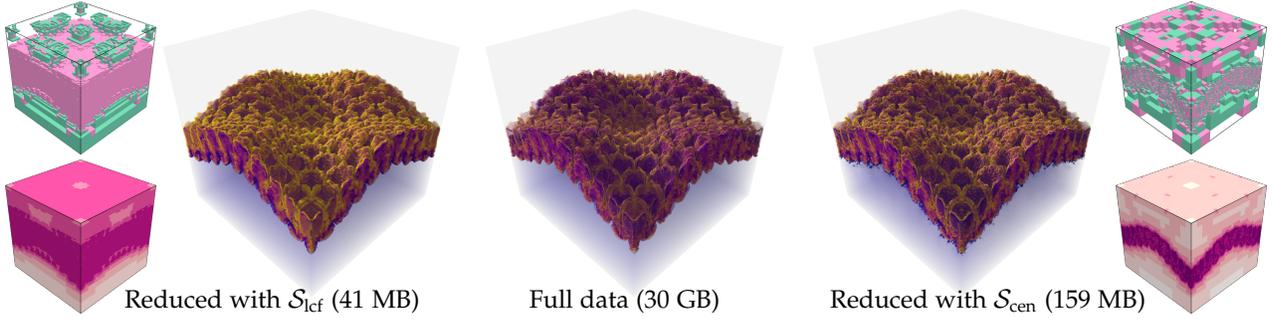
%}

Given a lossy data reduction scheme, we address the challenge of representing
the resulting data faithfully and without further loss of information, while
minimizing the number of cells and vertices in the representation as well as
the number of bits to store for each vertex.  Unlike existing multiresolution
approaches, we enable varying the precision spatially, \eg features of interest
can be represented with more precision bits than the rest.  Unlike common
compression techniques, we allow adaptive spatial refinement, \ie prevent
representation of large regions of uninteresting space.  In this way, our
framework leverages both types of reduction to provide gains in the memory
footprint not realizable by either type of approach individually.

Furthermore, our approach aims not necessarily for the best compression ratio,
but rather for the flexibility of generating, storing, and accessing data at
reduced resolution and precision.  Since future data will likely increase much
more in resolution than in precision, we anticipate that these unique
capabilities of our framework will become increasingly more essential to the
design, evaluation, and comparison of different data reduction strategies. This
work offers an important step toward a potential synergy between resolution
adaptivity and precision adaptivity in the future.

The ability to ``incrementally'' update the reduced representation through
streaming of partial data (\eg in a client-server setting or simply choosing
when to stop) offers significant benefits.  First, downstream processing can
start without waiting for a potentially long decompression step.  Second, with
fewer vertices and fewer bits per vertex, such downstream tasks can be
performed with improved memory efficiency and, thus, can finish significantly
faster to provide the user with approximate results that converge over time.
Therefore, we focus on scalar fields defined on uniform grids and introduce an
adaptive mesh that can be constructed \emph{incrementally} from
\emph{arbitrarily} ordered \textit{datastreams} (sequences of complete or
partial values).

Built upon a new type of tree, our representation --- \emph{Adaptive
Multilinear Meshes} (AMM), see \autoref{fig:teaser} --- utilizes rectangular
cuboidal cells to represent multilinear data. \mesh does not require a complete
global coverage of space but performs minimal refinement to allow isolated
regions of interest without needing all surrounding elements, which minimizes
the size of the representation.  \mesh representations can be created using an
easy-to-use, open-source tool and exported to the community-standard
VTK~\cite{vtk} meshes, making it straightforward to adopt for commonly used
visualization and analysis tasks, \eg with standard and generic tools, such as
Paraview~\cite{paraview} and VisIt~\cite{visit}, or with specific rendering
approaches~\cite{Morrical2019, Morrical_TVCG_2020, wald_ospray_2017}.

\mesh can be constructed through superpositioning piecewise multilinear
C\textsuperscript{0} fields in overlapping subgrids.  Specifically, we build
upon the recent work by Weiss and Lindstrom~\cite{Weiss:2016}, who use an
octree-based approach to encode data using tensor products of linear B-spline
wavelets~\cite{Cohen92}, where the basis functions are defined on piecewise
multilinear, spatially overlapping stencils.  Compared to their work, our
representation uses a more-flexible spatial hierarchy that further reduces the
mesh considerably, represents vertex values using a mixed-precision scheme, and
supports progressive refinement.  \mesh can also be created by directly reading
in arbitrary axis-aligned multilinear cells with values at the corresponding
corners, representing continuous as well as discontinuous functions.

Finally, \mesh can be constructed through incremental updates using
\emph{datastreams} that update the representation with additional resolution,
additional precision, or both. A key novelty in \mesh is that it ingests
datastreams with  \textit{arbitrary} orderings.  Prior work~\cite{Hoang:2019}
has shown that different analysis tasks may prefer different types of
datastreams --- some may need more resolution (\eg gradient computation)
whereas others more precision (\eg histogram computation).  All such
datastreams, and others that arbitrarily combine resolution and precision, are
supported by our representation.  In this way, \mesh offers a  framework to
evaluate in a consistent manner the impact and performance of different data
reduction strategies (see \autoref{fig:teaser}).

\para{Contributions.} 
We present \mesh, a new \emph{resolution-precision-adaptive representation}
that can ingest \emph{arbitrary data orderings} and provide an \emph{interface
to existing tools and algorithms}. Specifically, we make the following
contributions:
\begin{itemize}[leftmargin=*] \itemsep0mm

\item \mesh is a compact, adaptive representation of piecewise multilinear
scalar fields.  \mesh reduces the number of cells and vertices by supporting
(1) arbitrary axis-aligned splits through the center to create
\emph{rectangular cuboidal cells}, and (2) \emph{incomplete splitting of cells}
and their optimal resolution with lazy updates.  Using a tensor-product wavelet
basis, we extend the work of Weiss and Lindstrom~\cite{Weiss:2016} and reduce
the size of equivalent representations.

\item \mesh uses a mixed-precision, block-based encoding of values at the
vertices, providing superior data reduction than spatial adaptivity alone.

\item \mesh is the first-of-its-kind adaptive representation that can be
updated progressively using arbitrarily ordered datastreams,
%and can work with standard pipelines, 
offering new opportunities to explore dynamic and hybrid data reduction
strategies~\cite{Hoang:2019, Hoang2021}.

\item For wavelet transforms that require boundary conditions or are defined
only for square grids of certain dimensions, we present a novel
\emph{linear-lifting} method to extrapolate  the data to avoid discontinuities
and artificially large coefficients at the domain boundary, preventing
unnecessary mesh refinement.

\item We present an open-source implementation of the tool to create and export
\mesh to standard VTK meshes
(\href{https://github.com/llnl/amm}{github.com/llnl/amm}), allowing for its
wider adoption and use with state-of-the-art tools.

\end{itemize}

\vspace{-0.5em}
\section{Background and Related Work}

\para{Tree-based hierarchies,} such as k-d trees~\cite{Fogal10:gpu, Woodring11}
and octrees~\cite{JACKINS1980249}, are among the most popular
spatial-subdivision schemes due to their simplicity.  Octrees, in particular,
have found widespread adoption across diverse domains.  They are especially
useful when the data contains sparse details, so the smooth-varying regions can
be stored at coarser octree levels, thereby reducing storage, \eg using sparse
voxel octrees~\cite{Crassin09:gigavoxels,Gobbetti2008,large-scale-volume}.
Recent approaches have made modifications to traditional octrees to leverage
modern computational architectures.  For example,
OpenVDB~\cite{Museth:2013:vdb} increases the tree branching factor to ensure
that sibling nodes are stored contiguously in a cache-friendly layout;
SPGrid~\cite{Setaluri:2014} stores octree levels separately as sparse and
nonoverlapping grids, taking advantage of virtual memory handling capacities in
modern operating systems.  Similar to SPGrid, we store per-level vertices
separately and aggregate them on demand, but use hash tables instead of the
OS's virtual memory to handle sparsity.

\para{Adaptive mesh refinement (AMR)}~\cite{Berger1989:AMR,Dubey14} is another
popular class of multiresolution schemes, especially for simulations.  In
structured AMR, each resolution level consists of a set of nonoverlapping
uniform grids.  As the grids can be placed arbitrarily, fine-resolution grids
can be used to quickly resolve fine details.  An AMR mesh can be either
vertex-centered or cell-centered, depending upon where the data points are
stored.  Although the cell-centered approach is more common, visualizing the
resulting mesh requires preprocessing steps, such as remeshing and
stitching~\cite{Weber2001ExtractionOC, Moran11} and \textit{ad hoc}
interpolation~\cite{Ljung2006MultiresolutionII, Wang2018CPUIR}.  In contrast,
our representation is vertex-based, with a principled method for (multilinear)
interpolation.

\para{Wavelets} provide a rigorous framework for multiresolution decomposition
that is also amenable for data reduction and, thus, is commonly used in
visualization frameworks for large data~\cite{vapor, treib,
li-2017-spatiotemporal, woodring-11-revisiting}.  There also have been works
that study data simplification and approximation using wavelet-based
subdivision schemes~\cite{Bertram04, Gross96}, as well as representing
multilinear functions using a minimal number of mesh elements to reduce memory
footprints~\cite{Weiss:2016,SCI:Lin2004a}.

Linsen \etal~\cite{SCI:Lin2004a} subdivide cubes into simplices and use linear
interpolation for function reconstruction.  Weiss and
Lindstrom~\cite{Weiss:2016} later demonstrated that using multilinear
interpolation produces superior quality meshes with respect to approximation
error.  Conceptually, our representation is most similar to Weiss and
Lindstrom's approach, although \mesh  is a more general representation.  In
particular, our framework utilizes rectangular cuboidal cells as opposed to
cube-shaped cells (of a standard octree hierarchy) exclusively, thereby
significantly reducing memory requirements.

We use linear B-spline wavelets~\cite{Cohen92} to populate \mesh, since
multilinear interpolants are at the foundation of many visualization
techniques~\cite{Lorensen:1987:MC, CIGNONI2000, Nielson03, Ament10}.  Sparse
grids~\cite{zenger1991sparse, garcke2012sparse}, a common solution for
circumventing the curse of dimensionality when solving partial differential
equations, also form a piecewise multilinear multiresolution basis and would
fit into our hierarchical framework.

\para{Compression techniques.} Beside hierarchical decomposition, data
compression is effective at reducing data sizes.  Prominent compression methods
for scientific data, such as \zfp~\cite{zfp}, SZ~\cite{tao2017significantly},
and TTHRESH~\cite{ballester2019tthresh}, focus more on reducing data precision
through transform/prediction steps followed by quantization of the resulting
coefficients or residuals.  Compression techniques that reduce resolution have
also been investigated.  Recently, Ainsworth \etal~\cite{Ainsworth2018,
Ainsworth2019} presented a multigrid approach that offers compression at
different levels of the hierarchy.  Zhao \etal~\cite{ZhaoSZ} introduced a
multilevel spline-based approach for lossy compression.  However, these
techniques can support only a one-shot data reduction whereas \mesh is designed
to handle arbitrary datastreams and incremental updates.  Wavelet-based
compression~\cite{sbhp, speck, spiht, jpeg2k, li2019vapor, Hoang2021} often
produces progressive bitstreams that improve the data resolution as well as
precision.  However, such ideas do not provide any adaptive in-memory
representation that could be constructed from such streams.  This gap is filled
by \mesh, which complements these wavelet coders by offering a compact
in-memory representation for data approximations constructed from compressed
bitstreams.

It is not meaningful to directly compare compact meshes, such as the one by
Weiss and Lindstrom~\cite{Weiss:2016} or \mesh, with pure compression
techniques, since the former have overheads of embedded data structures needed
to support resolution adaptivity, and they also need to put more emphasis on
data access speed instead of data reduction alone.  Rather than focusing on
providing unbalanced comparisons, we consider \mesh and compression to be
complementary approaches and highlight the flexibility offered by \mesh that
may be used to leverage compression in the future.

\vspace{-0.5em}
\subsection{Octree and Regular Refinement}
\label{sec:octree}

\mesh is based on an advanced tree representation that builds upon the idea of
regular refinement of octrees, which are a common way of defining a spatial
hierarchy over regular grids.
With a slight abuse of notation, we use octree to refer to both octrees (in 3D)
and quadtrees (in 2D).
An \emph{octree} is a spatial hierarchy defined on a $d$-dimensional space; it
is defined as a collection of \emph{$d$-cubes}, which form the \emph{nodes} of
the tree.  Under regular refinement, a \emph{parent node} (a $d$-cube) is
decomposed into $2^d$ \emph{child nodes} (also  $d$-cubes), by splitting each
dimension in half.  The \emph{root node} covers the entire domain. The nodes
that have no children are referred to as \emph{leaf nodes}, whereas all others
are \emph{internal nodes}. The number of refinements needed to obtain a given
node from the root node defines the \emph{depth} (also referred to as the
\emph{level}) of the node.  An adaptive representation can be obtained by
selectively refining the nodes of interest.
When an octree is defined over a regular grid, such that the vertices of the
grid form the corners of the $d$-cubes, standard representations require the
\emph{size} of the domain (number of vertices in each dimension) to be
$2^L+1$, where $L$ defines the maximum depth of the hierarchy; the size of
a node at depth $\hl$ is given by $s_\hl = 2^{L-\hl}+1$. The midpoints of the
parent node and its facets (\ie edges, faces) form the vertices of the child
nodes, and each child contains one of the vertices of the parent node.
Given a regular cubical mesh defined by an octree, we refer to the
$k$-dimensional faces of the tree as \emph{primal cubes}, and to the
axis-aligned $(d-k)$-cubes defined by connecting the centers of their adjacent
$d$-cubes as \emph{dual cubes}.

\vspace{-0.5em}
\subsection{Multilinear Tensor-Product Wavelets} 
\label{sec:wavelets}

Wavelet transforms create multiscale data representations.  Such
representations are defined by translations and dilations of a \emph{wavelet
basis function}, $\wwb$, which extracts the detail at a given scale, and a
\emph{scaling basis function}, $\wsb$, which captures the coarse representation
after removing the details. Here, $\wwb$ is a high-pass filter whereas $\wsb$
is a low-pass filter that represents the wavelet bases across all remaining
scales.

\para{Lifting scheme for linear B-spline wavelet transform.}
Given a 1D uniform grid of length $2^L+1$, (discrete) wavelet transforms are
often implemented in the form of lifting schemes. The forward lifting transform
associated with linear B-spline wavelets consists of two phases. The first
phase, the \emph{w-lift}, defines the wavelet coefficient for every odd-indexed
vertex on the grid as the prediction residual of the average of its two
immediate neighbors from the given function value.  The second phase, the
\emph{s-lift}, is then optionally applied to preserve the mean of the function.
The s-lift updates the values of the even-indexed vertices by adding a weighted
sum of the values obtained in the w-lift.  Mathematically,
\mbox{w-lift} is \mbox{$\hat f(v_{2i+1}) =  f(v_{2i+1}) - \frac{1}{2} \left[f(v_{2i} )
+ f(v_{2i+2})\right]$}, and
\mbox{s-lift} is \mbox{$\hat f(v_{2i}) =  f(v_{2i}) + \frac{1}{4} \left[\hat
f(v_{2i-1} ) + \hat f(v_{2i+1})\right]$}, 
where, $f$ denotes the input function, $\hat f$ the wavelet (odd-indexed) and
scaling (even-indexed) coefficients, and $v_i$ denotes the indexed vertex in
the grid.
Multiple resolutions of wavelet transforms are obtained by recursively applying
lifting steps to the even-indexed vertices at each level. The inverse wavelet
transform simply inverts these operations.

When only the w-lift step is performed, the resulting wavelet bases satisfy the
Lagrange property, leading to an interpolation of the function, and the
corresponding wavelets are commonly called \emph{interpolating wavelets}. On
the other hand, using both the w-lift and the s-lift leads to
\emph{approximating wavelets}.
% because they preserve the mean value of the function. 
The wavelet synthesis bases for the interpolating and approximating wavelets
correspond to piecewise-linear spatial stencils with nodal values $\wwb^{I} =
[0, 1, 0]$ and $\wwb^{A} = \frac{1}{8} [0, -1, -2, 6, -2, -1, 0]$,
respectively, whereas the stencil corresponding to the scaling function for
both types is $\wsb = \frac{1}{2}[0, 1, 2, 1, 0]$.
Wavelet transforms exhibit the \emph{two-scale relation}~\cite{mallat}, \ie
both the scaling and the wavelet functions at a given scale can be expressed in
terms of (translated and dilated) scaling functions at the next finer scale.
When combined with the piecewise linear nature of the basis functions, the
spatial stencils described above produce a hierarchical representation in terms
of a regular grid, \ie a stencil at a given level can be described in terms of
the vertices at the next (finer) level.

\para{Multilinear tensor-product wavelets in 2D and 3D.}
These ideas can be generalized to higher dimensions by taking the tensor
product of the basis functions.  Shown by Weiss and
Lindstrom~\cite{Weiss:2016}, a basis function defined by the tensor product of
$k$ wavelets and $d-k$ scaling functions (for a $d$-dimensional regular grid)
is associated with the midpoint of a $k$-cube in the $d$-dimensional grid.  For
example, refer to \autoref{fig:stencils} and note that in 2D the tensor product
scaling functions $\wsb_x\wsb_y$ are associated with the vertices of the grid,
the wavelets $\wsb_x\wwb_y$ and $\wwb_x\wsb_y$ with the midpoints of edges, and
the wavelets $\wwb_x\wwb_y$ with midpoints of the square cells. Since the
scaling functions for linear B-spline wavelets correspond to linear
interpolation, their tensor product corresponds to multilinear interpolation.

\vspace{-0.25em}
\section{Adaptive Multilinear Meshes} \label{sec:amm}

\mesh provides flexible adaptivity in representing uniformly gridded scalar
data through a new type of spatial hierarchy that supports more-general
splitting operations as compared to existing tree-based hierarchies, such as
octrees and k-d trees. Our data structure enables significant reduction of the
size of the representation through two novelties: (1) \emph{rectangular
cuboidal cells}, which ensure a tree node is split only along the axes
required, and (2) \emph{improper nodes}, which facilitate partial splitting and
representation of a node.

We draw a distinction between \emph{nodes} of the tree, which define the
spatial hierarchy, and \emph{cells} of the resulting mesh, which are the leaf
nodes of the tree after resolving improper nodes (discussed in
\autoref{sec:amm:improper} and \autoref{sec:create:improper}) that lie
partially or completely inside the given domain (as discussed in
\autoref{sec:amm:treesize}, the tree may cover a larger spatial extent than
that of the given data). The corners of nodes/cells form the \emph{vertices} of
the tree/mesh and are associated with function values. The function can be
reconstructed anywhere using multilinear interpolation of vertices.

%% -----------------------------------------------------------------------------
\begin{figure}[!t]
\centering
\includegraphics[width=\linewidth]{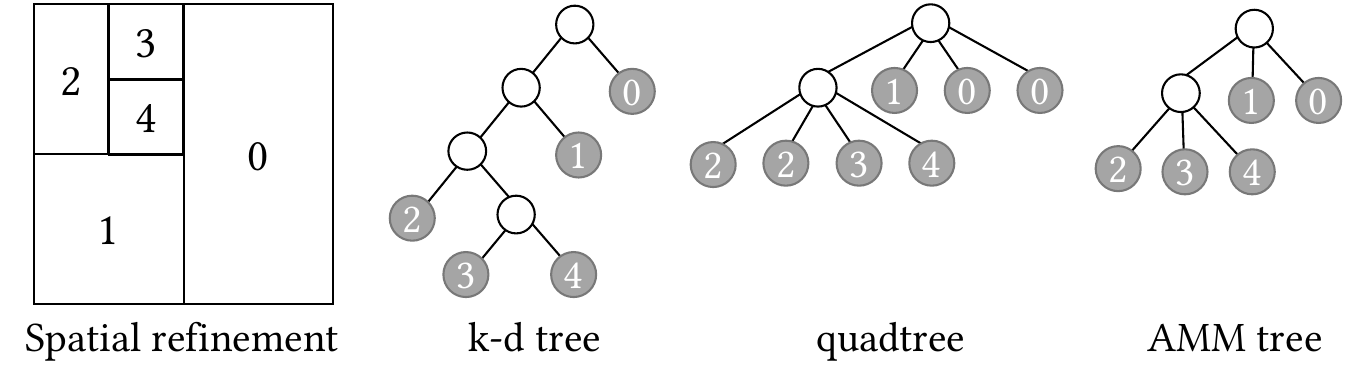}
\vspace{-2em}
\caption{\mesh's tree supports flexible axis-aligned split configurations,
which avoids artificially deep hierarchies (like with k-d trees), or an excessive
number of subdivisions (like with quad/octrees), which leads to redundant cells
and vertices. In this example, the leaf nodes (shaded) in all three trees
represent the same partitioning of space. \label{fig:comparetrees}}
\vspace{-1em}
\end{figure}

\subsection{Spatial Adaptivity Using a New Tree Data Structure} 
\label{sec:amm:tree}

Spatial hiearchies are commonly created as octrees or k-d trees. For the
former, each node (a $d$-cube) is either a leaf node (\ie no child nodes) or is
fully refined along $d$ axes (\ie $2^d$ child nodes). Hence, octrees provide
only one degree of freedom: no refinement or full refinement. Similarly, k-d
trees, which split a spatial region along a hyperplane, also offer only a
binary, axis-based hierarchy, usually alternating through the splitting
dimension. Both types of hierarchy are limited in their spatial adaptivity to
two configurations only.

In contrast, \mesh creates a spatial hierarchy that subdivides a $d$-cube
through its center with respect to any arbitrary combination of axes.  Hence,
the subdivision is restricted neither to a single axis (as in a k-d tree) nor
to all axes (as in an octree). This subdivision flexibility facilitates many
types of refinement configurations, ultimately allowing us to reduce size of
the resulting mesh (see \autoref{fig:comparetrees}).

%% -----------------------------------------------------------------------------
\subsubsection{Sizes of the Tree and Nodes} 
\label{sec:amm:treesize}

Similar to standard approaches, we enforce our spatial hierarchy to sizes that
are powers of two plus one and require the root node to represent a $d$-cube,
\ie equal sizes in all dimensions.
Given data extents $[X\times Y \times Z]$ in 3D, the spatial extent of the tree
and the \emph{size} (number of vertices in each dimension) of the root node are
$2^L+1$, where $L = \lceil \log_2 \left(\max(X,Y,Z)\right) \rceil $ is the
maximum depth of the tree.
We also borrow the regular refinement operation from octrees, which splits a
node (a $d$-cube) with size $s_\hl = 2^{L-\hl}+1$ is split into child nodes
that are also $d$-cubes of size $s_{\hl+1}$. Recall that $\hl$ is called the
depth of the node.

The overhead associated with expanding the spatial extent is negligible, since
when properly constructed (see \autoref{sec:create}), the regions outside the
original domain contain a small number of nodes. Furthermore, the expansion
offers opportunities for efficient storage and representation of the tree
(discussed in \autoref{sec:amm:storage}). Finally, using $2^L+1$ sizes also
offers a way to avoid the boundary artifacts, where otherwise arbitrary
refinement can happen near the boundary of the data domain (see
\autoref{sec:linearlifting}).

%% -----------------------------------------------------------------------------

\begin{figure}[!b]
\centering
\vspace{-1.5em}
\includegraphics[width=\linewidth]{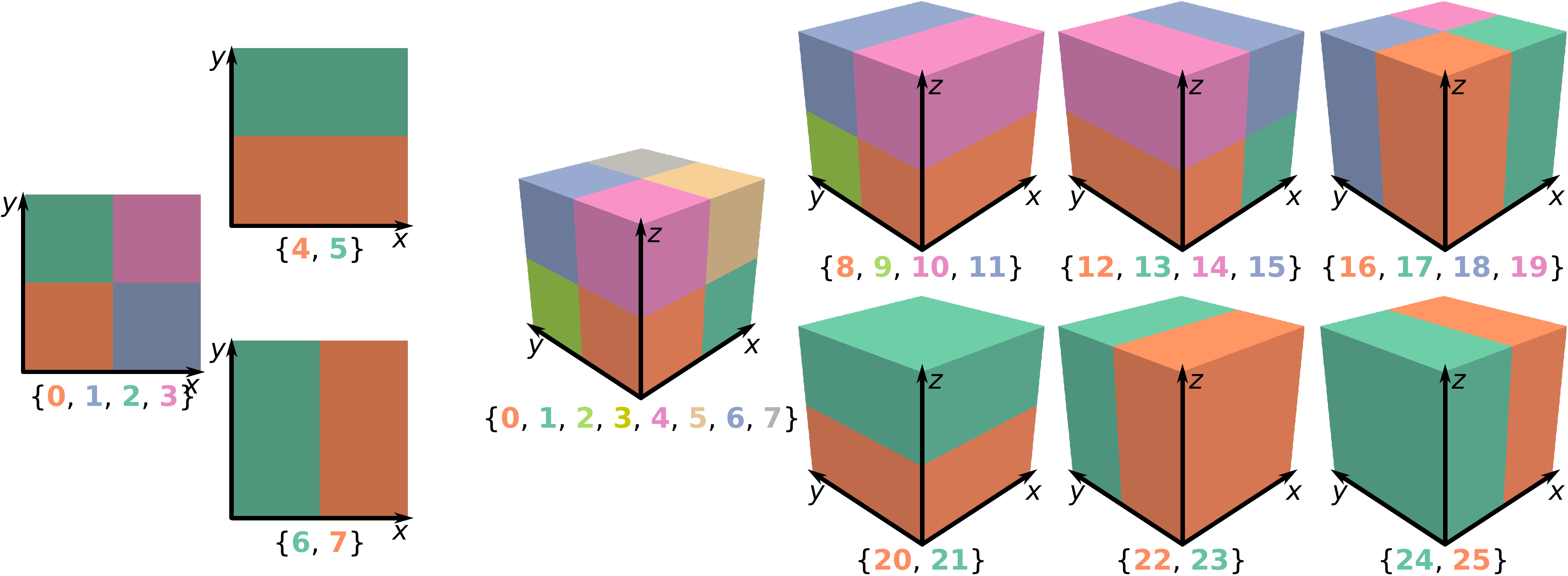}
\vspace{-1.5em}
\caption{\mesh provides flexible spatial adaptivity through arbitrary
combinations of axis-aligned subdivisions of a spatial cube (nodes of the
underlying tree). The figure shows the different types of child nodes (spatial
subdivisions) in 2D and 3D, and their color-corresponded ids.  \mesh can
support any combination of nonoverlapping child nodes, referred to as a
\emph{valid node configuration}, \eg \{0,4,18,25\} and \{10,11,12,13\} in 3D,
and, thus, represent more-general configurations of cells, reducing the memory
footprint.} \label{fig:childnodes}
\vspace{-0.4em}
\end{figure}

%% -----------------------------------------------------------------------------
\subsubsection{Rectangular Cuboidal Leaf Nodes}

Toward the goal of creating as few cells as possible, \mesh's leaf nodes are
allowed to be \emph{rectangular cuboidal} in shape whenever possible (\ie as
large as they can be).  On the other hand, internal nodes are required to have
equal sides (\ie $d$-cubes only) to favor simplicity and efficiency of
traversal.
Specifically, as illustrated in \autoref{fig:childnodes}, there can be 8 and 26
types of leaf nodes in 2D and 3D, respectively, as compared to 4 and 8 types of
``regular'' child nodes ($d$-cubes) for the 2D and 3D cases (quadtrees and
octrees). Intuitively, the types of rectangular nodes can be enumerated by
fixing $k$ dimensions that are not split and splitting the remaining $d-k$
dimensions and counting the combinations of axes that are split and not split.
Here, $k$ can be used to categorize the type of nodes. \mesh leaf nodes can be
\emph{type-0} (all $d$ sides equal to $s_{\hl+1}$), \emph{type-1} (one side,
$s_\hl$, longer than the other $d-1$ sides, $s_{\hl+1}$), and \emph{type-2}
(two equal sides, $s_\hl$, longer than the third, $s_{\hl+1}$); type-2 does not
exist in 2D. Formally, a penultimate level node in \mesh (a $d$-cube) may have
\mbox{$2^d + d \sum_{k=1}^{d-1} {2^{d-k}}$} types of child nodes. Here, the
first term, $2^d$, represents the regular child nodes ($d$-cubes of size
$s_{\hl+1}$), and the term in the summation captures child nodes that have $k$
``long'' dimensions (cuboids).

A \emph{valid node configuration} is a subdivision of a node into a set of
nonoverlapping child nodes of the same or different types. For example, in 2D,
child ids \{0,1,2,3\} (all type-0) and \{1,3,6\} (type-0 and type-1), and in
3D, \{10,11,12,13\} (all type-1) and \{0,4,18,25\} (type-0, type-1, and
type-2), are all valid configurations. Additionally, a leaf node's
configuration, \{\} (\ie no subdivision), is also counted as valid.
In this context, octrees and k-d trees support only two valid configurations
each --- either no refinement or full refinement, whereas \mesh allows $n_2=8$
in 2D and $n_3=146$ in 3D unique valid configurations. Both numbers can be
derived as 
\mbox{$n_d = 2 + \left(d - \sum_{p=1}^{d-1} 2^{-p}\right) n_{d-1}^2 $}, with
the base case of $n_0 = 0$.  Here, $n_{d-1}^2$ enumerates the possible
combinations of child nodes when a node is split along a single axes, the
multiplication by $d$ accounts for all $d$ axes, the expression in the
summation subtracts the redundancies, and the constant ``2'' adds the two
configurations for full refinement and no refinement (which get subtracted
while removing redundancies). The expression holds true for $n_1 = 2$ as well,
\ie for a 1D binary hiearchy, which has only two configurations --- refinement
and no refinement.

During the construction of \mesh, rectangular cuboidal leaf nodes are created
whenever possible.  In addition to the regular refinement that splits a node
into child nodes, \mesh also allows splitting rectangular cuboidal leaf nodes
into smaller sibling nodes, when needed.
For example, when a 3-cube is split along $x$ axis, only two leaf nodes, 24 and
25, are created, whereas, if the node is split along $x$ and $y$ axes, four
leaf nodes, \{16, 17, 18, 19\}, are created.  If the refinement next requires
creating leaf node 0, only leaf node 16 is split (along $z$), and the remaining
leaf nodes are left untouched, giving the new configuration as
\{0,4,17,18,19\}.  Each rectangular leaf node is handled independently, thus
preventing any unnecessary splits.

Furthermore, the two halves \{24,25\} of the node can be split along $y$ and
$z$, respectively, leading to the configuration \{13,15,16,18\}.
By allowing creation only of the cells explicitly needed for the
representation, \mesh's flexible spatial hierarchy reduces collateral memory
usage \autoref{fig:comparetrees}).
Finally, whereas creating similar decomposition using k-d trees is conceivable,
k-d trees allow splitting only one dimension at a time, artificially deepening
the hierarchy.

%% -----------------------------------------------------------------------------
\begin{figure}[!b]
\centering
\vspace{-1.4em}
\includegraphics[width=0.45\linewidth]{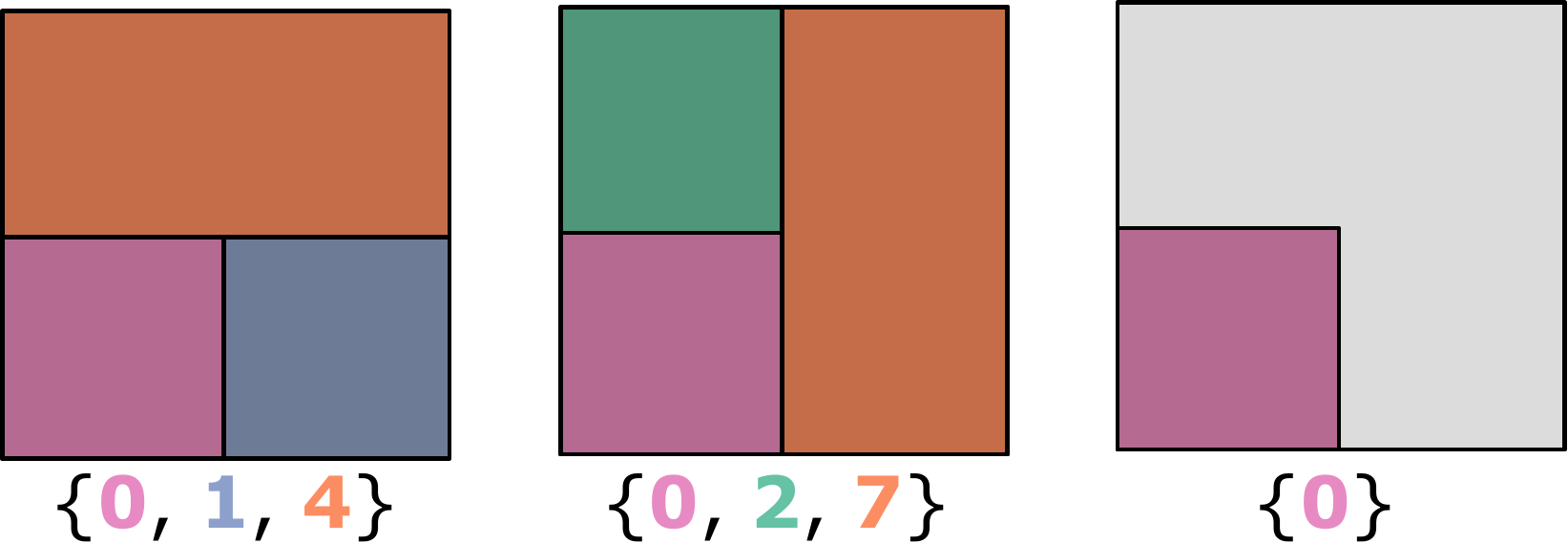}
\vspace{-0.5em}
\caption{\mesh uses \emph{improper nodes} (right) to preserve ambiguity in
refinement (left and middle) by refining a node partially. Here, the child
\{0\} could be created greedily by subdividing the node into \{0,1,4\} or
\{0,2,7\}, both of which could be suboptimal.  Instead, \mesh creates only
\{0\} and maintains the remaining region as unresolved until more data is
available to make the optimal choice.} \label{fig:vacuum}
\vspace{-0.4em}
\end{figure}

%% -----------------------------------------------------------------------------
\subsubsection{Improper Internal Nodes}
\label{sec:amm:improper}

Although rectangular cuboidal nodes provide highly flexible spatial adaptivity,
these additional degrees of freedom may lead to nonessential cells.  For
example, referring to \autoref{fig:vacuum}, consider splitting a 2D node to
create the child node \{0\}.  It is possible to split the given node in two
ways: splitting along $y$ and then $x$, creating configuration \{0,1,4\}, or
splitting along $x$ and then $y$, giving \{0,2,7\}.  Both these choices can
lead to redundant subdivision, depending upon the future requests, \eg
\{0,2,7\} is not optimal if child \{1\} is requested next because it will
create \{0,1,2,3\}, whereas \{0,1,4\} is already sufficient.  Rather than
making potentially unsuitable choices, \mesh defers the subdivision by creating
only the requested children.  We call a node that is not partitioned fully but
has at least one child an \emph{improper node}.

Partial subdivision within improper nodes creates a subset of child nodes that
can be used to represent noncuboidal shapes, \eg an L-shaped domain.  Improper
nodes can also be used to represent a region of interest without requiring a
complete coverage.  \autoref{fig:comparetrees} illustrates an example where
\mesh can reduce the size of the mesh using improper nodes.
Improper nodes are our solution to capture the degrees of freedom that have not
yet been resolved. By design, the ``unresolved'' portions of improper nodes do
not overlap with any existing child nodes. Therefore, the resulting
configurations are also considered valid, \eg \{0\} is a valid configuration.

Improper nodes represent a temporary state of \mesh to support incremental
construction and can be resolved by creating the missing child nodes on demand
to support intermediate queries without sacrificing incremental construction
(discussed further in \autoref{sec:create:improper}).

%% -----------------------------------------------------------------------------
\subsubsection{Pointerless Representation of the Tree} \label{sec:amm:storage}

\mesh uses a pointerless representation of its tree using the notion of
\emph{location codes}, which are commonly used for efficient representation of
octrees~\cite{Frisken2002, Gargantini:1982}.  Any node within an octree with a
maximum depth of $L$ can be uniquely encoded using $d\times L + 1$ bits, where
$d$ bits are used to locate a child with respect to (the location code of) its
parent.  In comparison, a location code in \mesh requires $d-1$ additional bits
per level to support rectangular cuboidal nodes, totaling $(2d - 1)\times L +
1$ bits per location code.
Using 64-bit unsigned integers for location codes, \mesh can represent a 3D
tree with $L=12$, \ie data sizes up to $4097^3$.

Location codes offer efficient tree traversal as they can be converted into the
coordinates of the vertex at the center of the node efficiently using bit shift
operations only.  For standard octrees, centers of (regular) nodes can be
reached by capturing and concatenating every $d$th bit of the location code
(from the right) to create the spatial coordinates. \mesh follows the same
procedure but also discards the $d-1$ additional bits for each level. Since all
internal nodes are regular nodes by design, all such discarded bits are zeros.
Given a location code, \mesh first looks at the $2d-1$ least significant bits,
which encode the child id of the node (with respect to its parent). For regular
nodes, the standard traversal is sufficient, whereas for rectangular nodes, the
standard traversal is used to first find the parent node (a $d$-cube), and then
a special case is used to find the correct child. \mesh uses location codes
also as search keys. Leaf nodes are not stored, and internal nodes are stored
as hash maps of location codes to an 8-bit unsigned integer that encodes its
node configuration.
\mesh makes extensive use of hashmaps and switch statements in the code to
manipulate node configurations efficiently. Given a valid node configuration
and a desired operation (\eg splitting along $z$ axis), the resulting
configuration is known and predefined in the code to optimize computational
cost.

Additionally, \mesh also stores the vertex values as hash maps from the index
to the value of the vertex.  Using sizes that are $2^L+1$ also offers a way to
efficiently encode vertex ids.  Whereas the simple and commonly used row-major
order requires multiplication and modulo operators to convert between
coordinates and index, we leverage the fact that each coordinate requires at
most $L+1$ bits and compute the index of a vertex $(x,y,z)$ efficiently using
bit-shifts, as \mbox{$(z << 2(L+1)) + (y << (L+1)) + (x)$}.

%% -----------------------------------------------------------------------------
\subsection{Precision Adaptivity using Blocks of Vertices} 
\label{sec:amm:mixedp}

To support mixed-precision representation, \mesh stores vertices as
\emph{blocks} of size $4\times4\times4$ in 3D and $8\times8$ in 2D.  Values are
stored in \emph{block floating-point} format~\cite{zfp}, where each value
$v_i=2^e\times m_i$ is expressed with respect to the largest exponent, $e$, in
the block.  For each block, an exponent is stored only once, and the mantissa
bits $m_i$ are stored in negabinary format (to avoid special handling of the
sign bit) to a fixed precision, \ie a fixed number of bits, $p$.  A block's
precision is defined as the smallest multiple of $8$ (bits) that captures all
nonzero bits in every $m_i$; each block may have a different precision.
\mesh also allows spatial adaptivity, \ie not all vertices in a block may
exist.  To identify which vertices exist, a 64-bit unsigned integer mask is
stored whose bits correspond to vertices in the block. Each block thus
comprises a bytestream that encodes the mantissa bits for all existing
vertices, the vertex mask (eight bytes), the common exponent $e$ (two bytes),
and the precision $p$ (one byte).
To index the blocks themselves, \mesh uses a hashmap indexed by blocks'
row-major indices.

If a new vertex is added to the block, the bytestream is recreated to insert
the new value at its correct location (in order of the local index in the
block).  We note, however, that updating bytestreams in this way is expensive.
For nonprogressive, state-of-the-art compression approaches, the encoding
usually happens with the availability of all data, and there is no need to
update the bytestream, which provides high throughput.  However, supporting
incremental updates poses computational challenges because the encoding is
dynamic and can change as new values in the block or new bits for existing
values are received.  In order to support progressive adaptivity, \mesh strives
to reduce the number of times the bytestream is recreated.  As mentioned, \mesh
represents function values with precision in multiples of bytes, not bits, so
that the cost of update is amortized over many bitplanes, \ie the bytestream
does not need to grow with each new bit.  We further amortize the computational
cost of update through an effective use of staging.
%(see \autoref{sec:create:precision}).

%% -----------------------------------------------------------------------------

\vspace{-0.5em}
\section{Construction of \mesh}

Although \mesh can also be created by simply reading in a collection of
multilinear cuboidal cells, in this section, we focus on an important way of
creating \mesh --- using tensor products of biorthogonal linear B-spline
wavelets~\cite{Cohen92}.

Here, \autoref{sec:stencils} describes the relevant properties of the
corresponding wavelet basis and draws comparison between \mesh and the
framework of Weiss and Lindstrom~\cite{Weiss:2016}.  \autoref{sec:create}
presents a practical algorithm to incrementally create \mesh through
arbitrarily-ordered datastreams of wavelet coefficients.  Finally,
\autoref{sec:linearlifting} discusses a new \emph{linear-lifting extrapolation}
scheme to expand the input data to a $2^L+1$ size without boundary artifacts.

%% -----------------------------------------------------------------------------
\begin{figure}[!b]
\centering
% ---------------------------------------------------------------------------------------
\vspace{-1.5em}
\includegraphics[width=\linewidth]{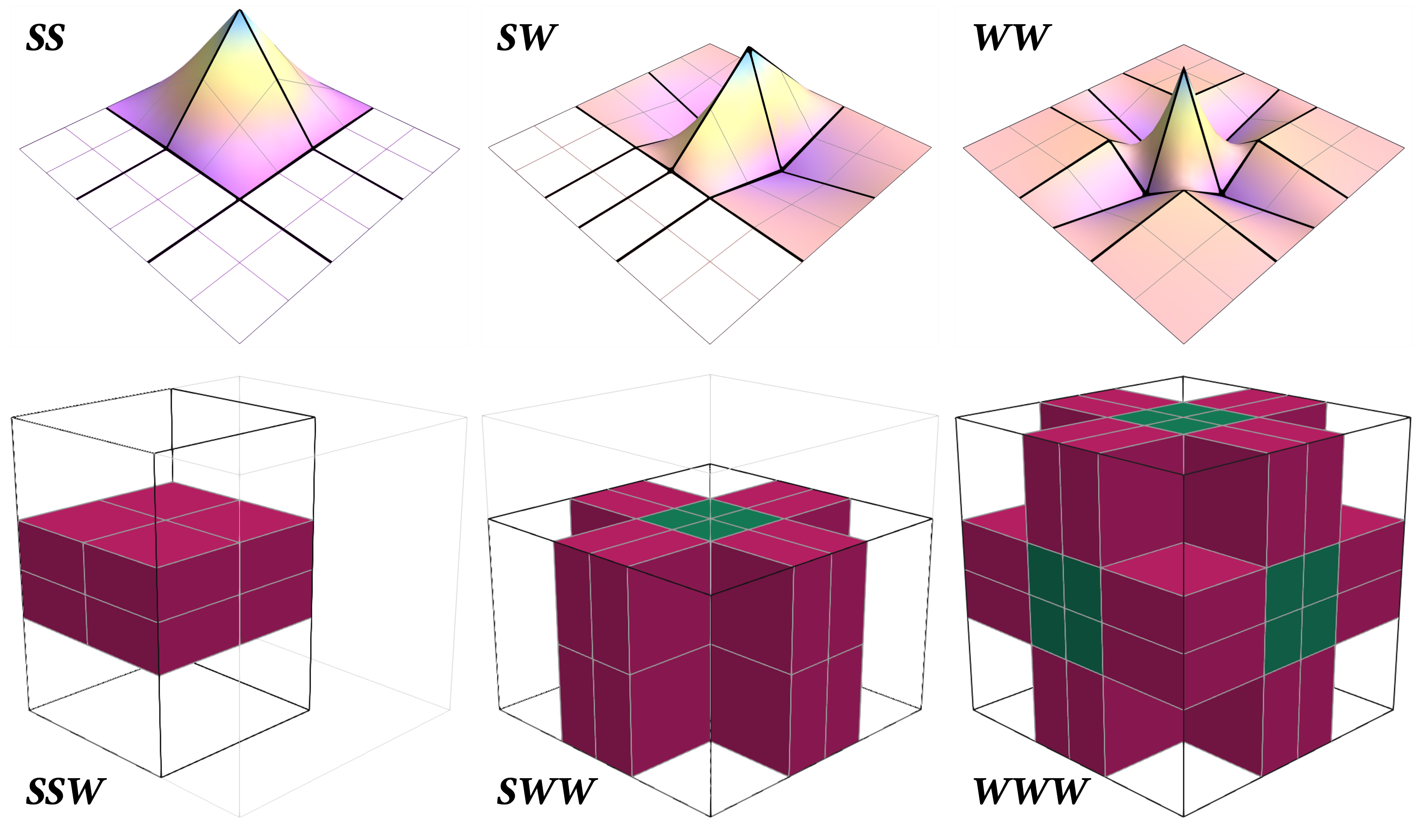}
\vspace{-1.75em}
\caption{Spatial stencils for bilinear and trilinear tensor-product B-spline
wavelets.  Top: The 2D stencils are shown with respect to an underlying regular
grid, given the cells at two adjacent levels (bold and light lines), with color
and height mapped to function value.  With respect to the coarser level, the
scaling stencil ($\wsb \wsb$) is associated with vertices, the wavelet stencils
($\wwb \wwb$) with faces, and mixed stencil ($\wsb \wwb$ and $\wwb \wsb$) with
edges.
Bottom: For easier visualization in 3D, only the rectangular cuboidal cells are
shown (type-1 = green, type-2 = magenta) within the bounding box of the stencil
(black). The scaling stencil ($\wsb \wsb \wsb$) is not
shown.}\label{fig:stencils}
%
% ---------------------------------------------------------------------------------------
%
\vspace{0.75em}
\begin{minipage}{0.52\linewidth}
\centering
\includegraphics[width=\linewidth]{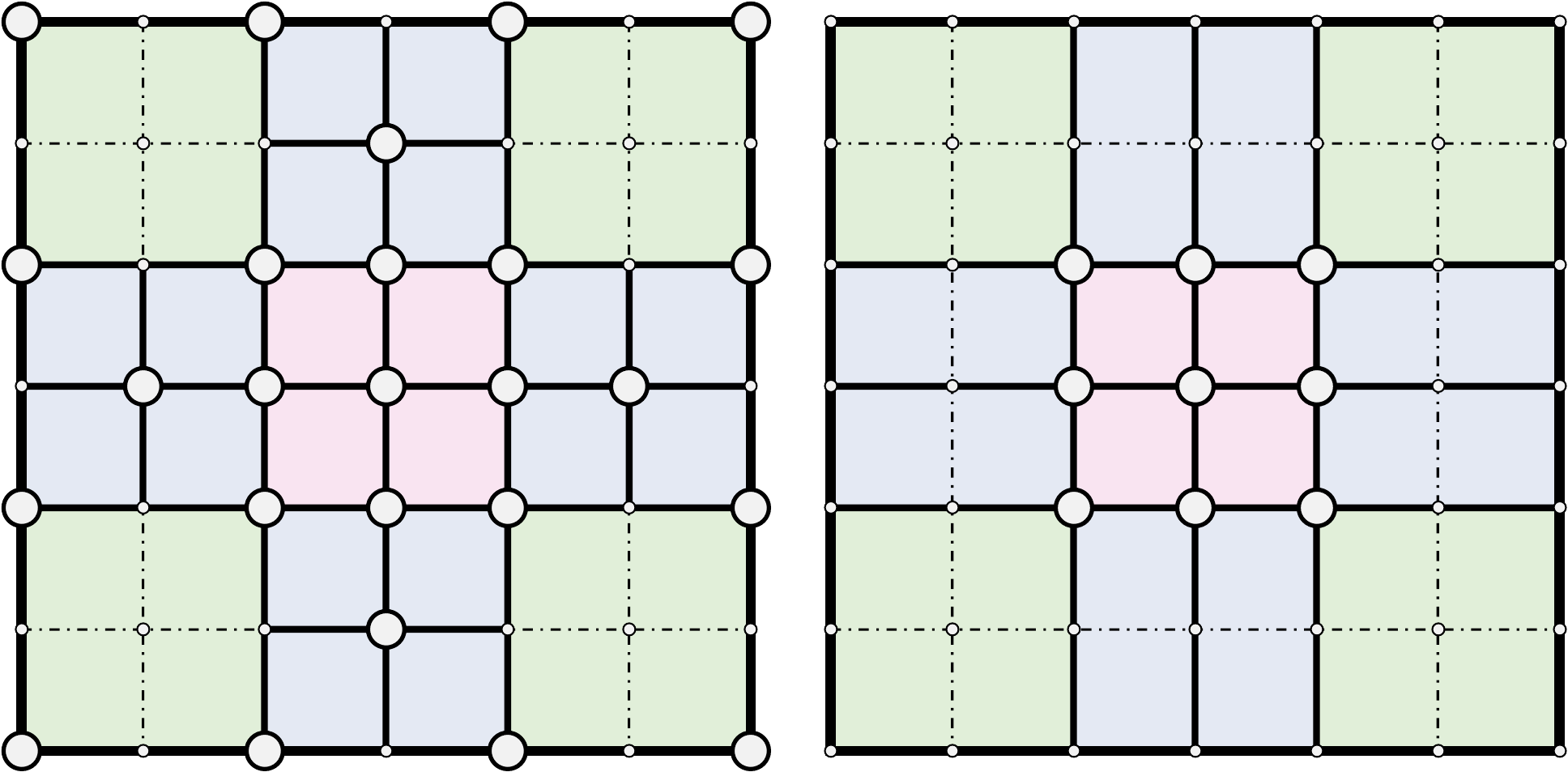}
\end{minipage}%
\hspace{1em}
\begin{minipage}{0.42\linewidth}
\centering
\setlength{\tabcolsep}{2.5pt}     % General space between columns (6pt standard)
\resizebox{\linewidth}{!}{\begin{tabular}{|l|r|rr|rr|}
\hline
& Total\,\,\,\, & \multicolumn{2}{c|}{\cite{Weiss:2016}} & \multicolumn{2}{c|}{\mesh}\\
& \#V & \#C & \#V & \#C & \#V  \\
\hline
\hline
$\wsb \wsb$ & 25 & 4 & 1 & 4 & 1  \\
$\wsb \wwb$ & 35 & 12 & 15 & 8 & 3  \\
$\wwb \wwb$ & 49 & 24 & 25 & 16 & 9  \\
\hline
$\wsb \wsb \wsb$ & 125 & 8 & 1 & 8 & 1  \\
$\wsb \wsb \wwb$ & 175 & 40 & 45 & 16 & 3  \\
$\wsb \wwb \wwb$ & 245 & 119 & 63 & 32 & 9  \\
$\wwb \wwb \wwb$ & 343 & 160 & 125 & 64 & 27  \\
\hline
\end{tabular}}
\end{minipage}
% ---------------------------------------------------------------------------------------
%
\vspace{-0.5em}
\caption{A $\wwb\wwb$ stencil (placed at the center vertex) as used by Weiss
and Lindstrom~\cite{Weiss:2016} (left) and \mesh (middle). Using rectangular
cells and other properties of the stencil, \mesh reduces the representation to
9 vertices and 16 cells (4 square green, 4 square pink, and 8 rectangular
blue), as compared to 25 vertices and 24 cells~\cite[Table 1]{Weiss:2016}.  The
associated table lists such comparisons for all types of stencils in 2D and
3D.} \label{fig:wlverts}\label{tab:wlverts}
% ---------------------------------------------------------------------------------------
%
\vspace{-0.5em}
\end{figure}

%% -----------------------------------------------------------------------------
\vspace{-0.5em}
\subsection{Wavelet Coefficients and Associated Stencils} \label{sec:stencils}

As shown in \autoref{fig:stencils}, 2D/3D tensor products of linear B-spline
wavelets are associated with \emph{``spatial stencils''} consisting of
multilinear cells at two adjacent levels of refinement. Superposition of such
stencils onto a spatial grid is equivalent to performing an inverse transform
(the lifting step), but using sparse wavelet coefficients.
Previously, Weiss and Lindstrom~\cite{Weiss:2016} leveraged these stencils to
create reduced and adaptive representations using a restricted set of wavelet
coefficients (filtered by magnitude). Nevertheless, since their hierarchy is
built upon a regular octree, they are unable to fully exploit the shape of
these stencils. Specifically, their representation cannot directly represent
the ``rectangular portions'' of these stencils and instead decomposes such
cells into sets of smaller, square cells. This additional decomposition implies
a finer mesh to represent sets of cells and vertices that can be trivially
interpolated and, therefore, do not provide any additional information.
We discuss how \mesh's flexible spatial hierarchy outperforms previous
work~\cite{Weiss:2016} by reducing the size of the representation.

As an example, consider the approximating wavelet stencil in 2D ($\wwb \wwb$),
illustrated in \autoref{fig:stencils}. When corresponding to a wavelet
coefficient at level $\hl$, the stencil is defined on a subgrid at level
$\hl+1$ and contains a total of $7\times7 = 49$ vertices.  Illustrated in
\autoref{fig:wlverts}, properties of these stencils can be exploited to reduce
the number of ``effective'' vertices (\ie those that need to be stored to
reconstruct the function).
According to Weiss and Lindstrom~\cite{Weiss:2016}, using the zero-valued
boundary and the multilinear interpolation of cells in the stencil, only $25$
effective vertices need to be represented.

Our framework further improves this reduction by allowing for rectangular cells
using \mesh.  Specifically, it becomes possible to represent the subdivisions
in the axis-aligned neighboring cells of the center point from $4\times4$ to
$4\times2$, which also removes the need to represent $8$ vertices. Our final
representation for this stencil, therefore, contains only $16$ cells and $9$
effective vertices. Unsurprisingly, the gains are substantially higher in 3D;
\autoref{fig:stencils} tabulates the reduction in every type of stencil using
our framework.

\para{``Stamping'' stencils to the mesh.}
Our approach for creating \mesh from a given set of wavelet coefficients
utilizes a ``stamping'' procedure for the corresponding spatial stencils.
Here, we use the ideas presented by Weiss and Lindstrom~\cite{Weiss:2016} to
identify the spatial context of a stencil (\ie cells to be created) using
iterators of the \emph{primal} and \emph{dual} $k$-cells of regular octree
refinement and other standard queries, such as neighhboring cells of a cell and
incident cells of a vertex.
However, the actual approach of creating these cells is different because of
the differences in the spatial hierarchy. To simplify the construction process,
\mesh exposes an API to create a cell and split a cell along any combination of
axes.

The second step in the stamping procedure is the addition of vertices.  Given
the center vertex of the stencil and its level in the wavelet hierarchy, it is
trivial to identify all vertices that need to be updated. As discussed in
\autoref{sec:create:verts}, stencil vertices are also ``stamped'' into a
staging phase and combined with the vertices from coarser and finer levels
later in the unstaging step.

%% -----------------------------------------------------------------------------
\vspace{-0.5em}
\subsection{A Practical Approach Toward Creating \mesh} \label{sec:create}

\begin{figure}[!b]
\vspace{-0.5em}
\includegraphics[width=\linewidth]{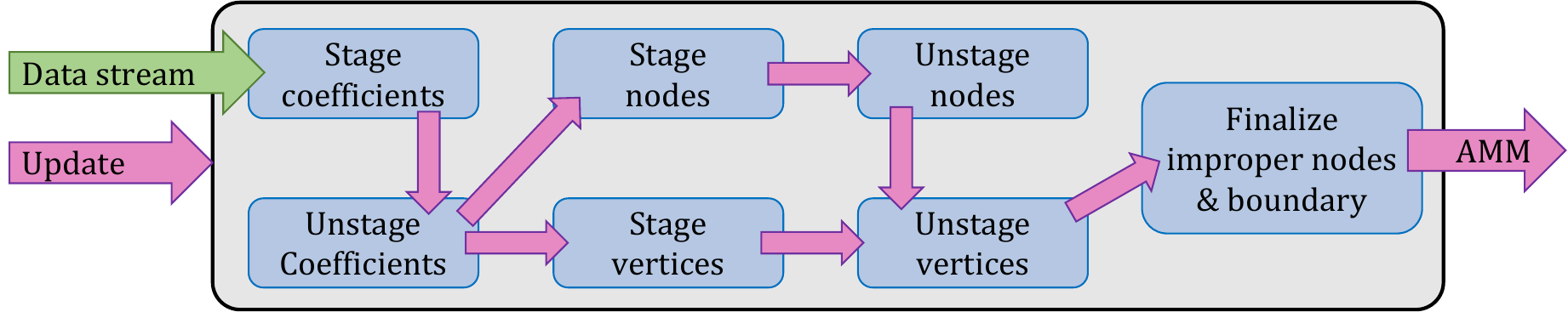}
\vspace{-1.5em}
\caption{For computational efficiency, construction of \mesh is performed using
three staging phases. Whereas \mesh continues to ingest and stage the incoming
datastream (green), this data is processed only when a request to update the
mesh is received (\eg a visualization query). The pink arrows follow the
movement of data within the construction pipeline after receiving the update
request.} \label{fig:stages}
\vspace{-0.5em}
\end{figure}

Our approach toward creating a resolution-precision-adaptive representation is
guided by four goals: (1) the smallest number of cells and vertices that can
represent the function faithfully, (2) speed of construction, (3) ability to
perform incremental updates through arbitrary data streams, and (4) ability to
represent the function using mixed-precision values.  To achieve these goals,
we utilize three staging phases (see \autoref{fig:stages}) to create \mesh.

%% -----------------------------------------------------------------------------

% -----------------------------------------------------------------------------
\vspace{-0.5em}
\subsubsection{Staging and Unstaging of Wavelet Coefficients}
\label{sec:create:wcoeffs}

The first challenge is posed by the arbitrary nature of data streams as
incoming coefficients may be in any order and may represent complete (all bits)
or partial (some bits) values.  In the case of partial values, updating the
mesh through stamping the corresponding stencils (tree traversals to create
nodes and update vertex values) every time a bit is received is wasteful.  Even
when complete values are to be received, the order of coefficients may require
excessive tree traversal, \eg updating a coarse node in a deep tree requires
traversing down and updating the corners of all child nodes.

To mitigate the cost of excessive traversals, \mesh collects incoming
coefficients without updating the tree. This staging is particularly useful for
collecting all bits of the same coefficient and then stamping the stencil only
once, preventing redundant operations. At the time of unstaging, these
coefficients are sorted coarse-to-fine before stamping to ensure that coarser
nodes are created first, preventing the need to propagate the values at the
corners of coarse nodes all the way down to the leaf nodes.
Finally, this staging process also acts as a filter by discarding any wavelet
coefficients whose stencils lie completely outside the data domain (to avoid
boundary artifacts, the wavelet transform may be performed on a larger, $2^L+1$
domain).

% -----------------------------------------------------------------------------
\vspace{-0.5em}
\subsubsection{Staging and Unstaging of Tree Nodes}

The next optimization our algorithm makes is to prevent unnecessary updates to
the spatial hierarchy, \ie creation and splitting of nodes. Since the stencils
of different wavelet coefficients overlap in their spatial extent, different
stencils may require updating the same nodes. For example, one stencil may
create a node, followed by another that splits it vertically, and another that
splits it horizontally.

Instead of processing the tree for each stencil, which requires tree
traversals, a few hash map queries, and interpolations to identify the values
of the split points, \mesh collects all such requests to identify the final
state of each node. At the time of unstaging (\ie when all stencils have been
accounted for), each node is processed only once, reducing the overall
computational cost.

% -----------------------------------------------------------------------------
\subsubsection{Staging and Unstaging of Tree Vertices} 
\label{sec:create:precision}
\label{sec:create:verts}

Whereas the coefficient staging reorders the received data from coarse-to-fine,
it can work only on the coefficients received since the last mesh update.
During incremental construction, one may already have a deep tree when coarse
coefficients are received. The resulting challenge is that any new updates
require propagating those changes down to the leaf nodes through recursive
interpolation within parent nodes, which quickly becomes computationally
prohibitive, especially when nodes at the coarse levels are updated frequently
in a deep and/or dense tree.

\mesh's vertex staging mitigates this cost by collecting the updates to vertex
values and storing them separately by their level in the tree because vertex
updates corresponding to a given level are additive. At the time of unstaging,
the tree is traversed down only once for each node updated in the current
stage. During this traversal, the corners of a given node are multilinearly
interpolated to add to the values of all corners of all child nodes --- the
corners of the child nodes that are also the corners of the parent node do not
need this update. Temporary caches are used to prevent duplicate updates to a
vertex for each of its incident nodes.

% -----------------------------------------------------------------------------
\para{Mixed-precision representation of vertex values.}
The unstaging of vertices naturally separates the vertex values that are
created/modified during the current staging/unstaging step from those that
existed before.  The final task in the vertex unstaging is to combine the two
sets of vertex values to give the final and correct values. At this time, if a
mixed-precision representation is requested, the vertex values aggregated from
the current stage are finalized into the bytestream representation of the
vertex values.

% -----------------------------------------------------------------------------
\vspace{-0.25em}
\subsubsection{Finalization of Improper Nodes and Boundary Nodes}
\label{sec:create:improper}

As described earlier, \mesh uses improper nodes to preserve unresolved degrees
of freedom.  However, since typical analysis and visualization tasks expect a
complete coverage of space, any improper nodes in the tree must be resolved
before preparing the mesh for use.  Improper nodes may be resolved by simply
creating additional child nodes (by design, these will be leaf nodes of the
tree) and associated vertices.  \mesh preserves the current state of the tree
by storing these additional leaf nodes and vertices in a separate and temporary
data structure that is used only to respond to the query at hand, without
sacrificing the improper nodes for incremental construction.

Furthermore, the underlying tree may also have a larger spatial extent than the
given data, in which case there may exist no leaf nodes whose boundary aligns
with the boundary of the data domain. In such cases, it is straightforward to
identify the leaves that exist across the data boundary and split them using
multilinear interpolation to create cells and vertices for the output mesh.
Similar to the case of improper nodes, the boundary leaf nodes and the
additional vertices are also stored in a temporary data structure to not affect
the state of the tree.  The \mesh API abstracts these temporary data
structures, allowing the application to use the mesh directly through cell and
vertex iterators and accessors.

%% -----------------------------------------------------------------------------

\definecolor{clsorange}{HTML}{d95f02}
\definecolor{clsgreen}{HTML}{1b9e77}
\definecolor{clsblue}{HTML}{1f78b4}
\definecolor{clspink}{HTML}{e7298a}
\definecolor{clsgray}{HTML}{aaaaaa}

\newcommand \lfnt[1] {{\texttt{#1}}}
\newcommand \lln[1] {\lfnt{#1}}
\newcommand \llw[1] {\textcolor{clsorange}{\lfnt{#1}}}
\newcommand \lls[1] {\textcolor{clsblue}{\lfnt{#1}}}
\newcommand \lll[1] {\textcolor{clspink}{\lfnt{#1}}}
\newcommand \llu[1] {\textcolor{clsgray}{\lfnt{#1}}}

\newcommand \wlftc[3] {\lfnt{(#1--#2--#3)}}

\vspace{-0.5em}
\subsection{Data Extrapolation Using a Linear-Lifting Scheme}
\label{sec:linearlifting}

Since wavelet transforms are computed for power-of-two sized grids, their
practical application often requires extending the signal to suitable lengths.
Whereas several approaches exist for this extension (\eg zero padding, linear
extrapolation, and symmetric extension), each is associated with different
types of boundary artifacts, such as discontinuity and nonsmoothness, that lead
to large wavelet coefficients, which may not be conducive to the application.
For \mesh, these artificial coefficients typically result in unnecessary
refinement, inflating the memory footprint significantly and needlessly.

Here, we present a new, \emph{linear-lifting} approach to extend the input
function at the boundary to avoid such artifacts and reduce the number of
unneccesary cells near the boundary.  Conceptually, we perform the usual
lifting steps everywhere in an extended domain of size $(2^L+1)^d$, but assign
values lazily to grid points outside the original domain, $[X \times Y \times
Z]$ (see \autoref{sec:amm:treesize}).
Denoted symbolically as \wlftc{*}{*}{*}, the w-lift step updates the value at
the center using the adjacent ones. With respect to the original and extended
domains, there exist four possible scenarios: \wlftc{x}{x}{x}, \wlftc{x}{x}{o},
\wlftc{x}{o}{o}, and \wlftc{o}{o}{o}, where \lfnt{x} represents a grid point
within the original domain whereas \lfnt{o} has an uninitialized value due to
being outside the original domain (but within the extended domain).
In the first case, all three relevant grid points exist within the original
domain and the standard w-lift can be applied. In the second case, we linearly
extrapolate the two known values to assign a value to the rightmost grid point
when needed, resulting in a zero-valued wavelet coefficient. For the third and
the fourth case, we set the wavelet coefficient to zero but defer setting the
value for the rightmost sample to an extrapolation step on some coarser level.
Furthermore, s-lift is applied only to values that have been initialized ---
standard step for the first case, but no effect for the remaining three.  With
this scheme, uninitialized grid points are given values such that when they are
used for lifting, the resulting wavelet coefficients are always zero.
\autoref{tab:lifting} illustrates our approach using a concrete 1D example.

\setlength{\tabcolsep}{3pt}     % General space between columns (6pt standard)
\begin{table}[!t]
\centering
%\vspace{-0.5em}
%\resizebox{0.88\columnwidth}{!}{\begin{tabular}{|l|rrrrrrrrr|}%rrrr|}
\resizebox{\columnwidth}{!}{%
\begin{tabular}{|l|rrrrrrrrr|}%rrrr|}
\hline
Input function & \lln{56} & \lln{8} & \lln{48} & \lln{44} & \lln{32} & \lln{8} & & &\\
\hline
\hline
Level 1: Extrapolate & \lln{56} & \lln{8} & \lln{48} & \lln{44} & \lln{32} & \lln{8} & \lll{-16} & & \\
Level 1: Forward w-lift & \lln{56} & \llw{-44} & \lln{48} & \llw{4} & \lln{32} & \llw{0} & \lln{-16} & &\\
Level 1: Forward s-lift & \lls{45} & \lln{-44} & \lls{38} & \lln{4} & \lls{33} & \lln{0} & \lls{-16} & &\\
\hline
Level 2: Extrapolate & \lln{45} & & \lln{38} & & \lln{33} & & \lln{-16} & & \lll{-65}\\
Level 2: Forward w-lift & \lln{45} & & \llw{-1} & & \lln{33} & & \llw{0} & & \lln{-65}\\
Level 2: Forward s-lift & \lls{45} & & \lln{-1} & & \lls{33} & & \lln{0} & & \lls{-65}\\
\hline
\hline
Coefficients stored in memory & \lln{45} & \lln{-44}& \lln{-1} & \lln{4} &
\lln{33} & & \lln{-16} & & \lln{-65}\\
\hline
\hline
Level 2: Insert w-coefficients & \lln{45} & &\llw{-1} & & \lln{33} & & \llw{0} & & \lln{-65}\\
Level 2: Inverse s-lift & \lls{45} & & \lln{-1} & & \lls{33} & & \lln{0} & & \lls{-65}\\
Level 2: Inverse w-lift & \lln{45} & & \llw{38} & & \lln{33} & & \llw{-16} & & \lln{-65}\\
\hline
Level 1: Insert w-coefficients & \lln{45} & \llw{-44} & \lln{38} & \llw{4} & \lln{33} & \llw{0} & \lln{-16} & \llw{0} & \lln{-65}\\
Level 1: Inverse s-lift & \lls{56} & \lln{-44} & \lls{48} & \lln{4} & \lls{32} & \lln{0} & \lls{-16} & \lln{0} & \lls{-65}\\
Level 1: Inverse w-lift & \lln{56} & \llw{8} & \lln{48} &
\llw{44} & \lln{32} & \llw{8} & \lln{-16} & \llw{-41} & \lln{-65}\\
\hline
\hline
Extrapolated function & \lln{56} & \lln{8} & \lln{48} &
\lln{44} & \lln{32} & \lln{8} & \lln{-16} & \lln{-41} & \lln{-65}\\
\hline
\end{tabular}}
\caption{Our linear-lifting approach extrapolates a 6-point function using two
levels of transform with integer arithmetic.  A (forward) lifting phase begins
with linear extrapolation (pink), followed by w-lift (brown) and s-lift (blue).
Inverse lifting extends the input function to 9 points, but only 7 coefficients
are stored for full reconstruction.  Note that the extrapolated function is
different from one obtained via simple linear extrapolation in the last two
elements.}
\label{tab:lifting}
%\vspace{-1em}
\end{table}

\begin{figure}[!t]
\vspace{-0.5em}
\centering
\begin{subfigure}[t]{0.3\linewidth}
\centering

\includegraphics[width=\linewidth]{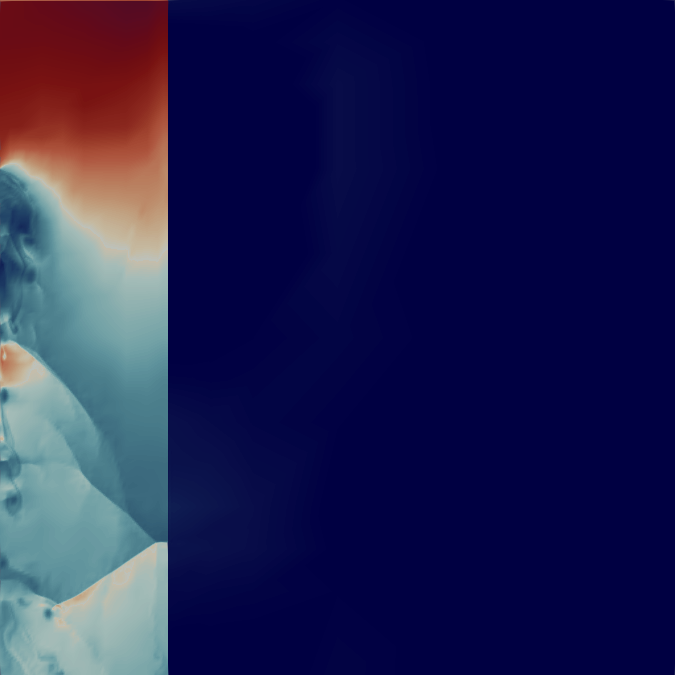}

\vspace{0.5em}
\includegraphics[width=\linewidth]{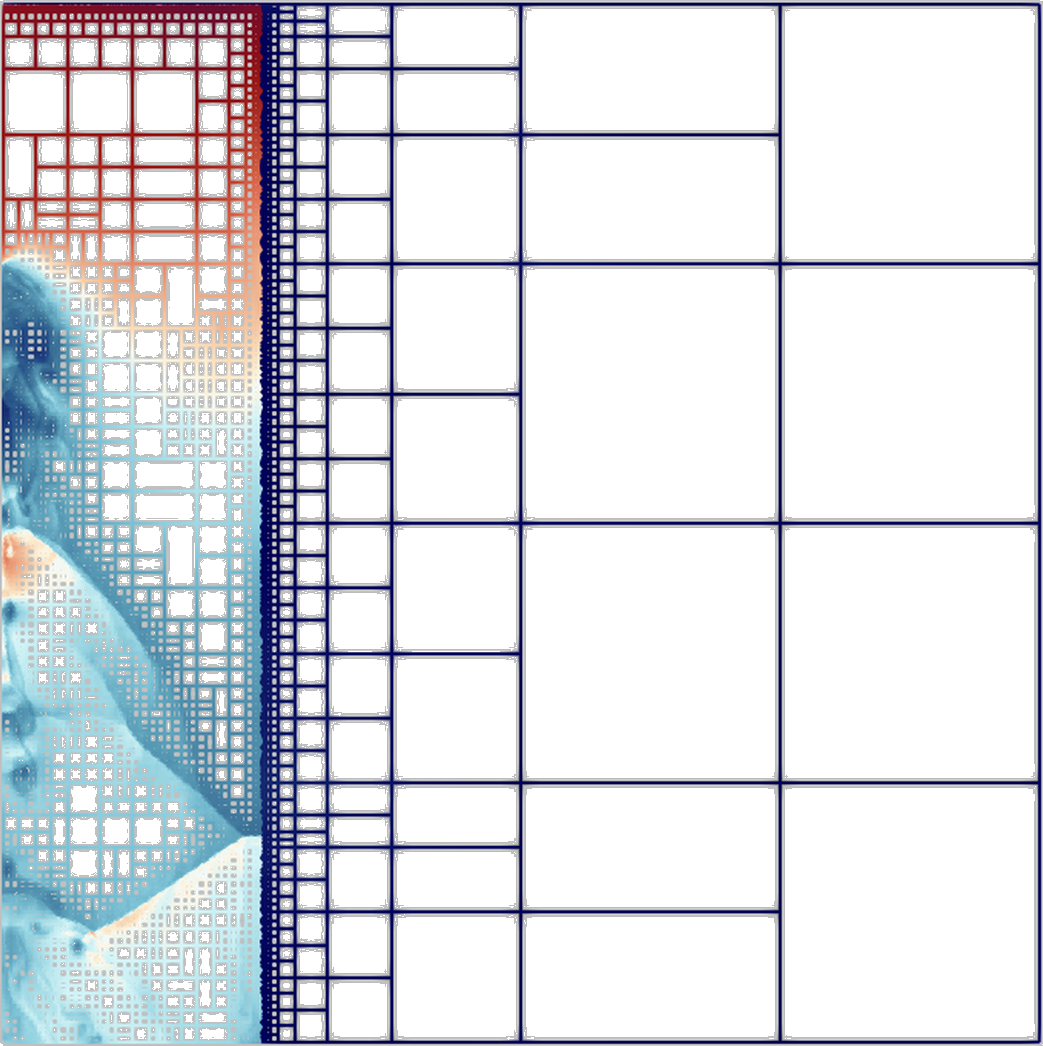}
\captionsetup{justification=centering}
\caption{Zero padding\newline(7997, 9918, 4664)}
\end{subfigure}%
\hfill
%
% PL: make center subfig wider to avoid caption break
\begin{subfigure}[t]{0.4\linewidth}
\centering
\includegraphics[width=0.75\linewidth]{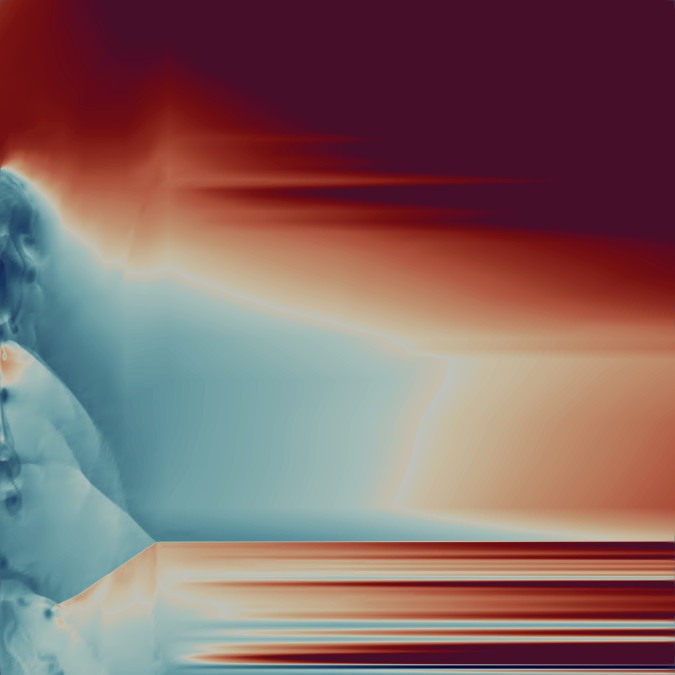}

\vspace{0.5em}
\includegraphics[width=0.75\linewidth]{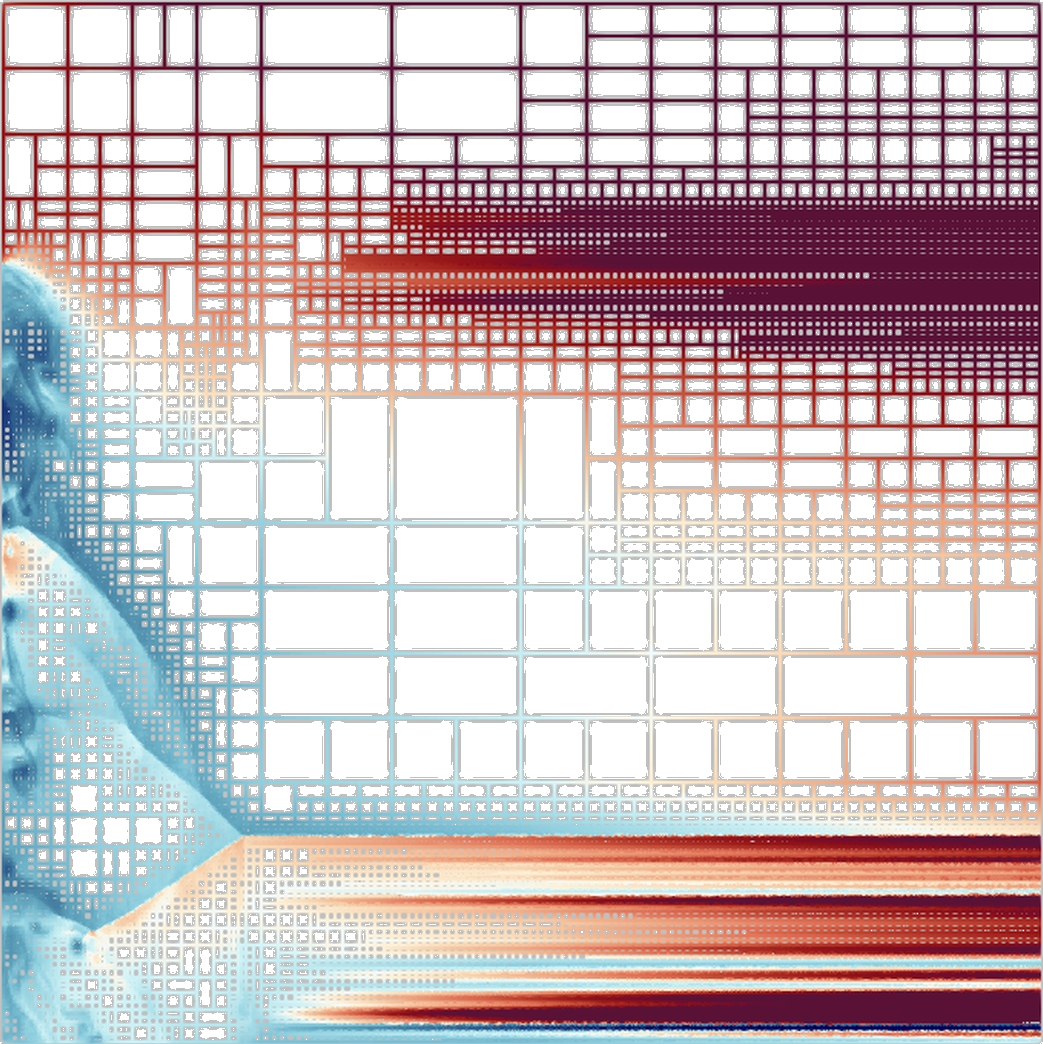}
\captionsetup{justification=centering}
\caption{Linear extrap.\newline(6263, 34742, 19007)}
\end{subfigure}%
\hfill
\begin{subfigure}[t]{0.3\linewidth}
\centering
\includegraphics[width=\linewidth]{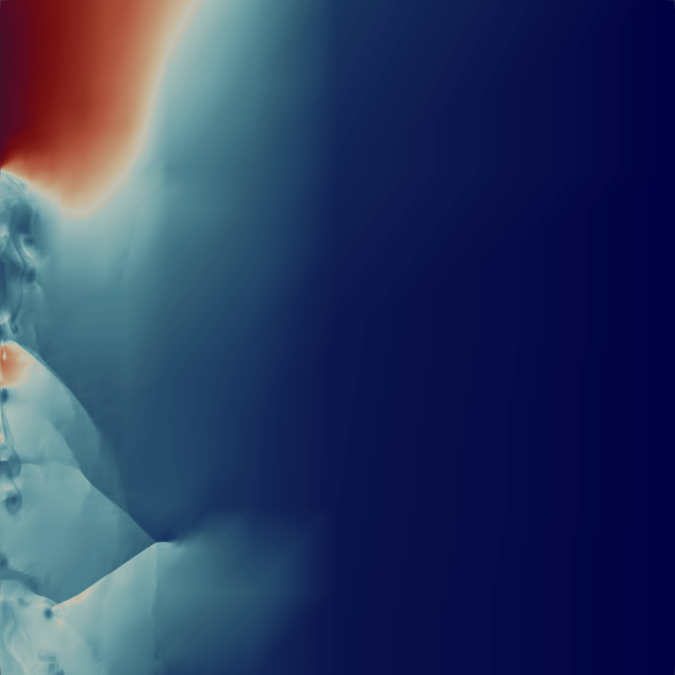}

\vspace{0.5em}
\includegraphics[width=\linewidth]{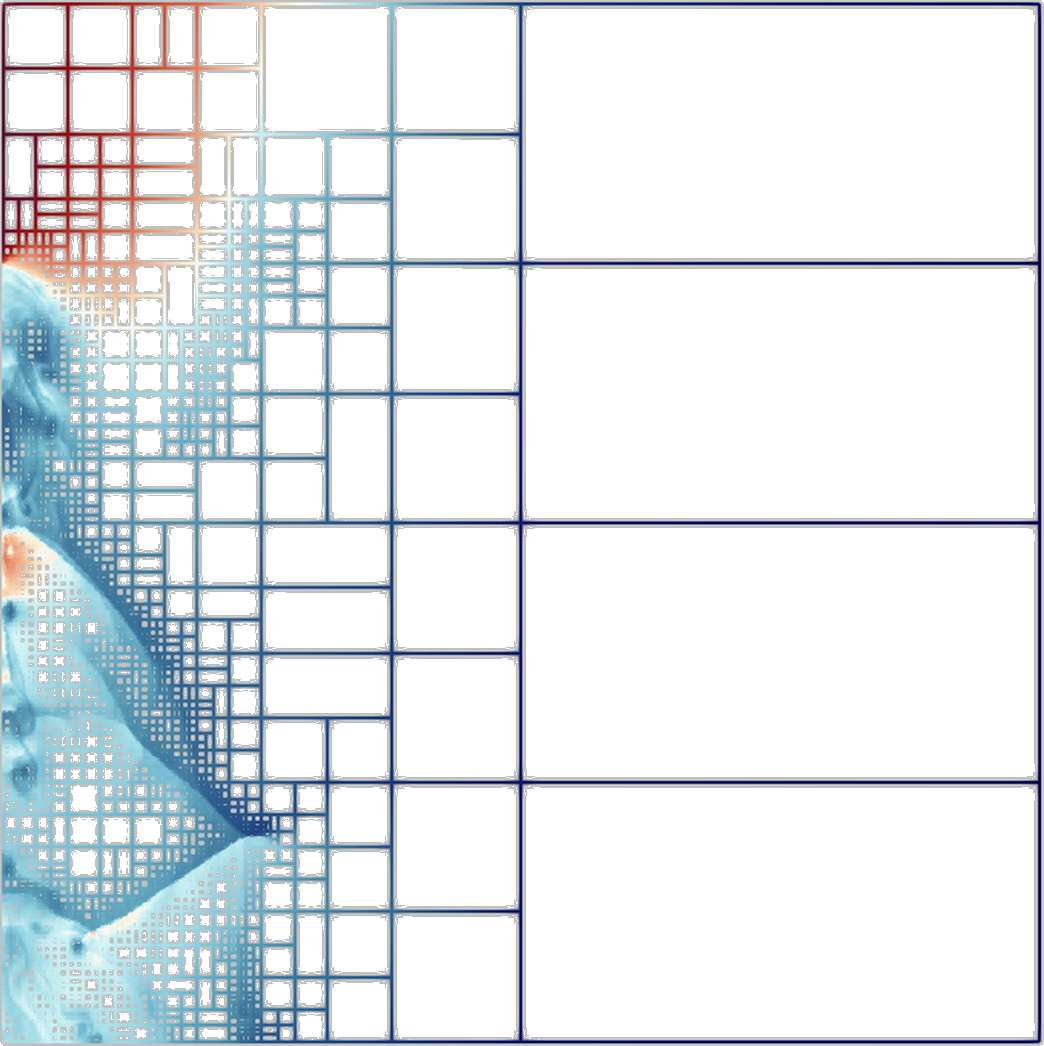}
\captionsetup{justification=centering}
\caption{Linear-lifting\newline(6263, 6342, 2965)}
\end{subfigure}%
%
%\vspace{-2mm}
\caption{Function extensions (top) and the corresponding meshes (bottom) for a
shockwave defined on $[256 \times 1024]$ domain and extrapolated to $[1025
\times 1025]$ using different methods. The associated metrics are number of
(cells, leaf nodes, internal nodes).
Zero padding introduces artificial discontinuities at the boundary of the input
domain (notice the vertical blue streak of finest-level cells). Linear
extrapolation maintains smoothness near the boundary, but can create
discontinuities farther out from the original domain. Our linear-lifting
approach avoids artificially large wavelet coefficients at the boundary and in
the extrapolated region.}
\label{fig:extrapolation}
\vspace{-1em}
\end{figure}

Our scheme differs from simple linear extrapolation in that it interleaves the
potential linear extrapolation steps described above with lifting steps.
Simple linear extrapolation does not ensure smoothness along dimensions
orthogonal to the one being extrapolated, while also failing to correct for the
nonlinear reconstruction introduced by s-lift steps.  In contrast, by
interleaving lifting steps with extrapolation steps across the hierarchy and
across spatial dimensions, our method ensures smoothness in the extended
function both across the boundary of the domain as well as across all
dimensions (see \autoref{fig:extrapolation}).  Naive linear extrapolation is
also entirely local as it depends only on the last two values near the domain
boundary, whereas our method extrapolates at multiple scales and therefore is
globally influenced.

In the example from \autoref{tab:lifting}, the extrapolated function is longer
than the original function by three elements unlike the traditional wavelet
transform, where the two functions have the same length.  However, in practice,
the inverse lifting steps are never performed, and thus the extrapolated
function is never explicitly computed or stored.  We do need to store the
potentially extrapolated value at each lifting step to support perfect
reconstruction, but this requires storing at most a single extra value along
each dimension of the original domain, as the same slot can be reused on the
next coarser transform level without compromising reconstruction using inverse
lifting steps.  Since the wavelet coefficients in the extrapolated domain are
all zero, in the resulting mesh, AMM's adaptivity allows the overhead (the
extra cells and vertices) to remain small, as seen
in~\autoref{fig:extrapolation}c.

%% -----------------------------------------------------------------------------
\vspace{-0.75em}
\section{Evaluation}

In our evaluation, we focus on the three design goals of \mesh: adaptivity in
spatial resolution, adaptivity in precision, and incremental construction, and
consider the size of the resulting representation, the reconstruction quality,
and the time it takes to compute \mesh.

We report the size of our adaptive representation in terms of the number cells
and vertices; we also report the approximate memory footprint of the
representation by counting eight bytes for each cell and vertex (to store
indices) and the number of bytes used (one to eight) to represent function
values, but ignore the additional overhead of data structures such as hash
maps.
To quantify the quality of a reconstructed function $\hat f$ against the
original function $f$, we measure the peak signal-to-noise ratio, PSNR $= 20
\log_{10} \left((f_{\max{}} - f_{\min{}})/2\right) 
- 10 \log_{10} \left(\sum{(f - \hat f)^2} / N\right)$, where $N$ is the total
  number of samples in the given data.

%% -----------------------------------------------------------------------------
\para{Datastreams.}
In this work, we consider six types of datastreams.
We borrow four datastreams from Hoang \etal~\cite{Hoang:2019} ---
\emph{``by-magnitude''} (\smag), \emph{``by-level''} (\slvl),
\emph{``by-bit-plane''} (\sbit), and \emph{``by-wavelet-norm''} (\swav).
Here, \smag is a \emph{resolution stream} since it transmits complete
coefficients, which are ordered by their magnitude. Streams \slvl, \sbit and
\swav are \emph{precision streams} since they may transmit bits of a
coefficient separately; \slvl orders coefficients by level in the wavelet
hierarchy (coarse to fine) but transmits the bits after discarding leading
zeros; \sbit picks the bits in order of bit plane (most significant to least
significant); and \swav combines the functionality of \sbit and \slvl to order
the bits based on both the bit plane and the wavelet basis on the subband of
the coefficient.
We consider two additional resolution streams ---
\emph{``by-level-coeff''}, (\ssbr), which transmits complete coefficients
ordered by level in the wavelet hierarchy; and \emph{``by-coeffient-energy''},
(\scen), which orders the coefficients based on both the magnitude of the
coefficient and the norm of the wavelet basis function for the corresponding
subband.

% naming of streams in our code
% 1:                                   by row major
% 2:                                   by subband row major
% 3:                                   by coeff wavelet norm
% 4:                                   by wavelet norm
% 5:                                   by level
% 6:                                   by bitplane
% 7:                                   by magnitude

%% -----------------------------------------------------------------------------
\para{Interfacing with other tools for visualization.}
Although several visualization tools support spatially adaptive
representations, precision adaptivity is generally not well supported due to
both software and hardware limitations, and the data is typically inflated
first to full-precision (floats or doubles) before visualization.
Whereas utilizing \mesh directly at reduced precision remains a task for the
future, currently \mesh facilitates analysis and visualization by providing a
simple interface to the visualization toolkit (VTK)~\cite{vtk}.
It is straightforward to output \mesh as a VTK unstructured grid, which can be
used with standard tools, such as Paraview~\cite{paraview}, VisIt~\cite{visit},
and OSPRay~\cite{wald_ospray_2017}, as well as to support hardware-accelerated
rendering~\cite{Morrical2019,Morrical_TVCG_2020}.

Volume renderings of the original, uniform-grid data in this paper are
generated using nanovdb~\cite{Museth_2021_Nanovdb}, and those of \mesh are
generated using the GPU-based unstructured volume rendering approach of
Morrical \etal~\cite{Morrical2019}, which employs fixed function tree traversal
units to efficiently locate and interpolate unstructured elements.
For this work, we improved the performance of this approach by leveraging the
properties of \mesh~--- axis-aligned cells with predictable sizes. Given
axis-aligned cells, we invert the per-vertex interpolants using analytical
voxel-trilinear interpolation instead of the root-finding method required for
general, curved hexahedra.  Likewise, memory bandwidth is reduced by reading
only two, diagonally opposite vertices instead of all eight as done for
arbitrary hexahedra.

Using these tools, we compare the performance of volume rendering for the
original data and \mesh in terms of the memory footprint and rendering time (as
milliseconds per frame). Whereas our current pipeline demonstrates the
computational benefits of \mesh (small meshes with axis-aligned cells) using
external tools, we envision a more-integrated visualization directly using
\mesh in the future.

%% -----------------------------------------------------------------------------
\vspace{-0.5em}
\subsection{Evaluation of Spatial Hierarchy}

\begin{figure}[!b]
\centering
\vspace{-1.5em}
\includegraphics[width=0.95\linewidth]{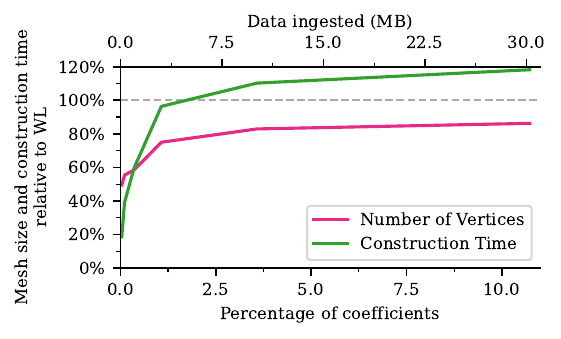}
\vspace{-1em}
\caption{In this example, \mesh improves the data reduction by 20\%--50\% as
compared to the framework of Weiss and Lindstrom~\cite{Weiss:2016}, while
taking up to 20\% more time to construct. However, in the relevant levels of
reduction ($<$2\% of the coefficients), \mesh provides significant
improvement.} \label{fig:compare:size}
\end{figure}

%% -----------------------------------------------------------------------------
First, we compare \mesh's adaptivity in spatial resolution to the recent work
of Weiss and Lindstrom (WL)~\cite{Weiss:2016}. To set up this comparison, we
note that the competing tool creates a reduced representation by filtering
wavelet coefficients by magnitude, akin to the \smag stream. Therefore, for
this comparison, \mesh was generated without the incremental mode, \ie
consuming all filtered data at once.

\begin{figure}[!b]
\vspace{-1.5em}
\centering
{\includegraphics[trim=510 130 230 260,clip,width=0.5\linewidth]{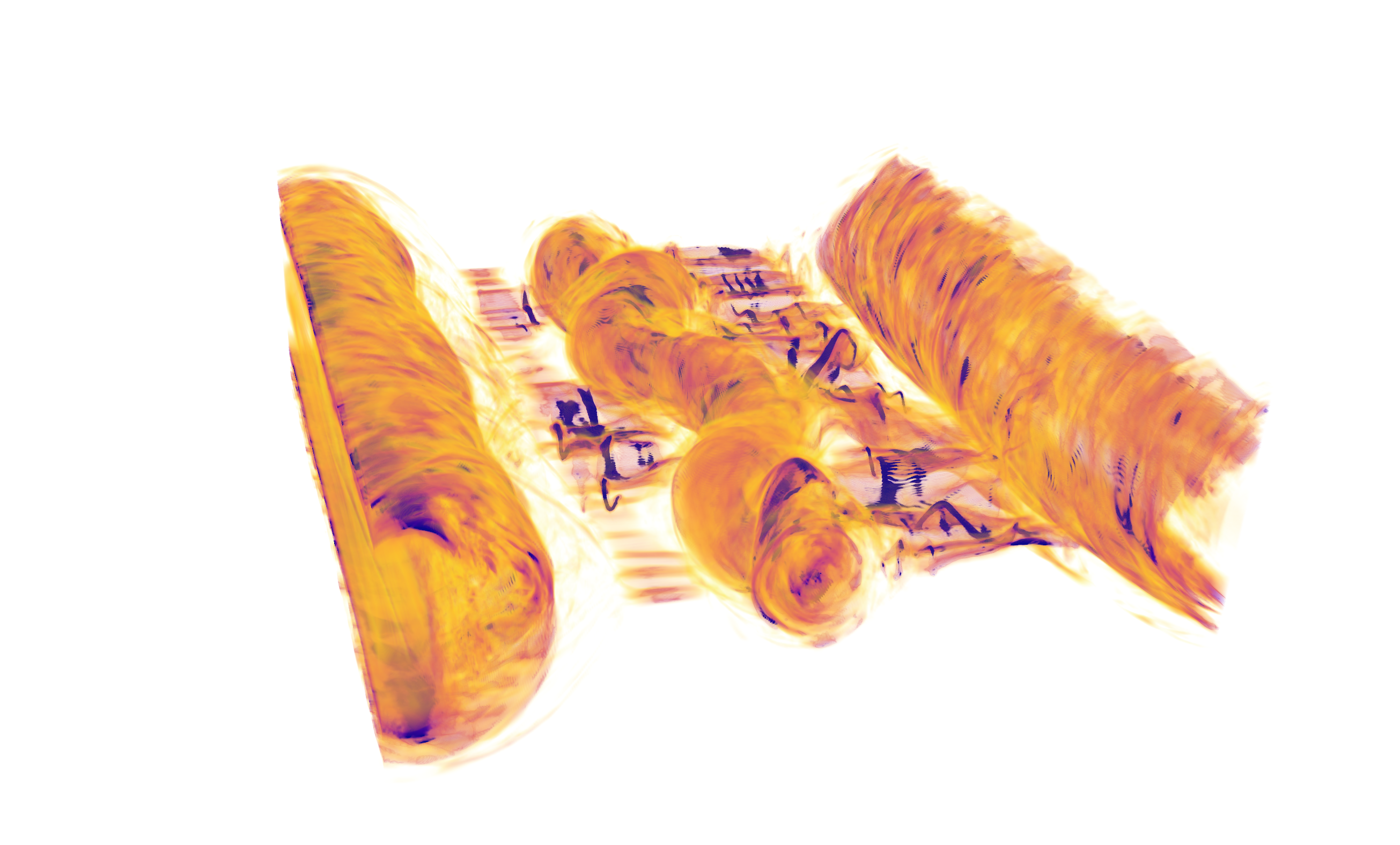}}
\hspace{-2mm}
{\includegraphics[trim=510 130 230 260,clip,width=0.5\linewidth]{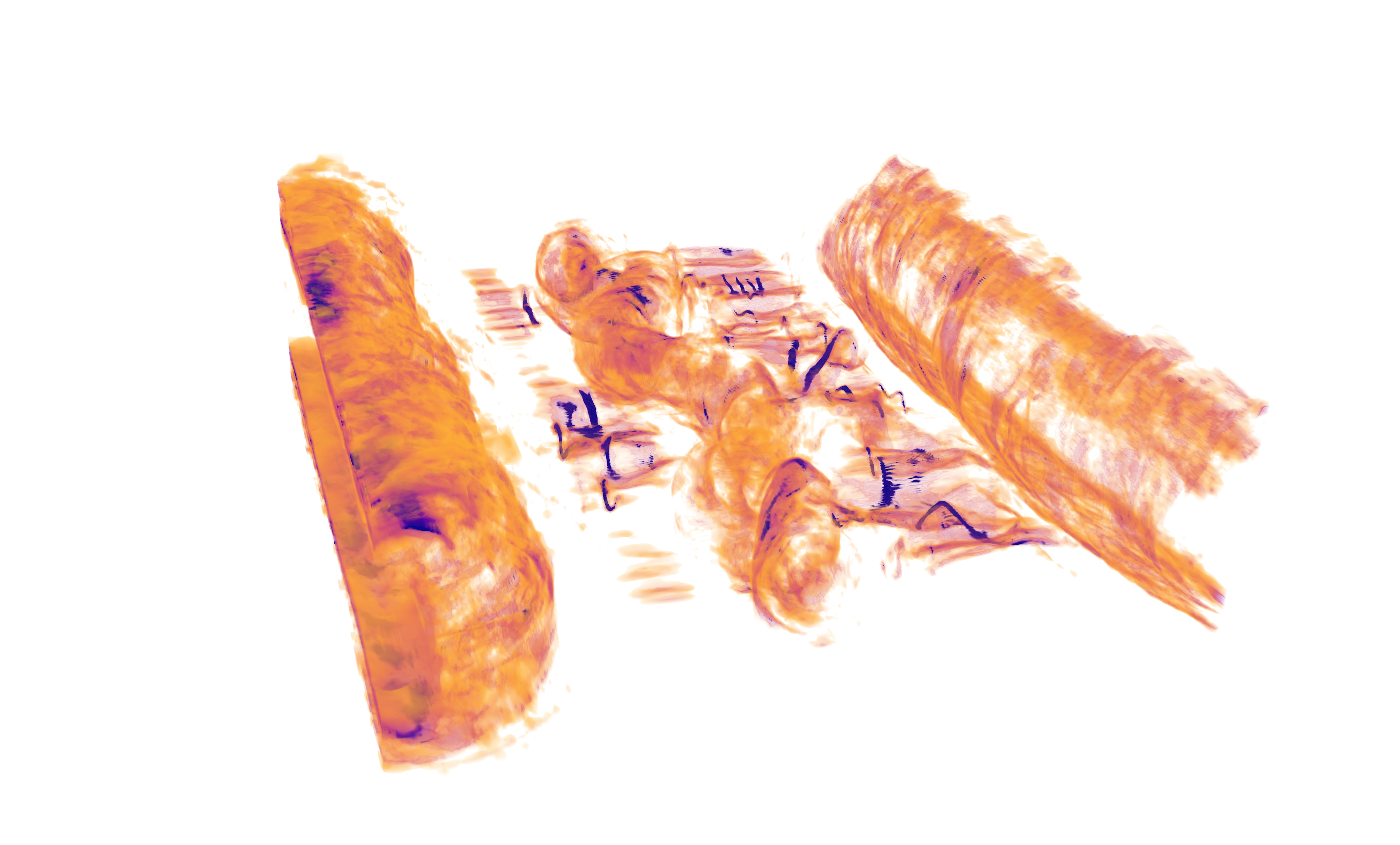}}
\vspace{-1.5em}
\caption{Volume rendering of the magnetic reconnection event dataset at full
resolution (left, 512~MB) and \mesh (right, 9.86~MB) are shown, highlighting
the flexibility of \mesh to handle different types of (dense vs.\ sparse)
datasets and still highlight important features therein.  \label{fig:app2}}
\end{figure}

\begin{figure*}[!b]
\centering
%
% ---------------------------------------------------------------------------
\begin{picture}(150,78)
\put(0,,0){\includegraphics[trim=5 200 10 220,clip,width=0.32\linewidth]
{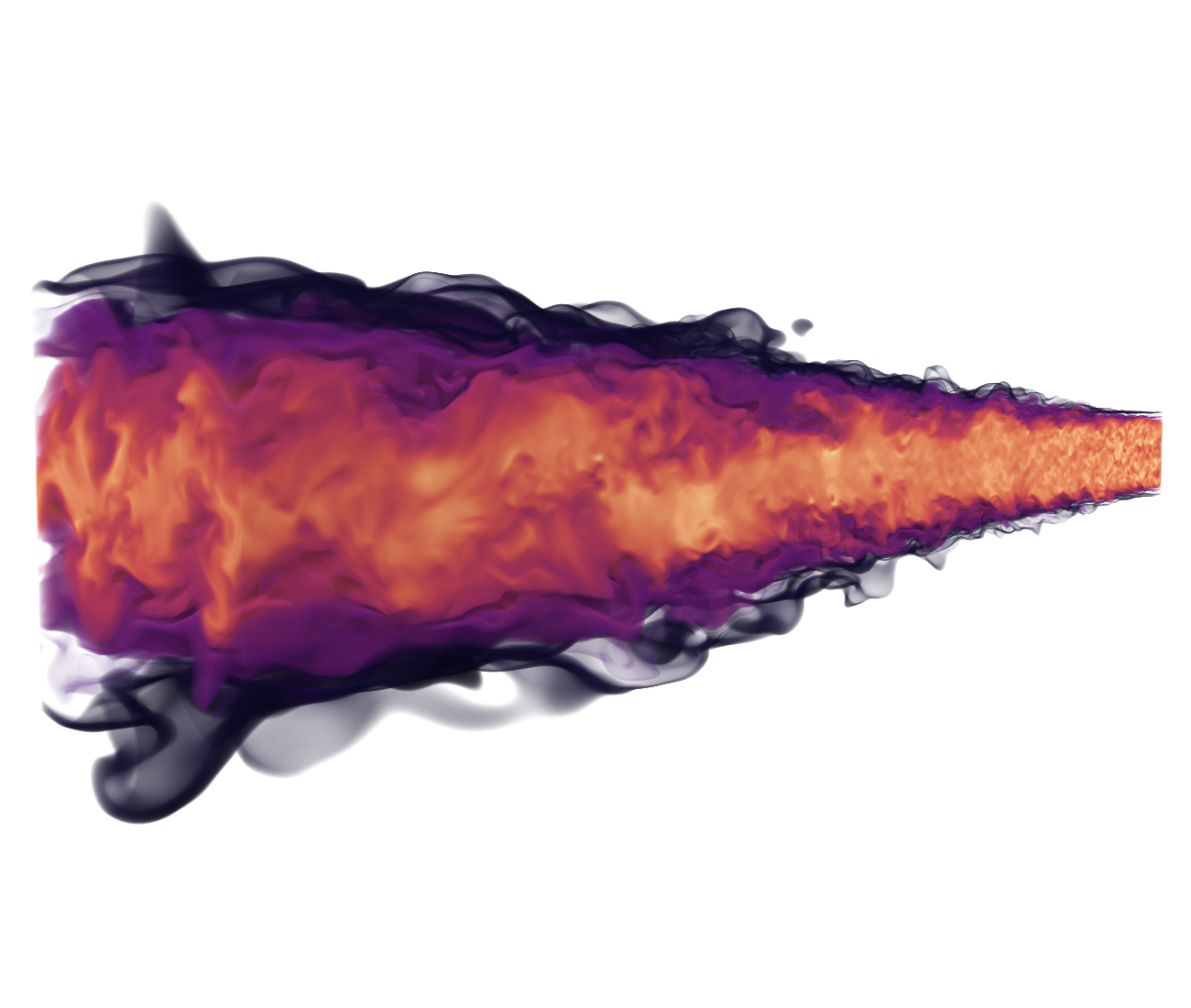}}
\put(70,5){Original Data (9.7 GB)}
\end{picture}
\hspace{1em}
\begin{picture}(150,78)
\put(0,,0){\includegraphics[trim=5 200 10 220,clip,width=0.32\linewidth]
{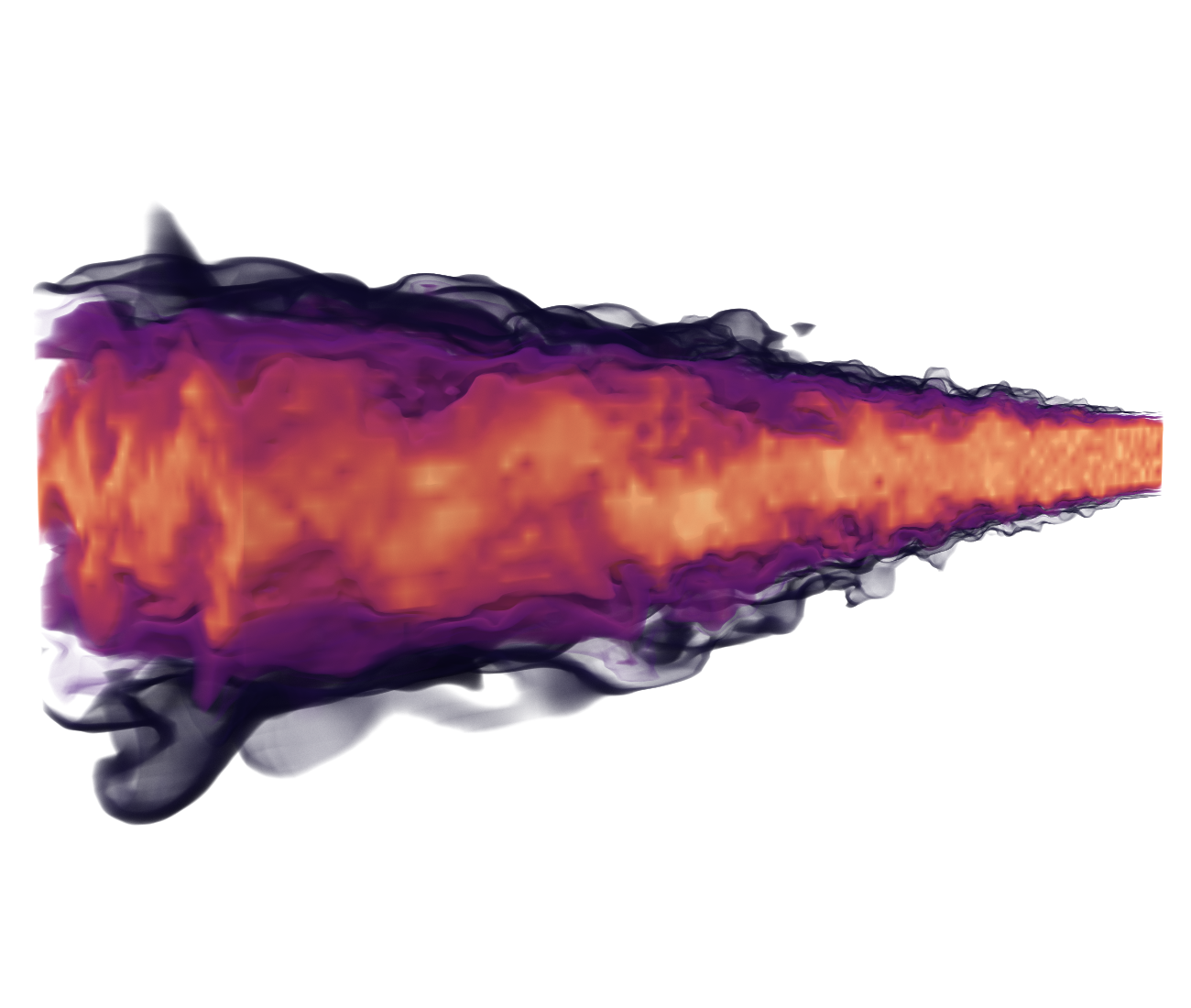}}
\put(102,5){\slvl (14.7 MB)}
\end{picture}
\hspace{1em}
\begin{picture}(150,78)
\put(0,,0){\includegraphics[trim=5 200 10 220,clip,width=0.32\linewidth]
{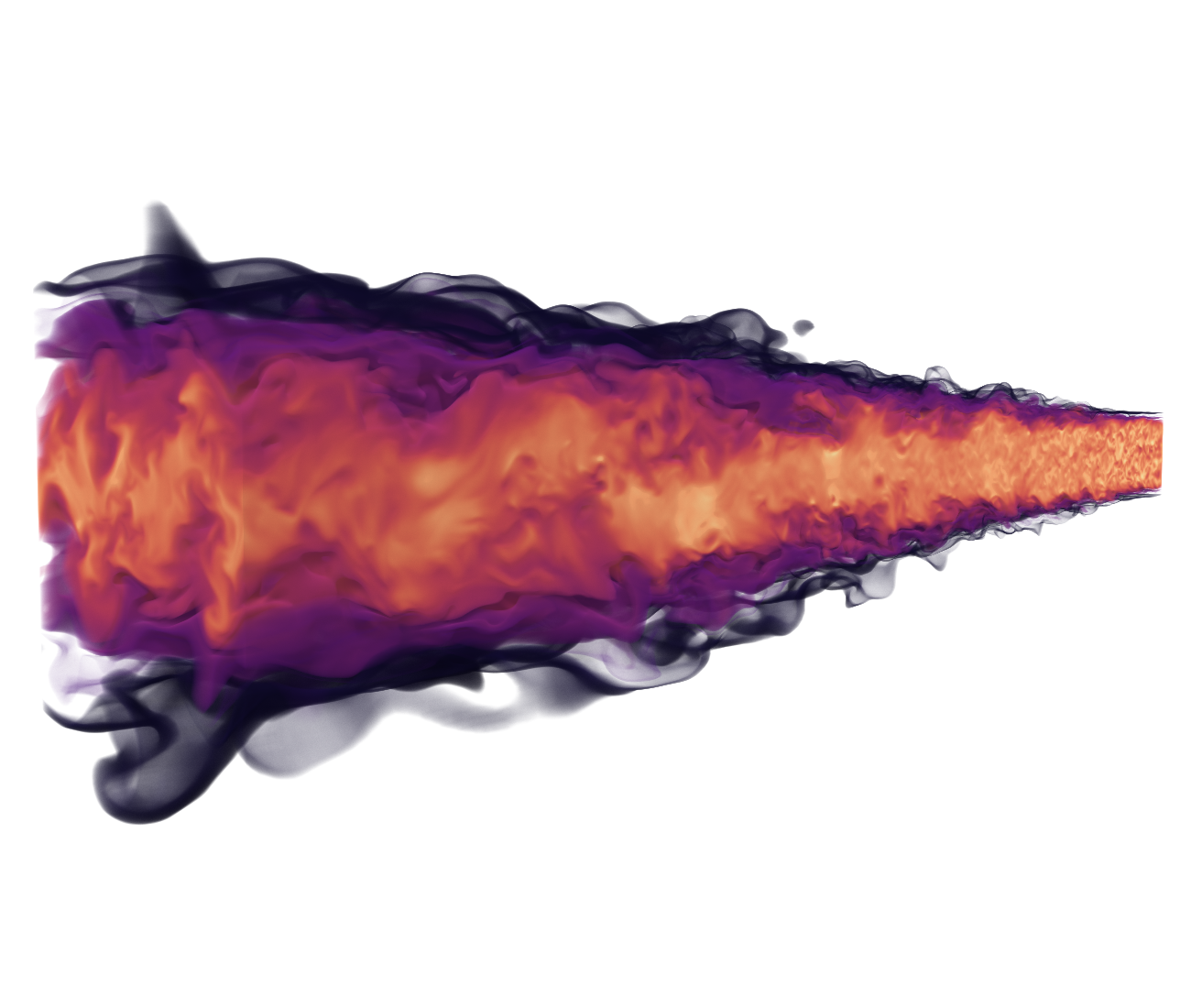}}
\put(102,5){\sbit (863 MB)}
\end{picture}

% ---------------------------------------------------------------------------

\vspace{0.1em}

% ---------------------------------------------------------------------------
\hspace{0.4em}
\includegraphics[width=0.33\linewidth]{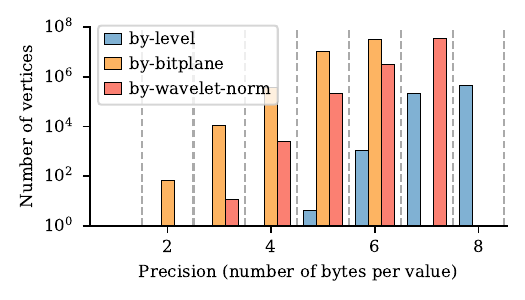}
\hfill
\includegraphics[trim=5 200 10 220,clip,width=0.32\linewidth]
{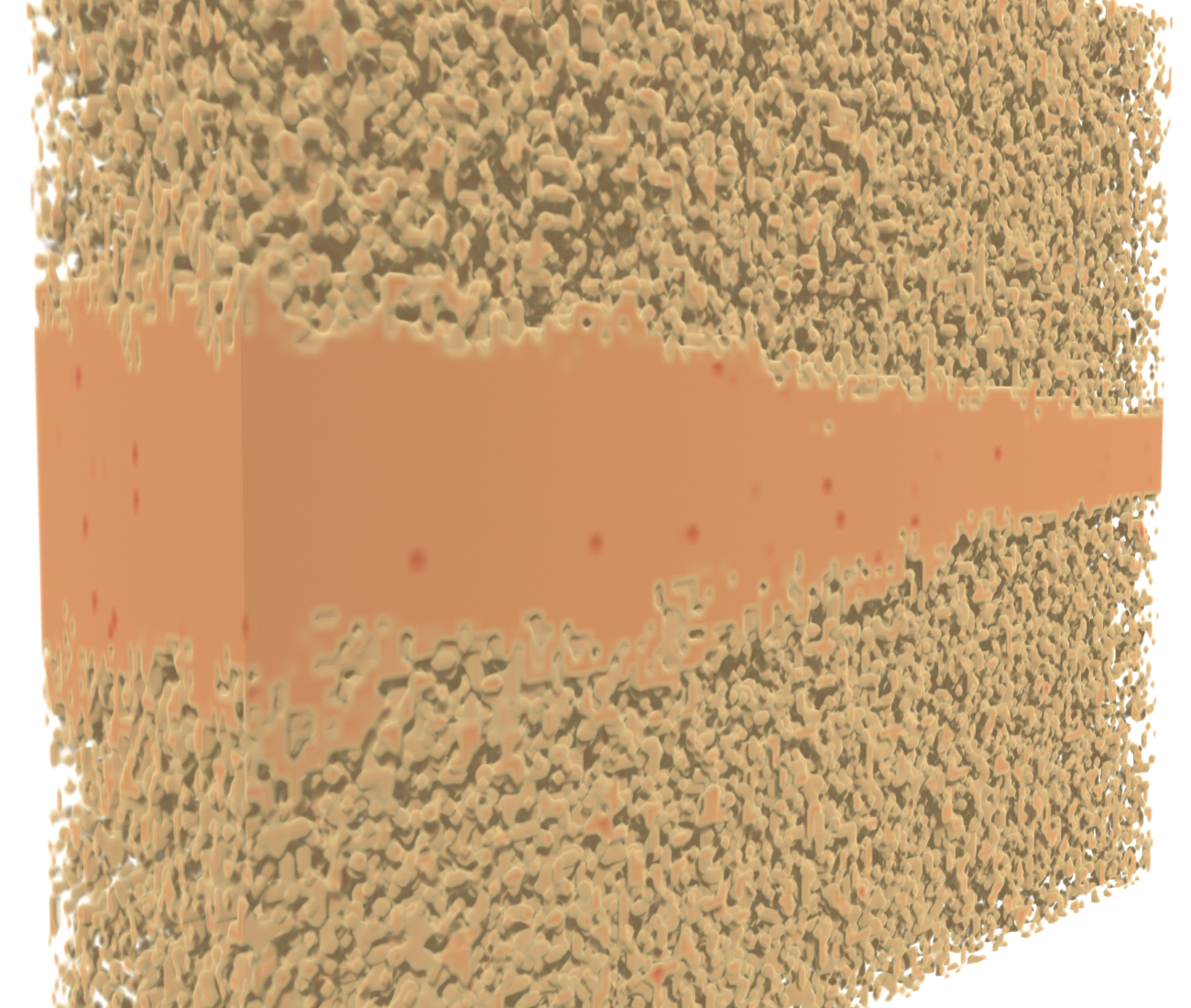}
\hfill
\includegraphics[trim=5 200 10 220,clip,width=0.32\linewidth]
{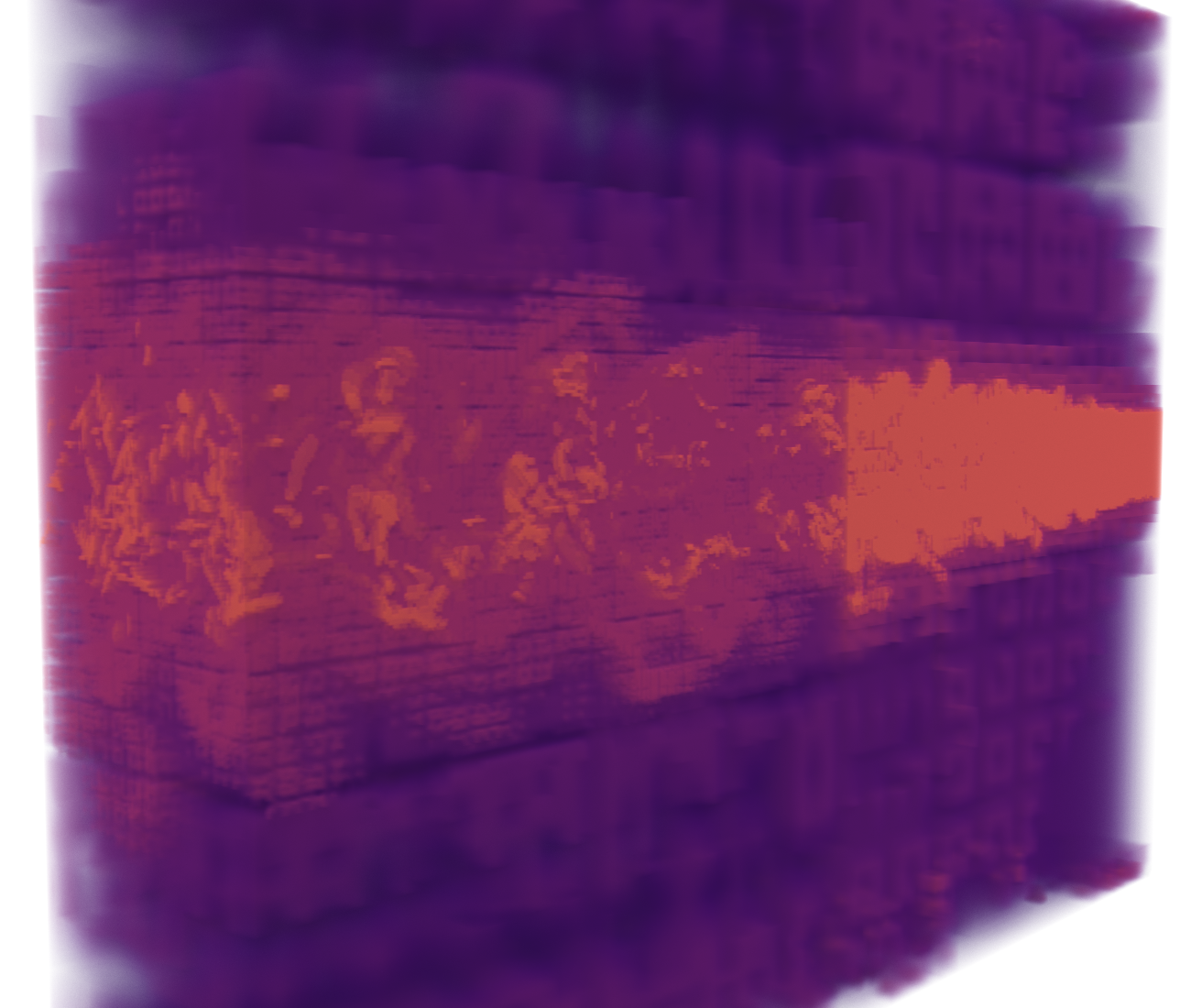}
%
% ---------------------------------------------------------------------------
%
\vspace{-0.5em}
\caption{Using mixed-precision format for vertex values, \mesh provides
additional savings in the representation. The figure compares reduced volumes
(upper row) with the original, and also provides analysis on the distribution
of precision (lower row) in the reduced data as a histogram and as volume
renderings (darker color means low precision).}
\vspace{-0.3em}
\label{fig:precision}
\end{figure*}
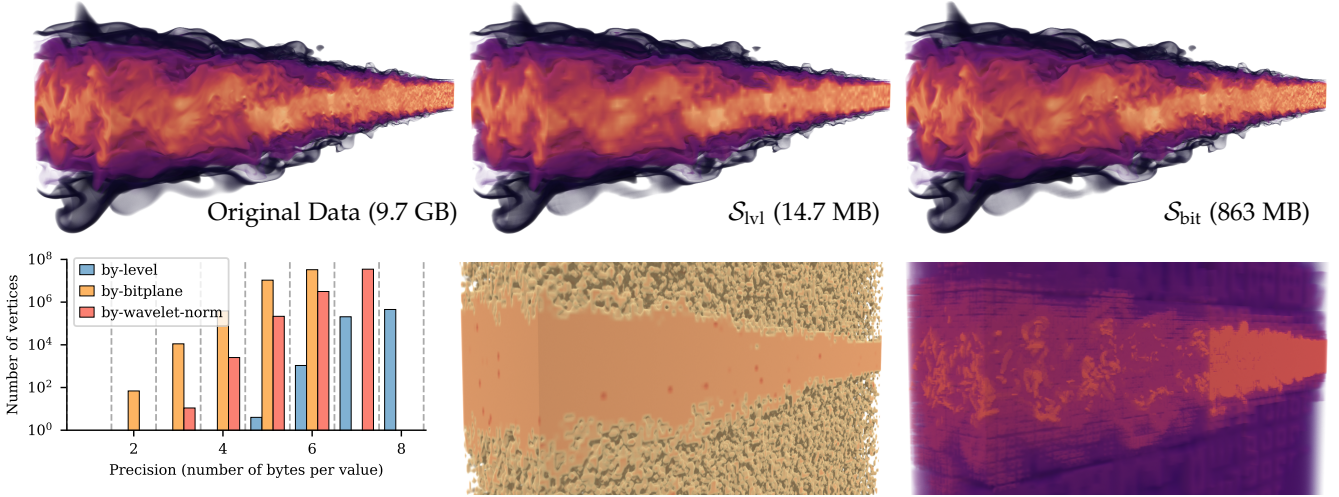

\begin{figure*}[!b]
\centering
\vspace{-0.75em}
\includegraphics[width=0.98\linewidth]{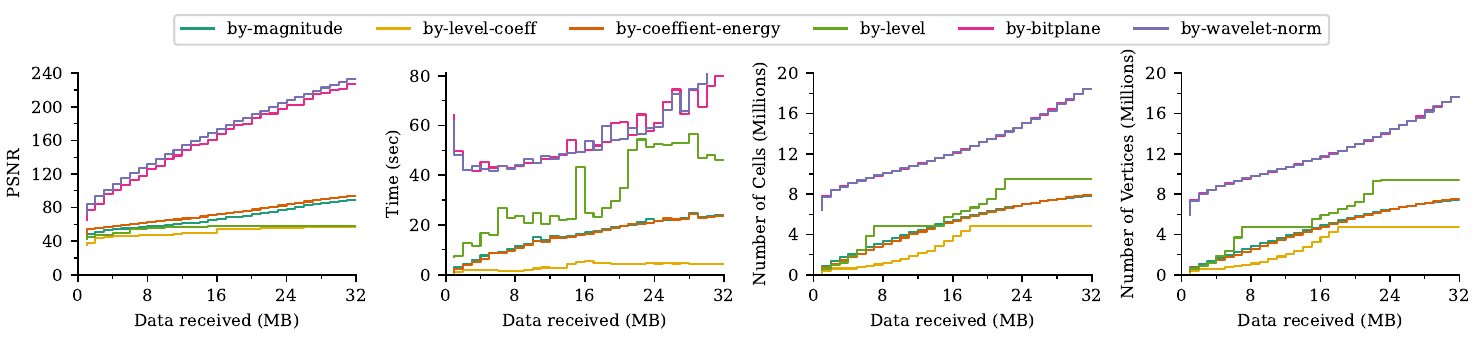}
\vspace{-0.75em}
\caption{\mesh provides a simple and consistent framework to create and compare
reduced representations using different resolution and precision streams. The
plots compare six streams and show that two precision streams
(\emph{by-bitplane} and \emph{by-wavelet-norm}) drastically outperform others
in terms of quality, but at commensurately large mesh sizes.}
\label{fig:streams:plots}
\end{figure*}

%% -----------------------------------------------------------------------------
To concisely present the comparisons, \autoref{fig:compare:size} reports the
relative size and time to compute \mesh with respect to WL.  Here, the
horizontal axis captures an unusually large range of coefficients --- given the
nature and size of this dataset, good quality reconstructions can be obtained
within the first 1--2\% of the coefficients (also shown by
WL~\cite[Fig.~9]{Weiss:2016}).  Within this more reasonable range, \mesh
produces about 20--50\% smaller meshes (not even counting cells) while taking
about the same time for construction. Once the mesh is refined further, \mesh's
gains stabilize, since at this time, there are fewer rectangular nodes, but it
still takes longer to process the more sophisticated hierarchy.

For comparisons, we added a ``standard hierarchy mode'' in \mesh, where
rectangular and improper nodes are  disallowed.  In the absence of a standard
output mesh from WL, the next result uses \mesh to create meshes both with and
without our proposed hierarchy, with the latter  acting as a proxy
for~\cite{Weiss:2016}.
Next, we reduce a $[512\times 512\times512]$ float32 dataset (512~MB), which
represents a magnetic reconnection event in relativistic
plasmas~\cite{PhysRevLett.113.155005}, to roughly the same footprint using both
techniques (about 10~MB each). Without using rectangular  nodes, one can
process only 8192 largest coefficients, giving a mesh with 9.94~MB footprint.
On the other hand, with \mesh's hierarchy, it is possible to utilize 23,040
coefficients with a 9.86~MB footprint, which allows capturing features of
interest in the data (see \autoref{fig:app2}).

Next, we consider the entropy field from a Richtmyer-Meshkov instability
simulation~\cite{Cohen2002} and use \mesh to compare the reduction obtained
using \ssbr and \scen. As shown in \autoref{fig:spatial:rm}, \ssbr (which
streams the data in order of wavelet subbands) sweeps through space rather
evenly, refining large regions to the same depth. On the other hand, \scen
(which scales the coefficients by wavelet function norm) allows capturing more
quickly the turbulent regions of interest. When streamed for 8~MB each, both
streams create different representation sizes, with \scen creating an almost
4$\times$ denser mesh.  Not only does \mesh provide improved data reduction due
to its subdivision flexibility, it also facilitates comparing different modes
of data reduction easily and through a consistent interface.
For this comparison, creating \mesh using \ssbr and \scen took 72.5~s and
86.1~s, respectively.
Volume rendering of the reduced meshes achieved 22.2~ms and 19.7~ms per frame,
with peak memory consumption 2.13 and 2.71~GB, respectively.  Comparing these
numbers with 31.9~ms per frame and 35.07~GB for the original data demonstrates
the computational benefits of \mesh.

%% -----------------------------------------------------------------------------
\vspace{-0.2em}
\subsection{Evaluation of Mixed-Precision Representation}

To demonstrate mixed-precision capabilities in \mesh, we consider the three
precision streams introduced above (\sbit, \slvl, and \swav) and use a
$[2025\times1600\times400]$ float64 dataset from the simulation of a lifted
flame.  We stream 4 MB data (partial coefficients) through these streams and
create \mesh.  Referring to the  histogram plot in \autoref{fig:precision}, we
note the distribution of precision captured by \mesh.  Whereas all streams show
a wide distribution, \sbit, in particular, represents several vertex values
using two bytes only.  Using \mesh, it is straightforward to also highlight the
differences between the precision distribution visually by rendering it as a
volume on the mesh (as shown in the figure for \slvl and \sbit).  Here, we
observe, unsurprisingly, that \slvl provides a wider spatial coverage but with
lower precision, whereas \sbit invests bits heavily based on spatial
resolution.  Finally, the top row of the figure visualizes the resulting
volumes and compares the reduced data with the original dataset at full
resolution and full precision.

%% -----------------------------------------------------------------------------
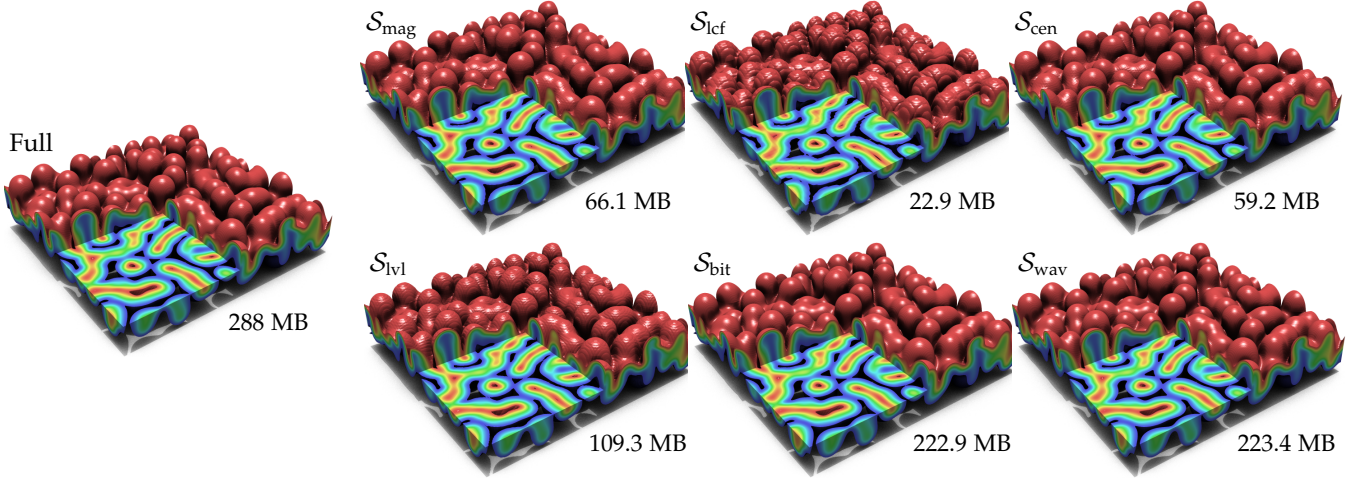
\begin{figure*}
%% -----------------------------------------------------------------------------
\begin{minipage}{0.24\linewidth}
\centering
{\begin{picture}(120,120)
\put(0,0){\includegraphics[trim=150 110 200 190,clip,width=\linewidth]
{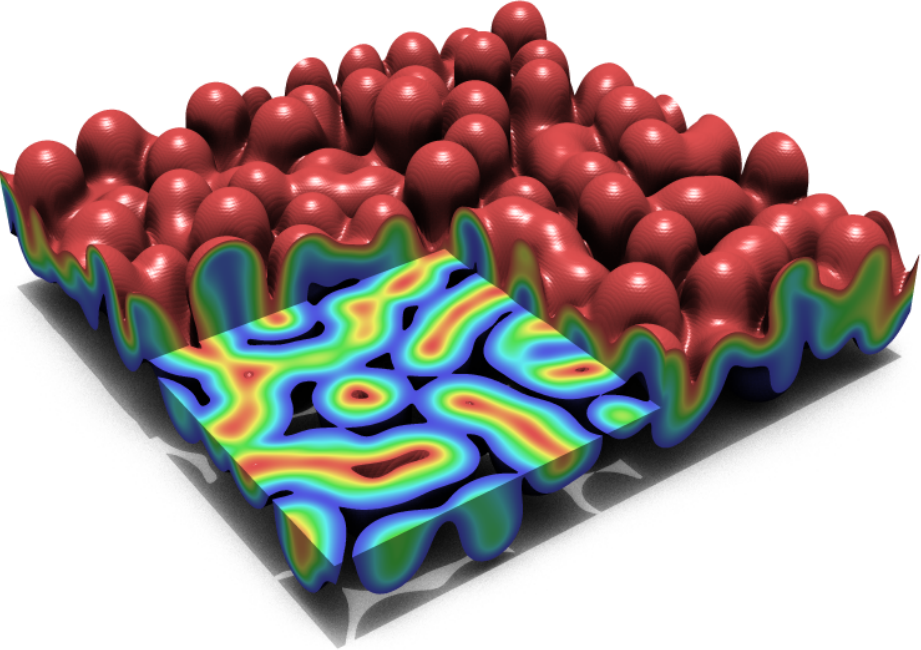}}
\put(2,78){Full}
\put(85,10){{\small{288 MB}}}
\end{picture}}
\end{minipage}%
\hfill
\begin{minipage}{0.76\linewidth}
\centering
\begin{picture}(120,120)
\put(0,0){\includegraphics[trim=150 110 200 190,clip,width=0.32\linewidth]
{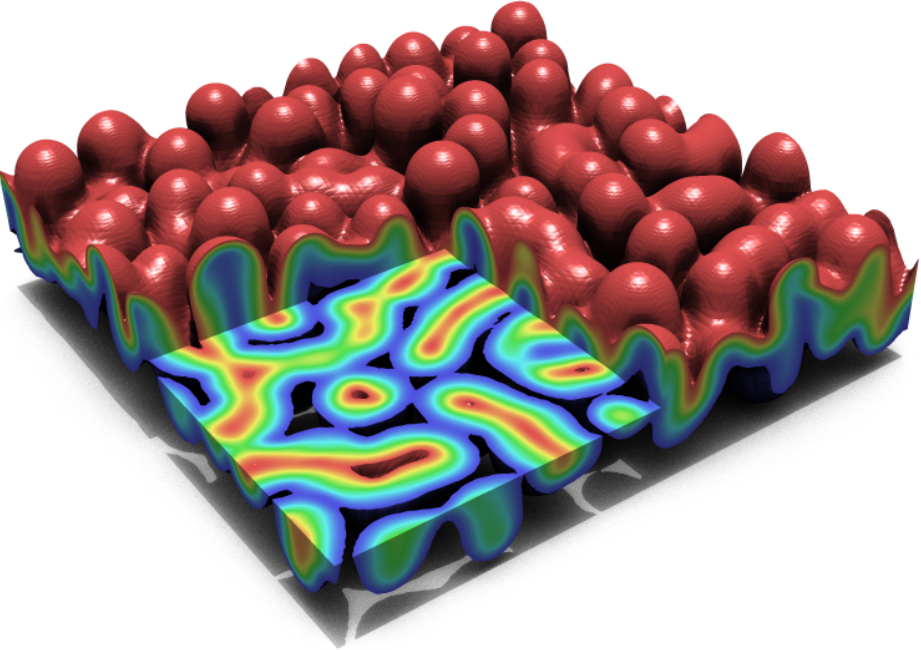}}
\put(2,78){\smag}
\put(85,10){{\small{66.1 MB}}}
\end{picture}
\begin{picture}(120,120)
\put(0,0){\includegraphics[trim=150 110 200 190,clip,width=0.32\linewidth]
{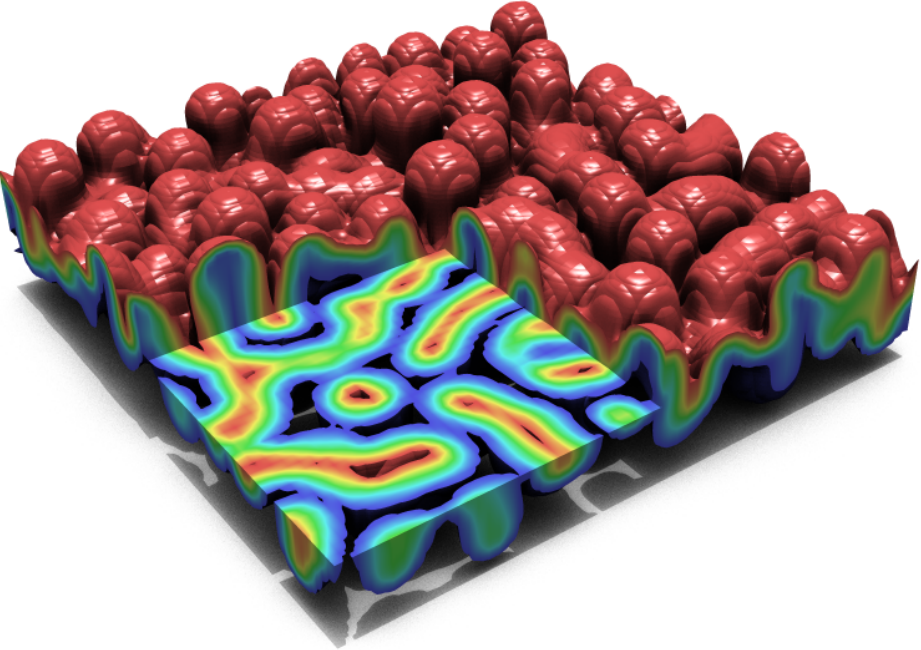}}
\put(2,78){\ssbr}
\put(85,10){{\small{22.9 MB}}}
\end{picture}
\begin{picture}(120,120)
\put(0,0){\includegraphics[trim=150 110 200 190,clip,width=0.32\linewidth]
{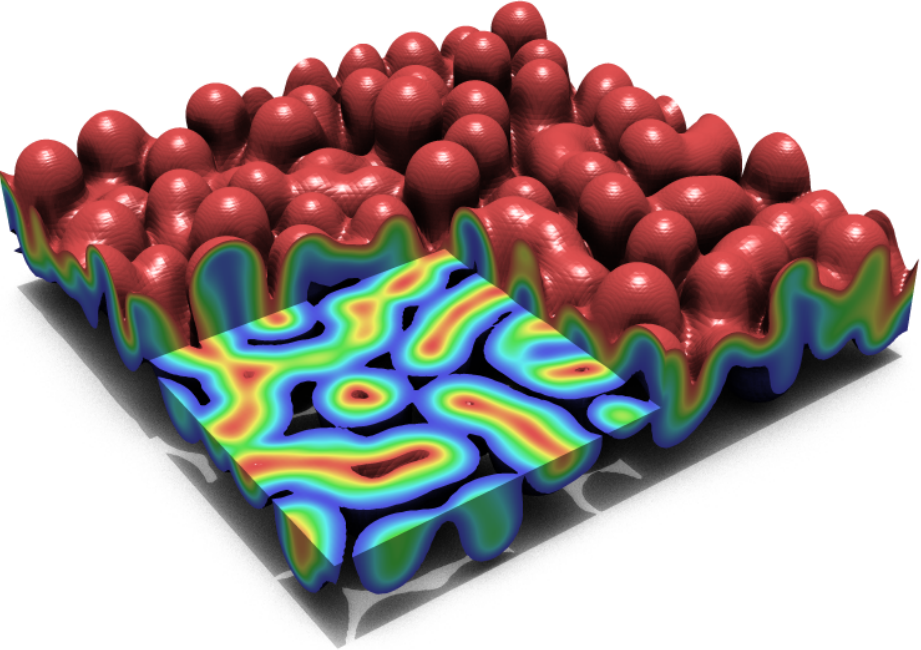}}
\put(2,78){\scen}
\put(85,10){{\small{59.2 MB}}}
\end{picture}
\vspace{-3em}

\begin{picture}(120,120)
\put(0,0){\includegraphics[trim=150 110 200 190,clip,width=0.32\linewidth]
{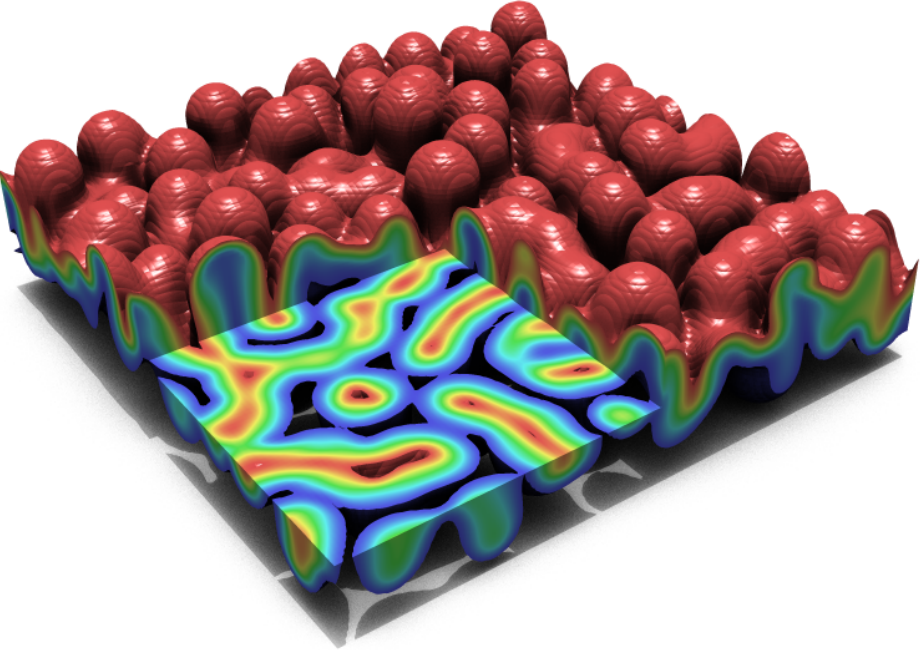}}
\put(2,78){\slvl}
\put(85,10){{\small{109.3 MB}}}
\end{picture}
\begin{picture}(120,120)
\put(0,0){\includegraphics[trim=150 110 200 190,clip,width=0.32\linewidth]
{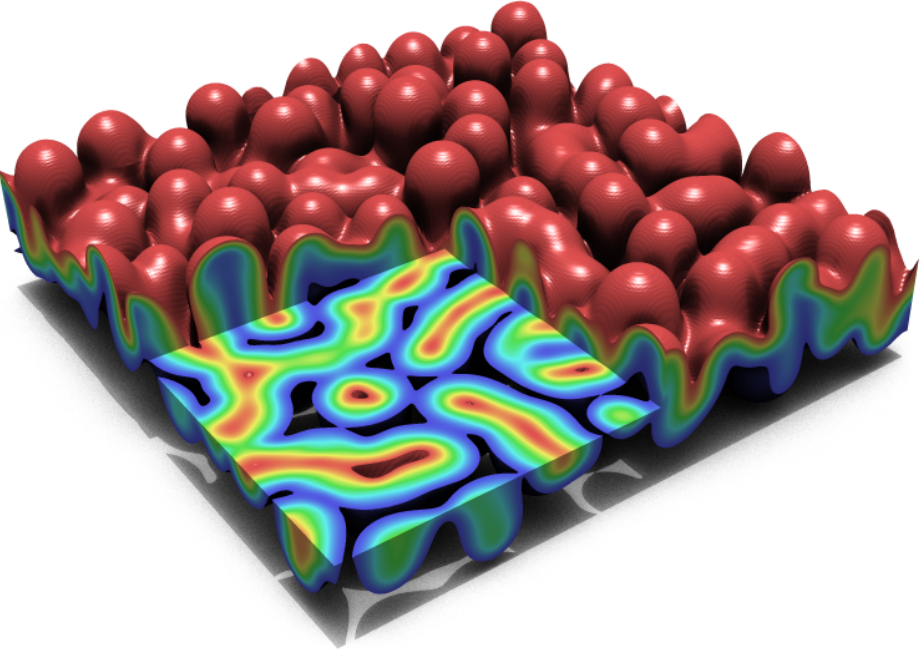}}
\put(2,78){\sbit}
\put(85,10){{\small{222.9 MB}}}
\end{picture}
\begin{picture}(120,120)
\put(0,0){\includegraphics[trim=150 110 200 190,clip,width=0.32\linewidth]
{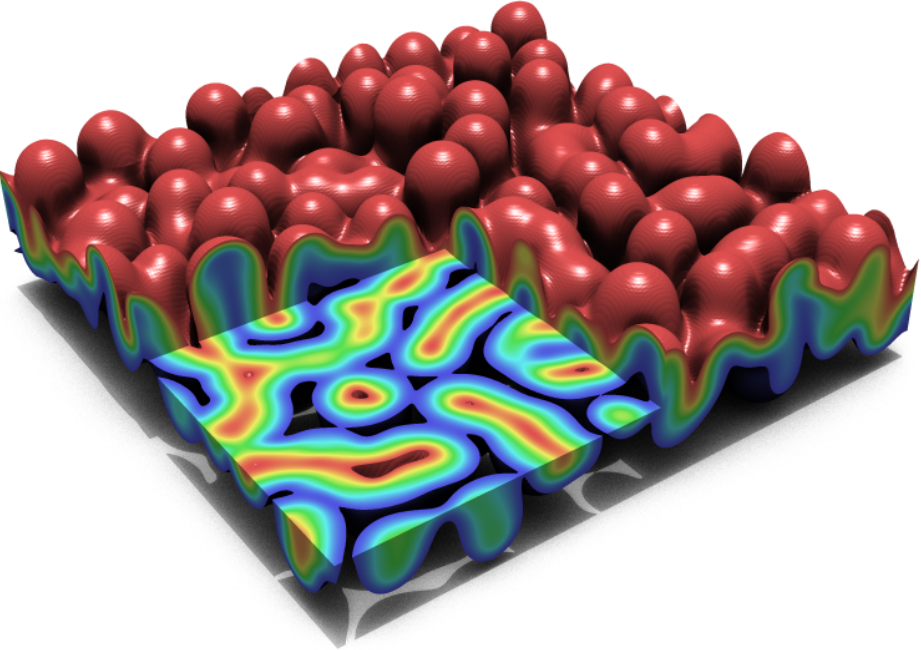}}
\put(2,78){\swav}
\put(85,10){{\small{223.4 MB}}}
\end{picture}
\end{minipage}%
\vspace{-0.5em}
\caption{Visual comparison of data reconstructed after streaming 8 MB for each
of the six streams against the original, full-resolution data. The numbers at
the bottom right provide the approximate memory footprint of the respective
meshes.}
\label{fig:streams:vis}
\vspace{-1em}
\end{figure*}

%% -----------------------------------------------------------------------------
\vspace{-0.3em}
\subsection{Evaluation of Incremental Updates}

Finally, we demonstrate the incremental construction of \mesh using different
types of streams. Here, we work with the Rayleigh-Taylor
instability~\cite{Cabot:2006aa} data defined using float64 values on a
$[384\times384\times256]$ grid. \mesh was created for a stream of 32 MB in
\emph{chunks} of 1 MB, \ie mesh update requests were sent after every 1 MB.
\autoref{fig:streams:plots} shows the evolution of the representation with
respect to these streams in terms of the reconstruction quality (PSNR), time to
update the mesh for each chunk, and the size of the resulting mesh (number of
cells and vertices). First, we note \smag and \scen provide almost equivalent
reconstruction and performance, because the two streams are quite similar and
differ by a scaling factor only, which appears to be nonconsequential for this
particular dataset. Next, we look at \ssbr and \slvl, which although providing
almost equivalent reconstruction quality, refine the mesh very differently.
Whereas \slvl creates larger mesh sizes and, therefore, is slower to compute,
the result indicates that \ssbr should be the choice of stream among the two.
Finally, we observe that, as expected, \sbit and \swav, both precision streams,
refine the mesh drastically from the beginning by transmitting the most
significant bits for most of the turbulent domain, thus not only creating a
dense mesh but also providing a much higher quality. By providing such analysis
in a consistent manner, \mesh facilitates, for the first time, comparing such
datastreams in a much more scalable manner than before~\cite{Hoang:2019}.
Finally, we leave the reader with \autoref{fig:streams:vis} to observe the
visual quality of the resulting representations for each of the streams.

%% -----------------------------------------------------------------------------

\section{Conclusion and Discussion}

In this paper, we present a \emph{resolution-precision-adaptive representation}
that can ingest \emph{arbitrary datastreams} progressively to provide an
\emph{interface to existing tools and algorithms}. By representing uniformly
sampled, scalar-valued data as piecewise multilinear functions using tensor
products of linear B-spline wavelets, we produce meshes based on a flexible
spatial hierarchy to reduce the size of the mesh and demonstrate faithful
reconstruction.  \mesh uses a mixed-precision representation of function values
to further reduce the memory footprint and to facilitate precision adaptivity.
Mixed-precision \mesh can also be used to trim down superfluous bits in the
(full-precision) coefficients received through spatial streams.  Currently,
\mesh's mixed-precision mode provides a lossless representation, but it is
straightforward to expand this framework to achieve lossy reduction by dropping
additional bits within acceptable error.

\mesh provides a VTK interface through output files, which still somewhat
introduces additional overhead in using \mesh representations. In the future,
we would like to expand our framework to expose an improved API with efficient
iterators, accessors, and common visualization queries, such as point queries
and contouring, which will facilitate leveraging this framework directly for
visualization.

The current implementation of \mesh is serial and CPU-based. Whereas it is easy
to conceive porting key operations to GPU kernels and/or to a distributed
algorithm, the primary challenge appears to be reducing data movement across
host and device or across MPI ranks, especially in a streaming setting where it
is not known a priori which parts of the tree will be updated. Addressing these
challenges offers interesting directions for future research.

We have also identified several opportunities for further improvement in the
performance of \mesh creation. Most notably, the stamping process for vertices
requires tree traversals and is the key bottleneck in further scaling. On the
other hand, the lifting process is a standard and efficient technique for
inverse transforms, but works with dense data (coefficients on uniform grid).
In the future, we will explore the possibilities of utilizing the lifting
approach for sparse data (restricted coefficients) and their stencils.

When compared to state-of-the-art compressors, such as \zfp or SZ, \mesh does
not offer competitive rate-distortion curves, since it does not currently
employ sophisticated compression to further reduce the memory footprint of the
adaptive mesh.  On the other hand, such compressors do not reduce the number of
vertices and cells and, thus, do not benefit the performance of downstream
tasks most of the time.  In contrast, by reducing the number of vertices and
cells in the representation, \mesh offers significant computational advantages
during traversal; current work demonstrates about 50\% faster volume rendering.
Compared to the technique closest to our own~\cite{Weiss:2016}, \mesh produces
significantly smaller meshes (20--50\% gain) at the same data quality.  An
important research direction for the future is to combine the best of \mesh and
compression, \eg by utilizing \zfp to replace the mixed-precision, block-based
representation of vertices currently used by \mesh.

Finally, a key enabling technology in \mesh is the support for incremental
updates using arbitrary sequences of data. By relaxing the assumption that the
data must be ordered in specific, predefined ways, we aim to position our
open-source tool (\href{https://github.com/llnl/amm}{github.com/llnl/amm}) as a
standard framework to explore and evaluate the reduction of data in the space
of precision and resolution, ultimately resulting in next-generation data
reduction techniques.

% % ---------------------------------------------------------------------------
% % use section* for acknowledgment
\ifCLASSOPTIONcompsoc
  \section*{Acknowledgments}
\else
  \section*{Acknowledgment}
\fi
This work was performed under the auspices of the U.S.\ Department of 
Energy (DOE) by Lawrence Livermore 
National Laboratory under Contract DE-AC52-07NA27344 and supported by the 
LLNL-LDRD Program under Project No. 17-SI-004. We thank Jeffrey Hittinger 
for insightful discussions during this project.
This work was funded in part by NSF OAC awards 2127548, 1941085, 2138811 NSF
CMMI awards 1629660, DOE award DE-FE0031880, and the Intel Graphics and
Visualization Institute of XeLLENCE, and oneAPI Center of Excellence.
LLNL-JRNL-771697.

%\clearpage
\bibliographystyle{IEEEtran}
\bibliography{IEEEabrv, paper}

\begin{thebibliography}{10}
\providecommand{\url}[1]{#1}
\csname url@samestyle\endcsname
\providecommand{\newblock}{\relax}
\providecommand{\bibinfo}[2]{#2}
\providecommand{\BIBentrySTDinterwordspacing}{\spaceskip=0pt\relax}
\providecommand{\BIBentryALTinterwordstretchfactor}{4}
\providecommand{\BIBentryALTinterwordspacing}{\spaceskip=\fontdimen2\font plus
\BIBentryALTinterwordstretchfactor\fontdimen3\font minus
  \fontdimen4\font\relax}
\providecommand{\BIBforeignlanguage}[2]{{%
\expandafter\ifx\csname l@#1\endcsname\relax
\typeout{** WARNING: IEEEtran.bst: No hyphenation pattern has been}%
\typeout{** loaded for the language `#1'. Using the pattern for}%
\typeout{** the default language instead.}%
\else
\language=\csname l@#1\endcsname
\fi
#2}}
\providecommand{\BIBdecl}{\relax}
\BIBdecl

\bibitem{DOE:2013}
J.~Chen, A.~Choudhary, S.~Feldman, B.~Hendrickson, C.~R. Johnson, R.~Mount,
  V.~Sarkar, V.~White, and D.~Williams, \emph{Synergistic Challenges in
  Data-Intensive Science and Exascale Computing: DOE ASCAC Data Subcommittee
  Report}, 2013.

\bibitem{Laney13}
D.~E. Laney, S.~Langer, C.~Weber, P.~Lindstrom, and A.~Wegener, ``Assessing the
  effects of data compression in simulations using physically motivated
  metrics,'' in \emph{Int. Conf. for High Perf. Computing, Networking, Storage,
  and Analysis (SC)}, 2013, pp. 1--12.

\bibitem{Iverson12}
J.~Iverson, C.~Kamath, and G.~Karypis, ``Fast and effective lossy compression
  algorithms for scientific datasets,'' in \emph{European Conf. on Parallel
  Processing}, 2012, pp. 843--856.

\bibitem{zfp}
P.~Lindstrom, ``Fixed-rate compressed floating-point arrays,'' \emph{IEEE
  Trans. on Vis. and Comp. Graph.}, vol.~20, no.~12, pp. 2674--2683, 2014.

\bibitem{Shekhar:1996:ODM}
R.~Shekhar, E.~Fayyad, R.~Yagel, and J.~F. Cornhill, ``Octree-based decimation
  of marching cubes surfaces,'' in \emph{Proc. of 7th Annual IEEE Vis.}, 1996,
  pp. 335--342.

\bibitem{Pascucci:2001:GSI}
V.~Pascucci and R.~Frank, ``Global static indexing for real-time exploration of
  very large regular grids,'' in \emph{Proc.\ of the 2001 ACM/IEEE Conf. on
  Supercomputing}, 2001, pp. 45--45.

\bibitem{Hoang:2019}
D.~Hoang, P.~Klacansky, H.~Bhatia, P.-T. Bremer, P.~Lindstrom, and V.~Pascucci,
  ``A study of the trade-off between reducing precision and reducing resolution
  for data analysis and visualization,'' \emph{IEEE Trans. on Vis. and Comp.
  Graph.}, vol.~25, no.~1, pp. 1193--1203, 2019.

\bibitem{Hoang2021}
D.~Hoang, B.~Summa, H.~Bhatia, P.~Lindstrom, P.~Klacansky, W.~Usher, P.-T.
  Bremer, and V.~Pascucci, ``Efficient and flexible hierarchical data layouts
  for a unified encoding of scalar field precision and resolution,'' \emph{IEEE
  Trans. on Vis. and Comp. Graph.}, vol.~27, no.~2, pp. 603--613, 2021.

\bibitem{vtk}
W.~Schroeder, K.~Martin, and B.~Lorensen, \emph{The Visualization Toolkit},
  4th~ed.\hskip 1em plus 0.5em minus 0.4em\relax Kitware, 2006.

\bibitem{paraview}
J.~Ahrens, B.~Geveci, and C.~Law, ``{ParaView: A}n end-user tool for large data
  visualization,'' \emph{Visualization Handbook}, 2005.

\bibitem{visit}
H.~Childs, E.~Brugger, B.~Whitlock, J.~Meredith, S.~Ahern, D.~Pugmire,
  K.~Biagas, M.~Miller, C.~Harrison, G.~H. Weber, H.~Krishnan, T.~Fogal,
  A.~Sanderson, C.~Garth, E.~W. Bethel, D.~Camp, O.~R\"{u}bel, M.~Durant, J.~M.
  Favre, and P.~Navr\'{a}til, ``{VisIt: A}n end-user tool for visualizing and
  analyzing very large data,'' in \emph{High Performance
  Visualization--Enabling Extreme-Scale Scientific Insight}, 2012.

\bibitem{Morrical2019}
N.~Morrical, W.~Usher, I.~Wald, and V.~Pascucci, ``Efficient space skipping and
  adaptive sampling of unstructured volumes using hardware accelerated ray
  tracing,'' in \emph{IEEE Vis. Conf. (VIS)}, 2019, pp. 256--260.

\bibitem{Morrical_TVCG_2020}
N.~Morrical, I.~Wald, W.~Usher, and V.~Pascucci, ``Accelerating unstructured
  mesh point location with {RT} cores,'' \emph{IEEE Trans. Vis. Comp. Graph.},
  2020.

\bibitem{wald_ospray_2017}
I.~Wald, G.~P. Johnson, J.~Amstutz, C.~Brownlee, A.~Knoll, J.~Jeffers,
  J.~G\"unther, and P.~Navr\'atil, ``{{OSPRay}} - {{A CPU Ray Tracing
  Framework}} for {{Scientific Visualization}},'' \emph{IEEE Trans. Vis. Comp.
  Graph.}, vol.~23, no.~1, pp. 931--940, 2017.

\bibitem{Weiss:2016}
K.~Weiss and P.~Lindstrom, ``Adaptive multilinear tensor product wavelets,''
  \emph{IEEE Trans. on Vis. and Comp. Graph.}, vol.~22, no.~1, pp. 985--994,
  2016.

\bibitem{Cohen92}
A.~Cohen, I.~Daubechies, and J.-C. Feauveau, ``Biorthogonal bases of compactly
  supported wavelets,'' \emph{Comm. on Pure and Appl. Math.}, vol.~45, no.~5,
  pp. 485--560, 1992.

\bibitem{Fogal10:gpu}
T.~Fogal, H.~Childs, S.~Shankar, J.~Kr\"uger, R.~D. Bergeron, and P.~Hatcher,
  ``Large data visualization on distributed memory multi-{GPU} clusters,'' in
  \emph{High Perf. Graph.}, 2010, p. 57–66.

\bibitem{Woodring11}
J.~Woodring, J.~Ahrens, J.~Figg, J.~Wendelberger, S.~Habib, and K.~Heitmann,
  ``In-situ sampling of a large-scale particle simulation for interactive
  visualization and analysis,'' in \emph{Proc. of the 13th Eurographics Conf.
  on Vis.}, 2011, pp. 1151--1160.

\bibitem{JACKINS1980249}
C.~L. Jackins and S.~L. Tanimoto, ``Oct-trees and their use in representing
  three-dimensional objects,'' \emph{Comput. Graph. Image Process.}, vol.~14,
  no.~3, pp. 249--270, 1980.

\bibitem{Crassin09:gigavoxels}
C.~Crassin, F.~Neyret, S.~Lefebvre, and E.~Eisemann, ``{GigaVoxels: R}ay-guided
  streaming for efficient and detailed voxel rendering,'' in \emph{Proc. of the
  Symp. on Interactive 3D Graph. and Games}, 2009, pp. 15--22.

\bibitem{Gobbetti2008}
E.~Gobbetti, F.~Marton, and J.~A. Iglesias~Guiti\'{a}n, ``A single-pass gpu ray
  casting framework for interactive out-of-core rendering of massive volumetric
  datasets,'' \emph{Vis. Comput.}, vol.~24, no.~7, pp. 797--806, 2008.

\bibitem{large-scale-volume}
J.~Beyer, M.~Hadwiger, and H.~Pfister, ``State-of-the-art in {GPU}-based
  large-scale volume visualization,'' \emph{Comp. Graph. Forum}, vol.~34,
  no.~8, p. 13–37, 2015.

\bibitem{Museth:2013:vdb}
K.~Museth, ``{VDB: H}igh-resolution sparse volumes with dynamic topology,''
  \emph{ACM Trans. Graph.}, vol.~32, no.~3, 2013.

\bibitem{Setaluri:2014}
R.~Setaluri, M.~Aanjaneya, S.~Bauer, and E.~Sifakis, ``{SPGrid}: A sparse paged
  grid structure applied to adaptive smoke simulation,'' \emph{ACM Trans.
  Graph.}, vol.~33, no.~6, 2014.

\bibitem{Berger1989:AMR}
M.~Berger and P.~Colella, ``Local adaptive mesh refinement for shock
  hydrodynamics,'' \emph{J. Comp. Phys.}, vol.~82, no.~1, pp. 64--84, 1989.

\bibitem{Dubey14}
A.~Dubey, A.~Almgren, J.~Bell, M.~Berzins, S.~Brandt, G.~Bryan, P.~Colella,
  D.~Graves, M.~Lijewski, F.~L\"{o}ffler, B.~O'Shea, E.~Schnetter,
  B.~Van~Straalen, and K.~Weide, ``A survey of high level frameworks in
  block-structured adaptive mesh refinement packages,'' \emph{J. Parallel
  Distrib. Comput.}, vol.~74, no.~12, p. 3217–3227, 2014.

\bibitem{Weber2001ExtractionOC}
G.~H. Weber, O.~Kreylos, T.~J. Ligocki, J.~M. Shalf, H.~Hagen, B.~Hamann, and
  K.~I. Joy, ``Extraction of crack-free isosurfaces from adaptive mesh
  refinement data,'' in \emph{Eurographics / IEEE VGTC Symp. on Visualization},
  2001.

\bibitem{Moran11}
P.~Moran and D.~Ellsworth, ``Visualization of {AMR} data with multi-level
  dual-mesh interpolation,'' \emph{IEEE Trans. Vis. Comp. Graph.}, vol.~17,
  no.~12, pp. 1862--1871, 2011.

\bibitem{Ljung2006MultiresolutionII}
P.~Ljung, C.~F. Lundstr{\"o}m, and A.~Ynnerman, ``Multiresolution interblock
  interpolation in direct volume rendering,'' in \emph{Proc.\ of the Eighth
  Joint Eurographics / IEEE VGTC Conf. on Vis.}, 2006, pp. 259--266.

\bibitem{Wang2018CPUIR}
F.~Wang, I.~Wald, Q.~Wu, W.~Usher, and C.~R. Johnson, ``{CPU} isosurface ray
  tracing of adaptive mesh refinement data,'' \emph{IEEE Trans. Vis. Comp.
  Graph.}, vol.~25, no.~1, pp. 1142--1151, 2019.

\bibitem{vapor}
S.~Li, S.~Jaroszynski, S.~Pearse, L.~Orf, and J.~Clyne, ``Vapor: A
  visualization package tailored to analyze simulation data in earth system
  science,'' \emph{Atmosphere}, vol.~10, no.~9, 2019.

\bibitem{treib}
M.~Treib, K.~Bürger, F.~Reichl, C.~Meneveau, A.~Szalay, and R.~Westermann,
  ``Turbulence visualization at the terascale on desktop pcs,'' \emph{IEEE
  Trans. Vis. Comp. Graph.}, vol.~18, no.~12, pp. 2169--2177, 2012.

\bibitem{li-2017-spatiotemporal}
S.~Li, S.~Sane, L.~Orf, P.~Mininni, J.~Clyne, and H.~Childs, ``Spatiotemporal
  wavelet compression for visualization of scientific simulation data,'' in
  \emph{IEEE Int. Conf. on Cluster Computing}, 2017, pp. 216--227.

\bibitem{woodring-11-revisiting}
J.~Woodring, S.~Mniszewski, C.~Brislawn, D.~DeMarle, and J.~Ahrens,
  ``Revisiting wavelet compression for large-scale climate data using jpeg 2000
  and ensuring data precision,'' in \emph{IEEE Symp. on Large Data Analysis and
  Vis.}, 2011, pp. 31--38.

\bibitem{Bertram04}
M.~Bertram, M.~A. Duchaineau, B.~Hamann, and K.~I. Joy, ``Generalized b-spline
  subdivision-surface wavelets for geometry compression,'' \emph{IEEE Trans.
  Vis. Comp. Graph.}, vol.~10, no.~3, pp. 326--338, 2004.

\bibitem{Gross96}
M.~H. Gross, O.~G. Staadt, and R.~Gatti, ``Efficient triangular surface
  approximations using wavelets and quadtree data structures,'' \emph{IEEE
  Trans. on Vis. and Comp. Graph.}, vol.~2, no.~2, pp. 130--143, 1996.

\bibitem{SCI:Lin2004a}
L.~Linsen, V.~Pascucci, M.~Duchaineau, B.~Hamann, and K.~Joy, ``Wavelet-based
  multiresolution with n-th-root-of-2,'' in \emph{Dagstuhl Seminar on Geometric
  Modeling}, 2004.

\bibitem{Lorensen:1987:MC}
W.~E. Lorensen and H.~E. Cline, ``{Marching Cubes: A} high resolution {3D}
  surface construction algorithm,'' \emph{SIGGRAPH Comput. Graph.}, vol.~21,
  no.~4, 1987.

\bibitem{CIGNONI2000}
P.~Cignoni, F.~Ganovelli, C.~Montani, and R.~Scopigno, ``Reconstruction of
  topologically correct and adaptive trilinear isosurfaces,'' \emph{Comput.
  Graph.}, vol.~24, no.~3, pp. 399--418, 2000.

\bibitem{Nielson03}
G.~M. Nielson, ``On marching cubes,'' \emph{IEEE Trans. Vis. Comp. Graph.},
  vol.~9, no.~3, pp. 283--297, 2003.

\bibitem{Ament10}
M.~Ament, D.~Weiskopf, and H.~Carr, ``Direct interval volume visualization,''
  \emph{IEEE Trans. Vis. Comp. Graph.}, vol.~16, no.~6, pp. 1505--1514, 2010.

\bibitem{zenger1991sparse}
C.~Zenger and W.~Hackbusch, ``Sparse grids,'' in \emph{Research Workshop of the
  Israel Science Foundation on Multiscale Phenomenon, Modelling and
  Computation}, 1991, p.~86.

\bibitem{garcke2012sparse}
J.~Garcke, ``Sparse grids in a nutshell,'' in \emph{Sparse grids and
  applications}.\hskip 1em plus 0.5em minus 0.4em\relax Springer, 2012, pp.
  57--80.

\bibitem{tao2017significantly}
D.~Tao, S.~Di, Z.~Chen, and F.~Cappello, ``Significantly improving lossy
  compression for scientific data sets based on multidimensional prediction and
  error-controlled quantization,'' in \emph{IEEE Int. Par. and Dist. Process.
  Symp.}, 2017, pp. 1129--1139.

\bibitem{ballester2019tthresh}
R.~Ballester-Ripoll, P.~Lindstrom, and R.~Pajarola, ``{TTHRESH}: Tensor
  compression for multidimensional visual data,'' \emph{IEEE Trans. on Vis. and
  Comp. Graph.}, vol.~26, no.~9, pp. 2891--2903, 2020.

\bibitem{Ainsworth2018}
M.~Ainsworth, O.~Tugluk, B.~Whitney, and S.~Klasky, ``Multilevel techniques for
  compression and reduction of scientific data--the univariate case,''
  \emph{Comput. Vis. Sci.}, vol.~19, no. 5-6, pp. 65--76, 2018.

\bibitem{Ainsworth2019}
------, ``Multilevel techniques for compression and reduction of scientific
  data---the multivariate case,'' \emph{SIAM Journal on Scientific Computing},
  vol.~41, no.~2, pp. A1278--A1303, 2019.

\bibitem{ZhaoSZ}
K.~Zhao, S.~Di, M.~Dmitriev, T.-L.~D. Tonellot, Z.~Chen, and F.~Cappello,
  ``Optimizing error-bounded lossy compression for scientific data by dynamic
  spline interpolation,'' in \emph{IEEE 37th Int. Conf. on Data Eng.}, 2021,
  pp. 1643--1654.

\bibitem{sbhp}
C.~Chrysafis, A.~Said, A.~Drukarev, A.~Islam, and W.~Pearlman, ``Sbhp-a low
  complexity wavelet coder,'' in \emph{IEEE Int. Conf. on Acoustics, Speech,
  and Signal Processing}, vol.~4, 2000, pp. 2035--2038.

\bibitem{speck}
W.~Pearlman, A.~Islam, N.~Nagaraj, and A.~Said, ``Efficient, low-complexity
  image coding with a set-partitioning embedded block coder,'' \emph{IEEE
  Trans. Circuits Sys. Video Tech.}, vol.~14, no.~11, pp. 1219--1235, 2004.

\bibitem{spiht}
A.~Said and W.~Pearlman, ``A new, fast, and efficient image codec based on set
  partitioning in hierarchical trees,'' \emph{IEEE Trans. Circuits Sys. Video
  Tech.}, vol.~6, no.~3, pp. 243--250, 1996.

\bibitem{jpeg2k}
A.~Skodras, C.~Christopoulos, and T.~Ebrahimi, ``The {JPEG} 2000 still image
  compression standard,'' \emph{IEEE Signal Process. Mag.}, vol.~18, no.~5, pp.
  36--58, 2001.

\bibitem{li2019vapor}
S.~Li, S.~Jaroszynski, S.~Pearse, L.~Orf, and J.~Clyne, ``Vapor: A
  visualization package tailored to analyze simulation data in earth system
  science,'' \emph{Atmosphere}, vol.~10, no.~9, p. 488, 2019.

\bibitem{mallat}
S.~Mallat, \emph{A wavelet tour of signal processing}.\hskip 1em plus 0.5em
  minus 0.4em\relax Academic Press, 2008.

\bibitem{Frisken2002}
S.~F. Frisken and R.~N. Perry, ``Simple and efficient traversal methods for
  quadtrees and octrees,'' \emph{J. Graphics Tools}, vol.~7, no.~3, pp. 1--11,
  2002.

\bibitem{Gargantini:1982}
I.~Gargantini, ``An effective way to represent quadtrees,'' \emph{Commun. ACM},
  vol.~25, no.~12, 1982.

\bibitem{Museth_2021_Nanovdb}
K.~Museth, ``Nanovdb: A gpu-friendly and portable vdb data structure for
  real-time rendering and simulation,'' in \emph{ACM SIGGRAPH Talks}, 2021.

\bibitem{PhysRevLett.113.155005}
F.~Guo, H.~Li, W.~Daughton, and Y.-H. Liu, ``Formation of hard power laws in
  the energetic particle spectra resulting from relativistic magnetic
  reconnection,'' \emph{Phys. Rev. Lett.}, vol. 113, 2014.

\bibitem{Cohen2002}
R.~H. Cohen, W.~P. Dannevik, A.~M. Dimits, D.~E. Eliason, A.~A. Mirin, Y.~Zhou,
  D.~H. Porter, and P.~R. Woodward, ``Three-dimensional simulation of a
  richtmyer–meshkov instability with a two-scale initial perturbation,''
  \emph{Phys. of Fluids}, vol.~14, no.~10, pp. 3692--3709, 2002.

\bibitem{Cabot:2006aa}
W.~H. Cabot and A.~W. Cook, ``Reynolds number effects on {Rayleigh}--{Taylor}
  instability with possible implications for type {Ia} supernovae,''
  \emph{Nature Phys.}, vol.~2, pp. 562--568, 2006.

\end{thebibliography}

% % ---------------------------------------------------------------------------
\newcommand \bionospace {\vskip -1.5\baselineskip plus -1fil}

% --------------------------------------------------------
\begin{IEEEbiography}
[{\includegraphics[width=1in,height=1.25in,clip,keepaspectratio]
{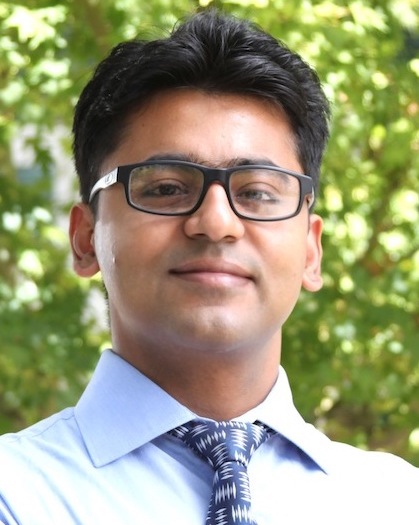}}] {Harsh Bhatia} 
is a Computer Scientist at the Center for Applied Scientific Computing at
Lawrence Livermore National Laboratory. His research spans broad areas of
visualization and computational topology, ML-based approaches, high-performance
computing, and workflow technologies, with special focus on scientific data.
Prior to joining LLNL, Harsh earned his Ph.D. from the Scientific Computing 
and Imaging Institute at The University of Utah in 2015, where he developed 
mathematical and combinatorial techniques for feature extraction for vector 
fields.
\end{IEEEbiography}
% --------------------------------------------------------
%\bionospace

% --------------------------------------------------------
\begin{IEEEbiography}
[{\includegraphics[width=1in,height=1.25in,clip,keepaspectratio]
{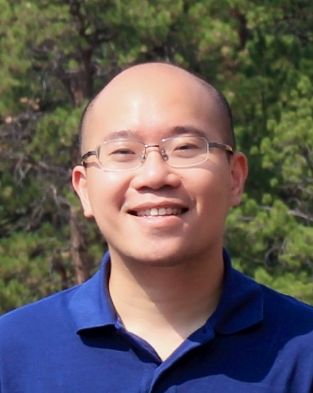}}] {Duong Hoang} 
is a Ph.D. student in computer science at the Scientific Computing and Imaging
(SCI) Institute, University of Utah. Before joining Utah, he obtained Bachelor
and Masters degrees in computer science from the National University of
Singapore. His research interests include data compression, scientific data
visualization and analysis, computer graphics and scientific computing.
\end{IEEEbiography}
% --------------------------------------------------------
\newpage

%
% --------------------------------------------------------
\begin{IEEEbiography}
[{\includegraphics[width=1in,height=1.25in,clip,keepaspectratio]
{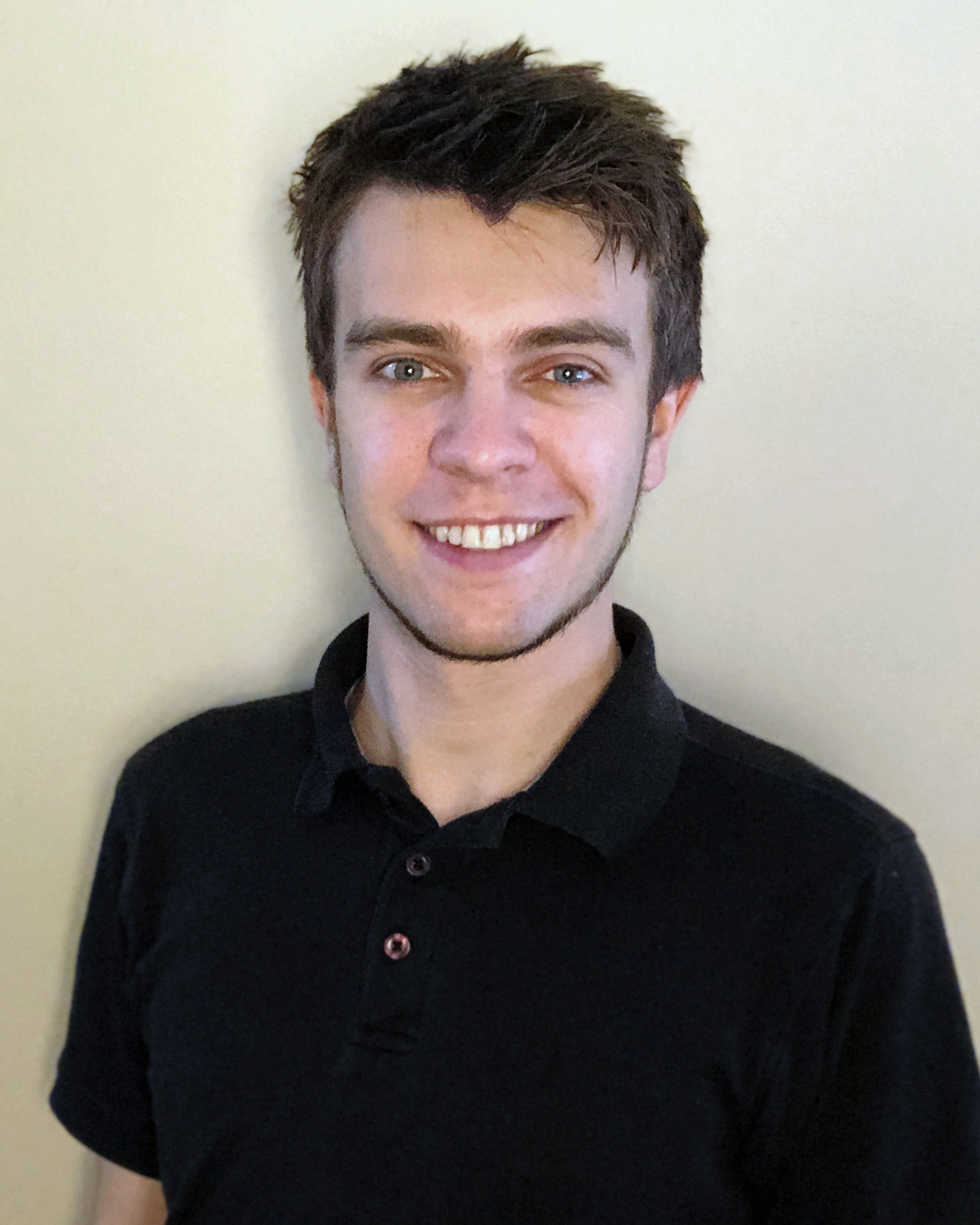}}] {Nate Morrical} 
is a Ph.D. student at the University of Utah, intern alumni from the NVIDIA
OptiX team and Pixar’s RenderMan group and a current member of the Scientific
Computing and Imaging Institute (SCI). His research interests include
high-performance ray-tracing frameworks and computing, scientific data
visualization, computational geometry, and real-time ray tracing. Before
joining SCI, Nate received his B.S. in computer science from Idaho State
University, where he researched interactive computer graphics and computational
geometry.
\end{IEEEbiography}
% --------------------------------------------------------
%\bionospace

%
% --------------------------------------------------------
\begin{IEEEbiography}
[{\includegraphics[width=1in,height=1.25in,clip,keepaspectratio]
{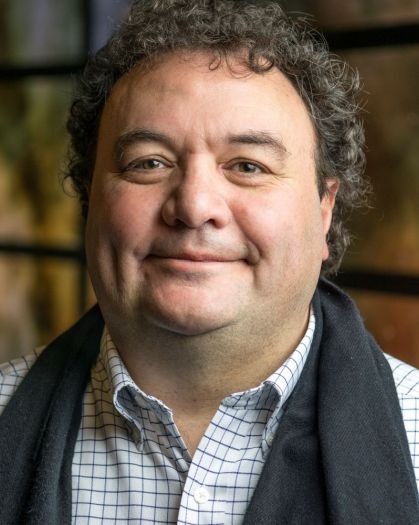}}] {Valerio Pascucci} 
received the EE laurea (master’s) degree from the University La Sapienza, Rome,
Italy, in December 1993, as a member of the Geometric Computing Group, and the
PhD degree in computer science from Purdue University, in May 2000. He is a
faculty member at the Scientific Computing and Imaging (SCI) Institute,
University of Utah. Before joining SCI, he served as a project leader at the
Lawrence Livermore National Laboratory (LLNL), Center for Applied Scientific
Computing (from May 2000) and as an adjunct professor at the Computer Science
Department of the University of California, Davis (from July 2005). Prior to
his tenure at CASC, he was a senior research associate at the University of
Texas at Austin, Center for Computational Visualization, CS, and TICAM
Departments.  He is a member of the IEEE.
\end{IEEEbiography}
% --------------------------------------------------------
%\bionospace

%
% --------------------------------------------------------
\begin{IEEEbiography}
[{\includegraphics[width=1in,height=1.25in,clip,keepaspectratio]
{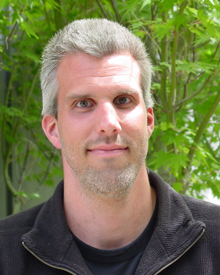}}]{Peer-Timo Bremer} 
is a member of technical staff and project leader at the Center for Applied
Scientific Computing (CASC) at the Lawrence Livermore National Laboratory
(LLNL) and Associated Director for Research at the Center for Extreme Data
Management, Analysis, and Visualization at the University of Utah. He also
serves on the research council of the Data Science Institute and as the ASCR
point of contact for Data science. His research interests include scientific
machine learning, large scale data analysis, topological techniques, and
visualization. Prior to his tenure at CASC, he was a postdoctoral research
associate at the University of Illinois, Urbana-Champaign. Peer-Timo earned a
Ph.D. in Computer science at the University of California, Davis in 2004 and a
Diploma in Mathematics and Computer Science from the Leibniz University in
Hannover, Germany in 2000. He is a member of the IEEE Computer Society and ACM.
\end{IEEEbiography}
% --------------------------------------------------------
%\bionospace

%
\begin{IEEEbiography}
[{\includegraphics[width=1in,height=1.25in,clip,keepaspectratio]
{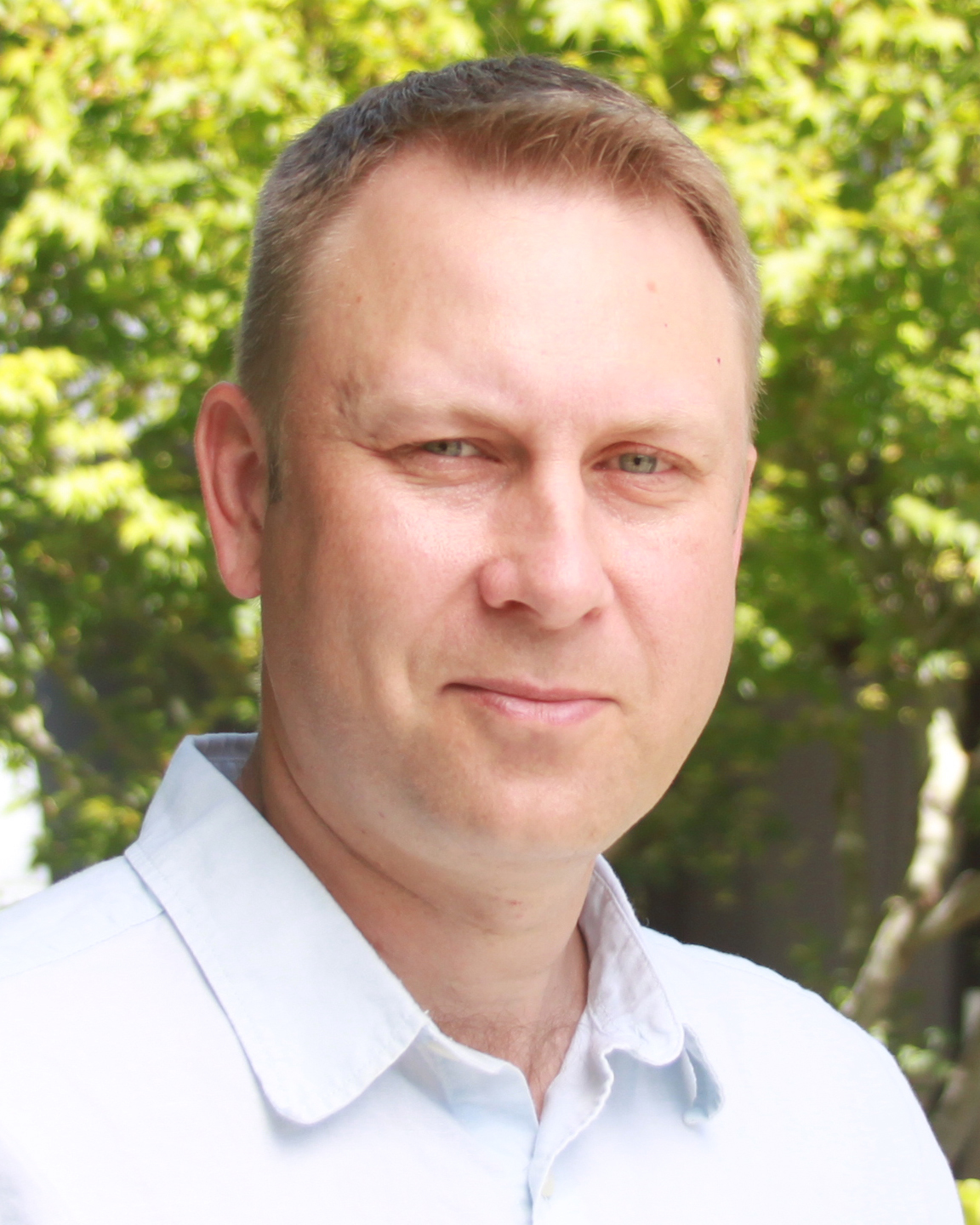}}]{Peter Lindstrom}
is a Computer Scientist at the Center for Applied Scientific Computing at
Lawrence Livermore National Laboratory, where he leads several research efforts
that span data compression, scientific visualization, and scientific computing.
He received a Ph.D. in computer science from Georgia Institute of Technology in
2000 and B.S. degrees in computer science, mathematics, and physics from Elon
University in 1994.  Peter previously served as Editor in Chief for Graphical
Models, Associate Editor for IEEE Transactions on Visualization and Computer
Graphics, and Paper Chair for IEEE VIS.  He is a Senior Member of IEEE.
\end{IEEEbiography}
% --------------------------------------------------------

% % biography section
% % 
% % If you have an EPS/PDF photo (graphicx package needed) extra braces are
% % needed around the contents of the optional argument to biography to prevent
% % the LaTeX parser from getting confused when it sees the complicated
% % \includegraphics command within an optional argument. (You could create
% % your own custom macro containing the \includegraphics command to make things
% % simpler here.)
% %\begin{IEEEbiography}[{\includegraphics[width=1in,height=1.25in,clip,keepaspectratio]{mshell}}]{Michael Shell}
% % or if you just want to reserve a space for a photo:

% \end{IEEEbiography}


% % insert where needed to balance the two columns on the last page with
% %\newpage


% % You can push biographies down or up by placing
% % a \vfill before or after them. The appropriate
% % use of \vfill depends on what kind of text is
% % on the last page and whether or not the columns
% % are being equalized.

% %\vfill

% % Can be used to pull up biographies so that the bottom of the last one
% % is flush with the other column.
% %\enlargethispage{-5in}
% ---------------------------------------------------------------------------

%\clearpage
%\appendix
%\section*{Supplementary Results}

\begin{figure*}[!t]
\centering
\includegraphics[width=0.48\linewidth]{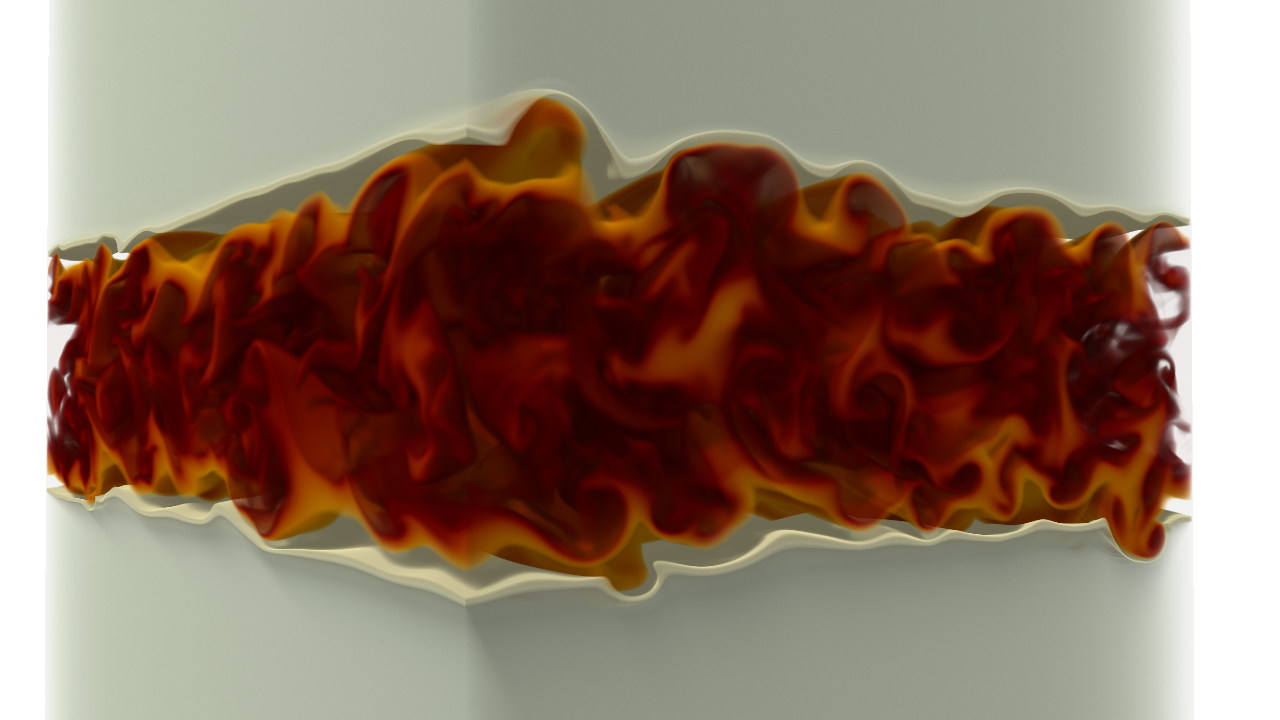}
\vspace{1em}

\includegraphics[width=0.48\linewidth]{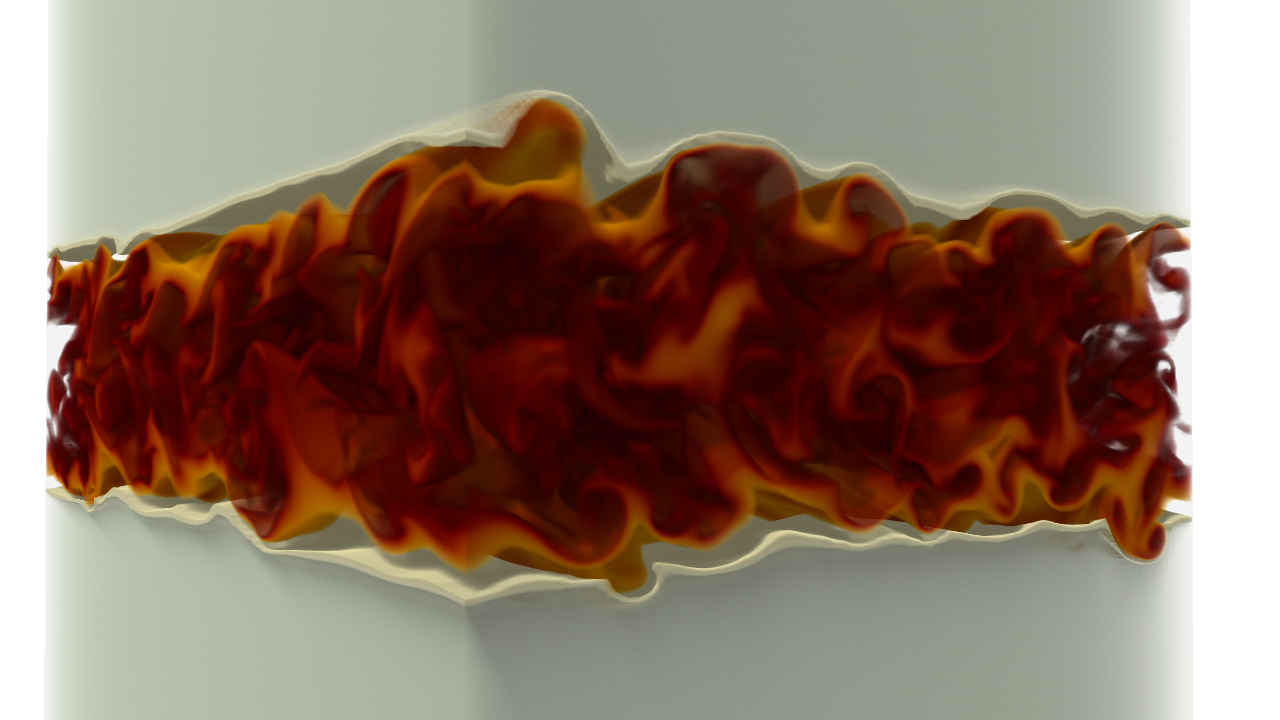}
\includegraphics[width=0.48\linewidth]{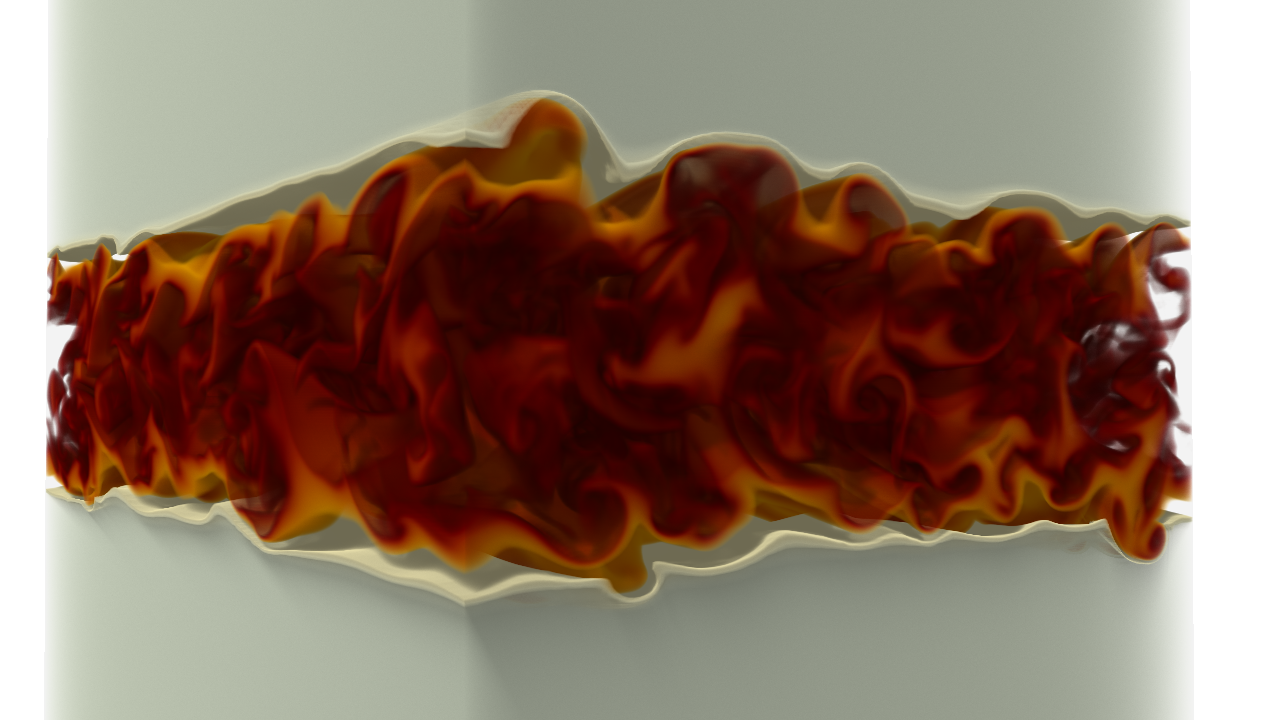}
\vspace{1em}

\includegraphics[width=0.48\linewidth]{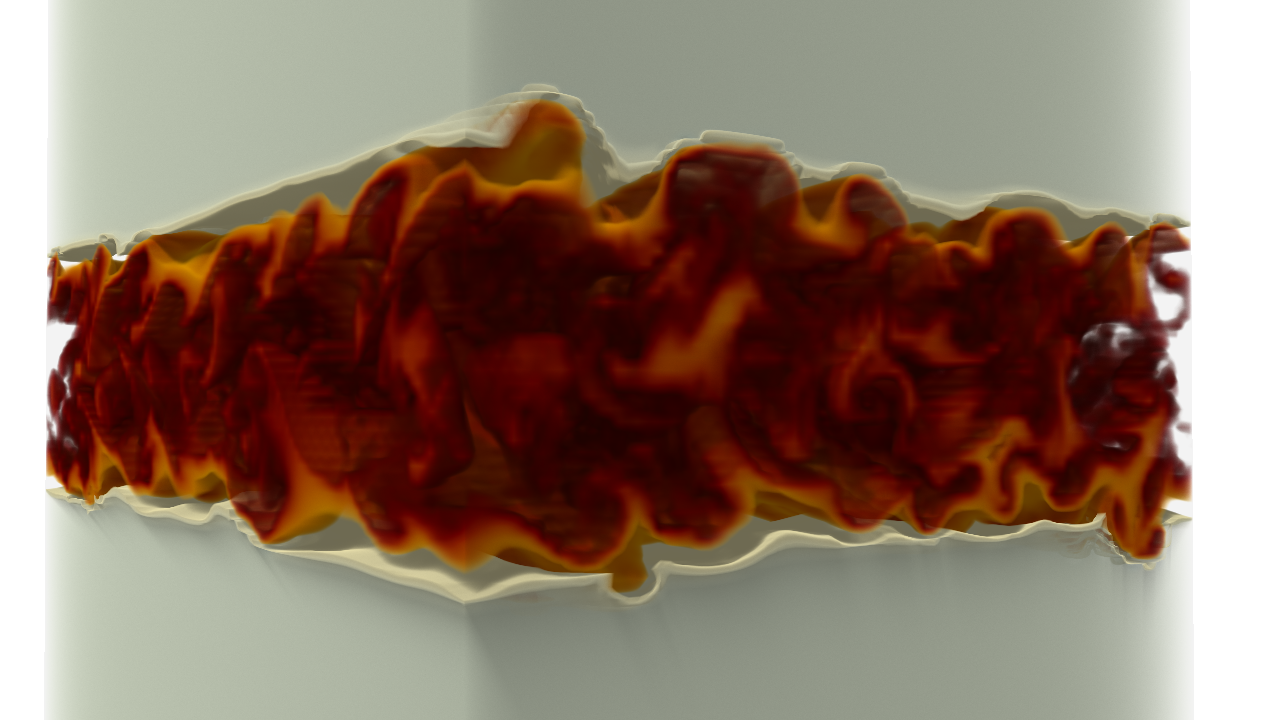}
\includegraphics[width=0.48\linewidth]{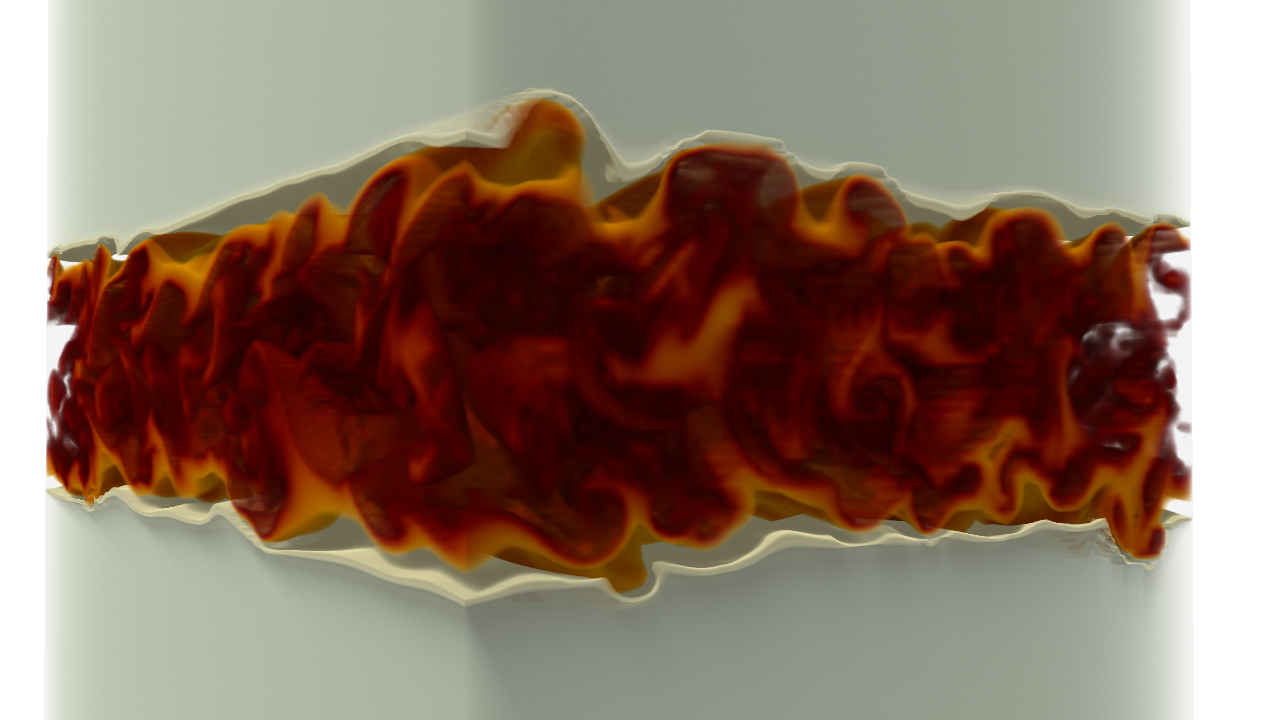}
\caption{Volume renderings of a turbulent jet flame dataset. The original data
at full resolution $[920 \times 1400 \times 72]$ is shown at the top. The
middle row shows the reduced meshes obtained after ingesting 8 MB and 16 MB,
respectively, using the \textit{by-coefficient-energy} stream. The bottom row
shows reduced meshes using the \textit{by-level-coeff} stream. Both streams
produce high quality visualizations, particularly after ingesting 16 MB.
\label{fig:app1}}
\end{figure*}

\end{document}